\documentclass[10pt,nopreprintline,authoryear]{elsarticle}
\usepackage[left=2.50cm, right=2.50cm, top=2.50cm, bottom=2.50cm]{geometry}
\usepackage{helvet}
\usepackage{amsmath, amsfonts, amssymb} 
\usepackage[english]{babel}
\usepackage{romannum}
\usepackage{array}
\newcolumntype{L}[1]{>{\raggedright\arraybackslash}p{#1}}

\usepackage{graphicx}   
\usepackage{caption} 
\usepackage{subcaption}
\usepackage{float}
 \usepackage{lineno}
\usepackage{color}

\usepackage{setspace}  
\usepackage{url}        
\usepackage{bm}         
\usepackage{multirow}
\usepackage{indentfirst}
\usepackage{booktabs}
\usepackage{algorithm}
\usepackage{algorithmic}
\usepackage{hyperref} 
\hypersetup{unicode, pdfauthor=author} 
\makeatletter
\def\ps@pprintTitle{%
  \let\@oddhead\@empty
  \let\@evenhead\@empty
  \let\@oddfoot\@empty
  \let\@evenfoot\@empty}
\makeatother

\usepackage{silence} 
\usepackage{cleveref}

\newcommand{\fref}[1]{Fig.~\ref{#1}}

\newcommand{\tref}[1]{Table~\ref{#1}}

\makeatletter
\renewcommand*\env@matrix[1][\arraystretch]{%
  \edef\arraystretch{#1}%
  \hskip -\arraycolsep
  \let\@ifnextchar\new@ifnextchar
  \array{*\c@MaxMatrixCols c}}
\makeatother

\begin{document}
\captionsetup[figure]{labelfont={bf},labelformat={default},labelsep=period,name={Fig.}}

\journal{Computers and Geotechnics}

\begin{frontmatter} 

\title{A 2.5D NURBS-Trace Infinite-Element Method for Moving-Load Wave Propagation and Soil--Structure Interaction in Semi-Infinite Ground}

\author[china1,china2]{Yanhui Zhong}
\author[china1,china2]{Hao Hong}
\author[china1,china2]{Bei Zhang}
\author[china1,china2]{Quansheng Zang\corref{cor1}}
\ead{qszang1991@zzu.edu.cn}
\author[Exeter]{Hussein Rappel}
\author[lux]{St\'ephane~P.~A.~Bordas}

\address[china1]{School of Water Conservancy and Transportation, Zhengzhou University, Zhengzhou 450001, China}
\address[china2]{National Local Joint Engineering Laboratory of Major Infrastructure Testing and Rehabilitation Technology, Zhengzhou 450001, China}
\address[Exeter]{Department of Engineering, Faculty of Environment, Science and Economy, University of Exeter, Exeter, UK}
\address[lux]{Department of Engineering, Faculty of Science, Technology and Medicine, University of Luxembourg, Luxembourg}

\cortext[cor1]{Corresponding author}


\date{}
\pagenumbering{arabic}
\begin{abstract}

For moving-load problems that are approximately invariant in geometry and material properties along the traveling direction, 2.5D analysis preserves all three displacement components while reducing computational cost relative to full three-dimensional discretization. This paper develops a 2.5D Non-Uniform Rational B-spline (NURBS)-trace infinite-element method (NBIEM) for wave propagation in linear viscoelastic semi-infinite geotechnical media within a coupled finite/infinite-element framework. The bounded near field is discretized by isogeometric analysis, whereas the exterior is represented by tensor products of the boundary NURBS basis and admissible outgoing or evanescent exponential radial functions. The two subdomains share the same discrete NURBS trace space and boundary control-point degrees of freedom, enforcing displacement continuity without projection or mortar variables. For the selected radial functions, the far-field stiffness and mass contributions are evaluated from closed-form radial moments, avoiding a finite radial cutoff and radial quadrature. Comparisons with closed-form half-space solutions verify the computed displacement responses in the sub-Rayleigh and super-compressional regimes and the stress responses in the intermediate super-shear but sub-compressional regime. Low-frequency studies assess the sensitivity of the selected exterior setting to radial parameters and artificial-boundary placement. Additional tests characterize complex-valued response accuracy, phase fidelity, computational cost, and the frequency-dependent working range of the default S-wave-informed exterior realization. Applications to layered media, track--subgrade systems, and buried structures demonstrate the method's ability to handle heterogeneous materials, multi-patch configurations, curved interfaces, and cover-depth-dependent geotechnical responses. The resulting framework provides a geometrically consistent and computationally efficient treatment of moving-load wave propagation and soil--structure interaction in semi-infinite domains.

\end{abstract}
            

\end{frontmatter}
\setcounter{page}{0}

\section{Introduction}
Ground vibration induced by moving loads is a persistent concern in transportation geotechnics and underground infrastructure engineering. Waves generated by high-speed trains, metro systems, heavy road traffic, and moving machinery can propagate through semi-infinite ground and affect the serviceability of adjacent buildings, sensitive facilities, pipelines, tunnels, and other buried structures. Field measurements have documented railway-induced ground vibration along operating railway lines \citep{CelebiKirtel-74,AlvesCostaCalcada-XX}. Road-traffic-induced ground vibration has been investigated through coupled vehicle--road and ground-response models \citep{MhannaSadekShahrour2012}. Underground-railway vibration and its transmission through the surrounding ground have also been examined numerically and analytically \citep{HeZhou-65,FengPaolucci-63}. For tunnels embedded in layered ground, analytical and theoretical models have clarified the effects of stratification, tunnel geometry, and moving excitation \citep{HeZhouDiGuoXiao2018,HuLiDengXie2020}.

Reliable prediction is therefore needed at the design and assessment stages. The modeling task is demanding because the response is intrinsically three-dimensional, whereas the ground domain is unbounded and must satisfy radiation conditions at infinity. For systems that are approximately invariant along the traveling direction, the 2.5D formulation provides an effective compromise between full three-dimensional discretization and overly restrictive two-dimensional plane-strain models. By applying a Fourier transform along the longitudinal direction, the moving-load problem is reduced to a sequence of two-dimensional cross-sectional problems in the wavenumber--frequency domain. Analytical formulations based on transfer matrices and cylindrical-to-plane wave transformations provide efficient solutions for circular tunnels in horizontally layered half-spaces \citep{HeZhouDiGuoXiao2018}. Related layered-ground models have also been developed for tunnel systems subjected to underground moving loads \citep{HuLiDengXie2020}. Their geometric and material assumptions, however, differ from those of general cross-sectional discretizations.

The treatment of the unbounded exterior domain is a central issue in 2.5D numerical modeling. In classical finite/infinite-element formulations, the near field is modeled by finite elements and the far field by infinite elements containing wave-propagation and decay functions \citep{YangLiu-56}. This construction is compact, but its accuracy depends on the selected radial approximation and on a finite/infinite-element discretization compatible with the frequency-domain formulation.
Alternative treatments enforce radiation through boundary coupling. Such coupling has been used in 2.5D models for layered half-spaces, tunnel vibration, and underground railway systems \citep{FrancoisSchevenels-57,JinThompson-XX,GhangaleArcos-76}. For saturated poroelastic ground, \citet{HeZhouDiShan2017} developed a regularized 2.5D FE--BE formulation for the dynamic interaction between longitudinally invariant structures and the surrounding soil, with the finite-element and boundary-element subdomains coupled through mechanical and hydraulic interface conditions. \citet{LiraviArcosGhangaleNooriRomeu2021} combined a local 2.5D FEM--BEM soil--structure model with an MFS-based field-recovery stage, using the interface response obtained from FEM--BEM to evaluate wave propagation at multiple soil observation points. More recently, \citet{LiraviAlizadehshirazKaewunruenNinic2026} proposed a wavenumber--frequency-domain FEM--SBM formulation, where SBM denotes the Singular Boundary Method, for the dynamic interaction of multiple structures and soil.

Scaled-boundary finite elements provide a complementary treatment and have been used for 2.5D ground-vibration analysis \citep{YaseriBazyarJavady-XX}. Recent applications have extended scaled-boundary formulations to three-dimensional coupled systems involving complex geometries and unbounded exterior domains. \citet{Wang2026Hydrodynamic} applied a three-dimensional SBFEM framework to hydrodynamic forces on circular piers in V-shaped deep-water canyons. For large-scale dam--reservoir--foundation interaction, \citet{Liu2026ImplicitExplicit} combined octree discretization with implicit--explicit time stepping. In another infinite-domain dynamic setting, \citet{Gao2025InfiniteFluid} investigated the free vibration of a laminated plate coupled with an infinite-domain fluid using a scaling-surface SBFEM based on boundary discretization and an analytical radial representation.

Perfectly matched layers provide a distinct absorbing-layer treatment for unbounded wave problems \citep{PledDesceliers-79,FrancoisSchevenelsPML-XX}. More recently, \citet{Zhang2026FluidSolid} developed a fluid--solid coupled SBPML formulation for transient-wave problems in unbounded media. Beyond elastodynamic applications, \citet{Huang2025ComplexHalfSpace} extended a modified SBFEM framework to steady-state seepage in complex geotechnical half-spaces with heterogeneous and multi-domain configurations. Collectively, these developments illustrate different strategies for representing unbounded media, coupled interfaces, and complex half-space geometries. For moving-load elastodynamics, the principal distinctions lie in the exterior approximation, interface coupling, radiation treatment, and computational cost of repeated wavenumber--frequency solutions.

Spline-based discretizations are relevant to this setting because NURBS basis functions can represent geometry and unknown fields within the same approximation space. Since the introduction of isogeometric analysis by \citet{HughesCottrell-61}, spline discretizations have been used in structural-vibration analysis \citep{CottrellReali2006Vibrations}. Their performance in time-harmonic acoustic problems has also been investigated \citep{KhajahAntoineBordas2019Acoustic}. High-order pollution and dispersion properties have received particular attention in wave-propagation studies \citep{DiwanMohamed2019Pollution,DengBehnoudfar-78}.
In frequency-domain geotechnical applications, geometric consistency must be considered together with phase fidelity, accumulated dispersion or pollution effects, boundary truncation, and spectral sampling. These issues are especially relevant for curved tunnel boundaries, layered interfaces, phase-sensitive propagation, and stress gradients near load paths or embedded structures.

Spline-based treatments of unbounded domains have also been developed in related settings. \citet{Beer2015MappedInfinite} introduced mapped infinite NURBS patches for isogeometric boundary-element analysis in geomechanics. In wave-propagation applications, \citet{VenasKvamsdalJenserud-XX} coupled isogeometric analysis with infinite elements for acoustic scattering. Isogeometric PML formulations have further demonstrated the compatibility of spline approximation with artificial absorbing layers in two-dimensional transient elastodynamics \citep{ChauBrun-68}. In the specific context of 2.5D railway-induced soil vibration, \citet{BJie-72} proposed an IGA/SBIGA formulation in which the bounded domain is discretized by IGA and the unbounded exterior by scaled-boundary isogeometric analysis.

Existing studies have therefore developed several 2.5D exterior-coupling strategies for longitudinally invariant geotechnical systems, including FE--BE, FEM--BEM--MFS, FEM--SBM based on the Singular Boundary Method, scaled-boundary, infinite-element, and absorbing-layer formulations. Separately, spline-based infinite patches, infinite elements, PMLs, and scaled-boundary discretizations have shown that NURBS representations can be used in unbounded-domain analysis. However, a domain-based NURBS-trace infinite-element implementation for three-component 2.5D geotechnical elastodynamics, in which the near- and far-field approximations share the same discrete NURBS trace and the selected radial matrix moments are evaluated in closed form, has not been systematically characterized for moving-load problems.

Building on established 2.5D exterior-coupling methods and previous spline-based treatments of unbounded domains, this study makes three main contributions. First, a three-component 2.5D Galerkin NURBS-trace infinite-element formulation is developed in which the bounded IGA domain and the semi-infinite exterior share an identical discrete NURBS trace and boundary control-point unknowns. The resulting domain-based elastodynamic weak form retains the full longitudinal-wavenumber-dependent strain and inertia coupling among the three displacement components. Second, for the selected outgoing or evanescent exponential radial functions, the moments entering the far-field stiffness and mass matrices are evaluated in closed form, avoiding a finite radial cutoff and radial numerical quadrature. Third, the formulation is systematically characterized through analytical displacement and stress benchmarks, low-frequency sensitivity studies of the selected radial realization, and direct comparisons with FEM/IEM and IGA/SBIGA in terms of amplitude, phase, and computational cost.

Subsequent applications to layered ground, track--subgrade systems, and buried structures demonstrate how the implementation handles material interfaces, multi-patch geometries, and curved internal boundaries. The remainder of the paper is organized as follows. Section~\ref{BasisFormulations} states the 2.5D continuum setting and moving-load convention. Section~\ref{IGAFormulation} develops the bounded IGA discretization and the proposed NURBS-trace infinite elements. Section~\ref{numerexamp} provides verification studies, boundary-sensitivity assessments, method comparisons, and numerical applications. Section~\ref{Discussion} discusses methodological implications and applicability, and Section~\ref{Conclusions} summarizes the main findings.
\section{2.5D continuum formulation under moving loads}  \label{BasisFormulations}
\subsection{Governing equations} \label{GoverningEq}
For the three-dimensional soil domain (Fig.~1), let $\mathbf{x}=(x,y,z)\in\Omega\subset\mathbb{R}^3$ denote the position vector and $t$ the time, and let $u_i(\mathbf{x},t)$ be the displacement components ($i=1,2,3$). The coordinate system is chosen such that $x$ is lateral, the positive $y$-direction is downward, and $z$ is the traveling direction. 
\begin{figure}[H]    
\centering    
\includegraphics[width=3in]{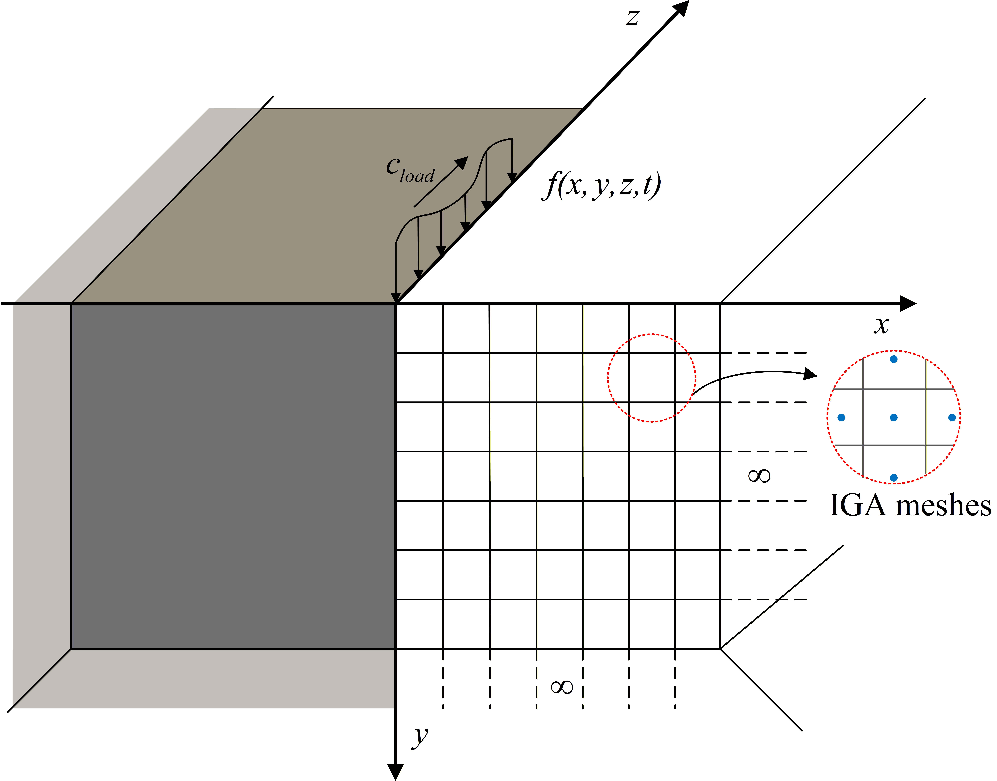}    
\caption{Problem definition for the three-dimensional semi-infinite soil domain under a moving surface load, including the coordinate system, traveling direction $z$, and cross-sectional discretization used in the 2.5D formulation.}
\label{fig1}
\end{figure}

The equations of motion are written as
\begin{equation}
\sigma_{ij,j}(\mathbf{x},t)+f_i(\mathbf{x},t)=\rho\,\ddot{u}_i(\mathbf{x},t),
\qquad i,j=1,2,3 ...
\label{eq:G1}
\end{equation}
where $\sigma_{ij}$ are the Cauchy stress components, $f_i$ are the body-force components per unit volume (often neglected for moving surface or tunnel excitations), $\rho$ is the mass density, $\sigma_{ij,j}$ denotes $\partial(\sigma_{ij})/\partial x_j$, and overdots denote partial derivatives with respect to time. Under the small-strain assumption, the infinitesimal strain tensor is defined by
\begin{equation}
\varepsilon_{ij}(\mathbf{x},t)=\frac{1}{2}\left(u_{i,j}(\mathbf{x},t)+u_{j,i}(\mathbf{x},t)\right)
\label{eq:G2}
\end{equation}
The linear constitutive law is expressed as
\begin{equation}
\sigma_{ij}(\mathbf{x},t)=D_{ijkl}^\ast\,\varepsilon_{kl}(\mathbf{x},t)
\label{eq:G3}
\end{equation}
where $D_{ijkl}^\ast$ is the fourth-order complex constitutive tensor. In this study, material dissipation is represented through complex moduli (hysteretic damping). Specifically, the complex Young's modulus is taken as
\begin{equation}
E^\ast = E\left(1+2\mathrm{i}\zeta\right)
\label{eq:E_complex}
\end{equation}
where $\zeta$ is a dimensionless loss parameter and $\mathrm{i}=\sqrt{-1}$. For an isotropic linear viscoelastic solid with hysteretic damping, the complex Lam\'e parameters are
\begin{equation}
\lambda^\ast=\frac{\nu E^\ast}{(1+\nu)(1-2\nu)},
\qquad
G^\ast=\frac{E^\ast}{2(1+\nu)}
\label{eq:lame_complex}
\end{equation}
where $\nu$ is Poisson's ratio. The constitutive tensor then simplifies to
\begin{equation}
D_{ijkl}^\ast=\lambda^\ast\,\delta_{ij}\delta_{kl}+G^\ast(\delta_{ik}\delta_{jl}+\delta_{il}\delta_{jk})       
\label{eq:D_tensor_iso} 
\end{equation}
where $\delta_{ij}$ is the Kronecker delta. Substituting Eqs. \eqref{eq:G2} and \eqref{eq:D_tensor_iso} into Eq. \eqref{eq:G1} yields the displacement-based Navier equation with complex moduli,
\begin{equation}
G^\ast\,u_{i,jj}(\mathbf{x},t)+(\lambda^\ast+G^\ast)\,u_{k,ki}(\mathbf{x},t)+f_i(\mathbf{x},t)=\rho\,\ddot{u}_i(\mathbf{x},t),
\qquad i=1,2,3 ...
\label{eq:Navier}
\end{equation}

The boundary $\Gamma=\partial\Omega$ is decomposed into $\Gamma_u$ and $\Gamma_t$ such that $\Gamma=\Gamma_u\cup\Gamma_t$ and $\Gamma_u\cap\Gamma_t=\varnothing$. The Dirichlet and Neumann boundary conditions are prescribed as
\begin{equation}
u_i(\mathbf{x},t)=\bar{u}_i(\mathbf{x},t)\ \text{on}\ \Gamma_u,
\qquad
\sigma_{ij}(\mathbf{x},t)\,n_j(\mathbf{x})=\bar{t}_i(\mathbf{x},t)\ \text{on}\ \Gamma_t
\label{eq:BCs}
\end{equation}
where $n_j$ is the outward unit normal to $\Gamma$, and $\bar{u}_i$ and $\bar{t}_i$ are the prescribed displacement and traction components, respectively. The initial conditions are
\begin{equation}
u_i(\mathbf{x},0)=u_i^0(\mathbf{x}),
\qquad
\dot{u}_i(\mathbf{x},0)=v_i^0(\mathbf{x})
\qquad \text{in }\Omega
\label{eq:ICs}
\end{equation}
where $u_i^0$ and $v_i^0$ denote the initial displacement and velocity fields. For a semi-infinite soil domain, the solution must additionally satisfy radiation or decay conditions as $\|\mathbf{x}\|\to\infty$, which will be addressed in the unbounded-domain treatment detailed in Section~\ref{IGAFormulation}. 

\subsection{2.5D reduction using Fourier transform and wavenumber-domain equations}

The transformation from a full three-dimensional problem to a 2.5D formulation relies on the geometric invariance of the road--soil system along the traveling direction ($z$-axis). Since the material properties and cross-sectional geometry do not vary with $z$, the linear viscoelastic dynamic response can be effectively decomposed using spectral methods.

We adopt a Fourier transform approach to decouple the spatial coordinate $z$ and time $t$. The displacement field $u_i(\mathbf{x},t)$ is expressed as a superposition of harmonic waves. Assuming time-harmonic dependence $e^{\mathrm{i}\omega t}$ and spatial dependence $e^{-\mathrm{i}kz}$, the displacement ansatz is

\begin{equation}
u_i(x,y,z,t)=\hat{u}_i(x,y;k,\omega)\,e^{-\mathrm{i}k z}\,e^{\mathrm{i}\omega t},
\qquad i=1,2,3 ...
\label{eq:2p5d_ansatz}
\end{equation}
Substitution of Eq. \eqref{eq:2p5d_ansatz} into the governing equations derived in Section \ref{GoverningEq} reduces the three-dimensional problem to a family of two-dimensional boundary-value problems on the $(x,y)$ cross-section, parameterized by $(k,\omega)$. Specifically, the differential operators are mapped to the wavenumber--frequency domain as follows:
\begin{equation}
\frac{\partial(\cdot)}{\partial z}\rightarrow -\mathrm{i}k(\cdot),\qquad
\frac{\partial^2(\cdot)}{\partial z^2}\rightarrow -k^2(\cdot),\qquad
\frac{\partial^2(\cdot)}{\partial t^2}\rightarrow -\omega^2(\cdot)
\label{eq:2p5d_map}
\end{equation}
This transformation reduces the original 3D problem to a family of 2D boundary value problems defined over the cross-section $(x,y)$, reducing the computational burden relative to full three-dimensional discretization.

To retain consistency with this 2.5D framework, the moving loads must also be transformed into the wavenumber--frequency domain. The excitation caused by a convoy of vehicles moving at a constant moving-load speed $c$ is typically modeled as a separable function of space and time. Consider a general load component defined by
\begin{equation}
F(x,y,z,t)=\psi(x,y)\,\phi(z-ct)\sum_{r=1}^{n} T_{r}\,e^{\mathrm{i}\omega_{r} t}
\label{eq:moving_load_general}
\end{equation}
where $\psi(x,y)$ represents the transverse distribution of the load, $\phi(z-ct)$ describes the longitudinal footprint moving with the vehicle, and $T_r$ denotes the amplitude of the $r$-th harmonic component with angular frequency $\omega_r$.

For clarity, we derive the transformation for a single harmonic component $(T, \omega_0)$. Here $\omega=2\pi f$ is the analysis angular frequency and $\omega_0=2\pi f_0$ is the load-oscillation angular frequency. The temporal Fourier transform is formally defined as:
\begin{equation}
\tilde{F}(x,y,z,\omega)=\frac{1}{2\pi}\int_{-\infty}^{+\infty}F(x,y,z,t)\,e^{-\mathrm{i}\omega t}\,\mathrm{d}t
\label{eq:FT_time_def}
\end{equation}
Substituting the load expression into Eq. \eqref{eq:FT_time_def} leads to the following.
\begin{equation}
\tilde{F}(x,y,z,\omega)=\frac{T}{2\pi}\,\psi(x,y)\int_{-\infty}^{+\infty}\phi(z-ct)\,e^{-\mathrm{i}(\omega-\omega_0)t}\,\mathrm{d}t
\label{eq:FT_step1}
\end{equation}
To evaluate this integral, it is convenient to switch to a moving coordinate system. Introducing the change of variables $\tau=z-ct$, with $t = (z-\tau)/c, \qquad dt = -d\tau/c$, separates the oscillating term from the moving frame. The substitution reverses the integration limits, canceling the negative sign from the differential. The expression thus becomes:
\begin{equation}
\tilde{F}(x,y,z,\omega)=\frac{T}{c}\,\psi(x,y)\,e^{-\mathrm{i}(\omega-\omega_0)z/c}
\left[ \frac{1}{2\pi}\int_{-\infty}^{+\infty}\phi(\tau)\,e^{\mathrm{i}(\omega-\omega_0)\tau/c}\,\mathrm{d}\tau \right]
\label{eq:FT_step2}
\end{equation}
The bracketed term is the spatial Fourier transform of the longitudinal load function. By defining the equivalent longitudinal wavenumber $k$ as
\begin{equation}
k=\frac{\omega-\omega_0}{c}
\label{eq:k_def}
\end{equation}
and letting $\tilde{\phi}(k)$ denote the spatial transform pair of $\phi(\tau)$,
\begin{equation}
\tilde{\phi}(k)=\frac{1}{2\pi}\int_{-\infty}^{+\infty}\phi(\tau)\,e^{-\mathrm{i}k\tau}\,\mathrm{d}\tau
\label{eq:phi_FT_def}
\end{equation}
Eq. \eqref{eq:FT_step2} simplifies to a compact form:
\begin{equation}
\tilde{F}(x,y,z,\omega)=\frac{T}{c}\,\psi(x,y)\,e^{-\mathrm{i}k z}\,\tilde{\phi}(-k)
\label{eq:load_freq_domain_final}
\end{equation}
Eqs. \eqref{eq:k_def} and \eqref{eq:load_freq_domain_final} show a fundamental characteristic of the moving load problem: the temporal frequency $\omega$ and the longitudinal wavenumber $k$ are not independent but are strictly coupled through the vehicle speed $c$. This coupling represents the Doppler shift effect inherent in moving source problems. Consequently, for a given load-oscillation angular frequency $\omega_0$ and speed $c$, the response only needs to be computed at specific wavenumbers, significantly reducing the computational cost compared to a full 3D analysis. The total response for the multi-harmonic excitation in Eq. \eqref{eq:moving_load_general} is then obtained by linear superposition of these individual components.

Accordingly, a linear time history can be reconstructed by solving the required frequency--wavenumber components and applying inverse Fourier synthesis to the resulting complex responses. The longitudinal wavenumber is not sampled independently, because each frequency uses the wavenumber defined by Eq.~\eqref{eq:k_def}.

\section{Isogeometric formulation for \texorpdfstring{the}{the} 2.5D moving-load problem}\label{IGAFormulation}
\subsection{NURBS representation of the cross-section and displacement field}
The isogeometric construction is performed on the $(x,y)$ cross-section $\Omega_{2D}$; the longitudinal direction is represented by the wavenumber $k$. The B-spline and NURBS definitions needed for the geometry and displacement spaces are summarized below to fix the notation used in the proposed infinite-element construction.

Let the open knot vector be
\begin{equation}
\boldsymbol{\Xi}=\{\xi_1,\xi_2,\ldots,\xi_{n_b+p+1}\}
\end{equation}
where $p$ is the polynomial degree and $n_b$ is the number of one-dimensional basis functions. The Cox--de Boor recursion starts from
\begin{equation}
B_{i,0}(\xi)=
\begin{cases}
1, & \xi_i\le \xi<\xi_{i+1}\\
0, & \text{otherwise}
\end{cases}
\label{eq:Bspline_p0}
\end{equation}
and for $p\ge 1$,
\begin{equation}
B_{i,p}(\xi)=\frac{\xi-\xi_i}{\xi_{i+p}-\xi_i}\,B_{i,p-1}(\xi)
+\frac{\xi_{i+p+1}-\xi}{\xi_{i+p+1}-\xi_{i+1}}\,B_{i+1,p-1}(\xi)
\label{eq:Bspline_recursion}
\end{equation}
Non-empty knot spans define the elements. A knot of multiplicity $m$ gives $C^{p-m}$ continuity; endpoint multiplicity $p+1$ makes an open-knot B-spline basis interpolatory at both ends.

Given positive weights $\{w_i\}_{i=1}^{n_b}$, the rational extension used to represent conic sections is
\begin{equation}
N_{i,p}(\xi)=\frac{w_i B_{i,p}(\xi)}{\displaystyle\sum_{\hat{i}=1}^{n_b} w_{\hat{i}} B_{\hat{i},p}(\xi)}
\label{eq:NURBS_1D}
\end{equation}
A NURBS curve with control points $\mathbf{P}_i$ is
\begin{equation}
\mathbf{S}(\xi)=\sum_{i=1}^{n_b} N_{i,p}(\xi)\,\mathbf{P}_i
\label{eq:NURBS_curve}
\end{equation}

For the cross-section, tensor products in the parametric coordinates $(\xi,\eta)$ give the geometric mapping
\begin{equation}
\mathbf{x}_{\perp}(\xi,\eta)=\sum_{A=1}^{n_{cp}} R_A(\xi,\eta)\,\mathbf{P}_A,
\qquad \mathbf{x}_{\perp}=(x,y)
\label{eq:geom_map}
\end{equation}
where $R_A(\xi,\eta)$ are the bivariate NURBS basis functions, $\mathbf{P}_A$ are the control points, and $n_{cp}$ is their total number. The same $R_A$ are used for the displacement approximation.

As in other wave discretizations, the element scale must resolve the smallest wavelength of interest to control dispersion and phase error. Spline order and continuity can be selected independently, while repeated knots reduce continuity at corners and material interfaces. These properties are valuable because the cross-sectional system is solved repeatedly over the required $(k,\omega)$ pairs.

Knot repetition changes continuity but does not make tensor-product refinement fully local. The present implementation therefore uses compatible multi-patch refinement to concentrate resolution near load paths, material interfaces, and curved internal boundaries while preserving a conforming NURBS trace on artificial-boundary segments.

The resulting displacement interpolation, geometric Jacobian, and physical derivatives are stated together with the 2.5D strain operator in the following subsection; standard NURBS differentiation follows \citet{HughesCottrell-61}.

\subsection{Isogeometric discretization of the bounded cross-section in the wavenumber--frequency domain}
\label{subsec:IGA_bounded}

For each prescribed pair $(k,\omega)$ arising from the 2.5D reduction in Section~\ref{BasisFormulations}, the governing equations are posed on the bounded cross-section $\Omega_{2D}\subset\mathbb{R}^2$ with physical coordinates $\mathbf{x}_\perp$. An isogeometric discretization is adopted so that the cross-sectional geometry and the displacement amplitudes are represented by the same spline basis. This is particularly attractive in wave propagation analysis, where higher-order smooth bases can alleviate dispersion and pollution errors, and where knot multiplicity provides a direct mechanism to control continuity when geometric corners or material interfaces must be represented.

Let $(\xi,\eta)\in\widehat{\Omega}$ be the parametric coordinates of the NURBS surface defining the cross-section. For a given $(k,\omega)$, the displacement amplitudes are approximated as
\begin{equation}
\hat{\mathbf{u}}(\mathbf{x}_\perp;k,\omega)
=
\sum_{A=1}^{n_{cp}} R_A(\xi,\eta)\,\mathbf{d}_A(k,\omega),
\qquad
\mathbf{d}_A=
\begin{bmatrix}
d_{xA} & d_{yA} & d_{zA}
\end{bmatrix}^{T}
\label{eq:IGA_u_interp}
\end{equation}
where $\hat{\mathbf{u}}=[\hat{u}_x,\hat{u}_y,\hat{u}_z]^T$ and $\mathbf{d}_A$ collects the three displacement degrees of freedom at control point $A$. Introducing the global vector $\mathbf{d}$ by stacking all $\mathbf{d}_A$, the approximation takes the matrix form
\begin{equation}
\hat{\mathbf{u}}=\mathbf{N}\mathbf{d},
\qquad
\mathbf{N}=
\begin{bmatrix}
R_1\mathbf{I}_3 & R_2\mathbf{I}_3 & \cdots & R_{n_{cp}}\mathbf{I}_3
\end{bmatrix}
\label{eq:IGA_N_matrix}
\end{equation}
where $\mathbf{I}_3$ is the $3\times 3$ identity matrix.

Spatial derivatives with respect to physical coordinates follow from parametric derivatives through the Jacobian of the geometric mapping. The Jacobian matrix is defined as
\begin{equation}
\mathbf{J}(\xi,\eta)=
\begin{bmatrix}
\displaystyle \frac{\partial x}{\partial \xi} & \displaystyle \frac{\partial x}{\partial \eta} \\[6pt]
\displaystyle \frac{\partial y}{\partial \xi} & \displaystyle \frac{\partial y}{\partial \eta}
\end{bmatrix},
\qquad
J=\det\mathbf{J}
\label{eq:IGA_Jacobian}
\end{equation}
Accordingly, for each basis function $R_A$, the physical derivatives are obtained as
\begin{equation}
\begin{bmatrix}
\displaystyle \frac{\partial R_A}{\partial x} \\[6pt]
\displaystyle \frac{\partial R_A}{\partial y}
\end{bmatrix}
=
\mathbf{J}^{-T}
\begin{bmatrix}
\displaystyle \frac{\partial R_A}{\partial \xi} \\[6pt]
\displaystyle \frac{\partial R_A}{\partial \eta}
\end{bmatrix}
\label{eq:IGA_grad_transform}
\end{equation}

With the harmonic dependence $e^{-\mathrm{i}k z}e^{\mathrm{i}\omega t}$, differentiation with respect to $z$ is replaced by multiplication with $-\mathrm{i}k$. Using engineering strains, the strain vector is defined as:
\begin{equation}
\boldsymbol{\varepsilon}=
\begin{bmatrix}
\varepsilon_{xx} & \varepsilon_{yy} & \varepsilon_{zz} &
\gamma_{xy} & \gamma_{yz} & \gamma_{xz}
\end{bmatrix}^{T}
\label{eq:strain_vector_def}
\end{equation}
with components
\begin{equation}
\varepsilon_{xx}=\frac{\partial \hat{u}_x}{\partial x},\quad
\varepsilon_{yy}=\frac{\partial \hat{u}_y}{\partial y},\quad
\varepsilon_{zz}=-\mathrm{i}k\,\hat{u}_z,\quad
\gamma_{xy}=\frac{\partial \hat{u}_x}{\partial y}+\frac{\partial \hat{u}_y}{\partial x},\quad
\gamma_{yz}=\frac{\partial \hat{u}_z}{\partial y}-\mathrm{i}k\,\hat{u}_y,\quad
\gamma_{xz}=\frac{\partial \hat{u}_z}{\partial x}-\mathrm{i}k\,\hat{u}_x
\label{eq:2p5d_strains}
\end{equation}
Substituting Eq. \eqref{eq:IGA_u_interp} into Eq. \eqref{eq:2p5d_strains} yields
\begin{equation}
\boldsymbol{\varepsilon}=\mathbf{B}(k)\mathbf{d},
\qquad
\mathbf{B}(k)=
\begin{bmatrix}
\mathbf{B}_1(k) & \mathbf{B}_2(k) & \cdots & \mathbf{B}_{n_{cp}}(k)
\end{bmatrix}
\label{eq:B_global}
\end{equation}
where the $6\times 3$ block associated with control point $A$ is
\begin{equation}
\mathbf{B}_A(k)=
\begin{bmatrix}
R_{A,x} & 0        & 0 \\
0        & R_{A,y} & 0 \\
0        & 0        & -\mathrm{i}k\,R_A \\
R_{A,y} & R_{A,x} & 0 \\
0        & -\mathrm{i}k\,R_A & R_{A,y} \\
-\mathrm{i}k\,R_A & 0        & R_{A,x}
\end{bmatrix}
\label{eq:B_block}
\end{equation}
with $R_{A,x}=\partial R_A/\partial x$ and $R_{A,y}=\partial R_A/\partial y$ computed from Eq. \eqref{eq:IGA_grad_transform}.

In the wavenumber--frequency domain, material dissipation is represented through complex-valued moduli. The stress--strain relation is written in Voigt form as
\begin{equation}
\boldsymbol{\sigma}=
\mathbf{D}^\ast \boldsymbol{\varepsilon},
\qquad
\boldsymbol{\sigma}=
\begin{bmatrix}
\sigma_{xx} & \sigma_{yy} & \sigma_{zz} & \tau_{xy} & \tau_{yz} & \tau_{xz}
\end{bmatrix}^{T}
\label{eq:constitutive_voigt}
\end{equation}
with
\begin{equation}
\mathbf{D}^\ast=
\begin{bmatrix}
\lambda^\ast+2G^\ast & \lambda^\ast       & \lambda^\ast       & 0     & 0     & 0 \\
\lambda^\ast       & \lambda^\ast+2G^\ast & \lambda^\ast       & 0     & 0     & 0 \\
\lambda^\ast       & \lambda^\ast       & \lambda^\ast+2G^\ast & 0     & 0     & 0 \\
0                  & 0                  & 0                  & G^\ast & 0     & 0 \\
0                  & 0                  & 0                  & 0     & G^\ast & 0 \\
0                  & 0                  & 0                  & 0     & 0     & G^\ast
\end{bmatrix}
\label{eq:Dstar_matrix}
\end{equation}

A Galerkin discretization in the wavenumber--frequency domain yields, for each $(k,\omega)$,
\begin{equation}
\bigl[\mathbf{K}(k)-\omega^2\mathbf{M}\bigr]\mathbf{d}(k,\omega)=\mathbf{f}(k,\omega)
\label{eq:discrete_system_kw}
\end{equation}
where $\mathbf{K}(k)$ is complex-valued through $\mathbf{D}^\ast$ and $\mathbf{M}$ is real-valued. The global matrices are assembled from element contributions. Denoting the parametric element domain by $\widehat{\Omega}_e$, the element stiffness and mass matrices are computed as
\begin{equation}
\mathbf{K}_e(k)=\int_{\widehat{\Omega}_e}\mathbf{B}_e^{H}(k)\,\mathbf{D}^\ast\,\mathbf{B}_e(k)\,J\,\mathrm{d}\xi\,\mathrm{d}\eta,
\qquad
\mathbf{M}_e=\int_{\widehat{\Omega}_e}\rho\,\mathbf{N}_e^{T}\mathbf{N}_e\,J\,\mathrm{d}\xi\,\mathrm{d}\eta
\label{eq:Ke_Me_param}
\end{equation}
where $(\cdot)^{H}$ denotes the conjugate transpose, and $\rho$ is the mass density. Numerical integration is performed by Gaussian quadrature of sufficiently high order to capture the rational character of the NURBS basis functions and the required accuracy in wave propagation computations.

For the corresponding lossless reference material with real-valued $\mathbf{D}$, the $k$-dependence of $\mathbf{B}(k)$ implies a quadratic decomposition of the stiffness matrix,
\begin{equation}
\mathbf{K}(k)=\mathbf{K}_0+\mathrm{i}k\,\mathbf{K}_1+k^2\mathbf{K}_2
\label{eq:K_decomp}
\end{equation}
where $\mathbf{K}_0$, $\mathbf{K}_1$, and $\mathbf{K}_2$ are $k$-independent matrices assembled from the corresponding $k^0$, $k^1$, and $k^2$ contributions in $\mathbf{B}(k)$. The decomposition in Eq. \eqref{eq:K_decomp} remains valid when complex moduli are used, while Hermitian symmetry is generally lost.

The load vector $\mathbf{f}(k,\omega)$ corresponds to the consistent forces induced by the transformed traction boundary conditions on $\Gamma_t$,
\begin{equation}
\mathbf{f}(k,\omega)=\int_{\Gamma_t}\mathbf{N}^{H}\hat{\mathbf{t}}(\mathbf{x}_\perp;k,\omega)\,\mathrm{d}\Gamma
\label{eq:load_vector_kw}
\end{equation}
where $\hat{\mathbf{t}}(\mathbf{x}_\perp;k,\omega)$ is the traction amplitude in the wavenumber--frequency domain. In the numerical examples, the Fourier-normalization and longitudinal-footprint factors in Eq.~\eqref{eq:load_freq_domain_final} are absorbed into the prescribed harmonic coefficient $P$, which denotes a total force amplitude in N rather than a force per unit length. For a $y$-directed load at $\mathbf{x}_{\perp0}$, the entries are assembled as $f_{A,y}=N_A(\mathbf{x}_{\perp0})P$, and the partition-of-unity property gives $\sum_A f_{A,y}=P$.

\subsection{2.5D NURBS-trace infinite elements for enforcing radiation conditions}
\label{subsec:IGA_infinite}

For each prescribed pair $(k,\omega)$ arising from the 2.5D reduction in Section~\ref{BasisFormulations}, the cross-sectional problem is posed on $\Omega_{2D}\subset\mathbb{R}^2$ with physical coordinates $\mathbf{x}_\perp=(x,y)$. The bounded near field is denoted by $\Omega_n$ and is coupled to the semi-infinite exterior $\Omega_f$ through the artificial boundary $\Gamma_b$. Thus,
\begin{equation}
\Omega_{2D}=\Omega_n\cup\Omega_f,
\qquad
\Gamma_b=\partial\Omega_n\cap\partial\Omega_f .
\end{equation}
In the discrete formulation, the artificial boundary is represented by trace segments $\Gamma_b^q$ satisfying $\Gamma_b=\cup_q\Gamma_b^q$. The near field is discretized by the NURBS-based isogeometric formulation described in Subsection~\ref{subsec:IGA_bounded}, while the exterior contribution associated with each trace segment is represented by infinite elements. Their role is to approximate the exterior response by combining the boundary trace with radial functions that satisfy the required outgoing or decay behavior.

The following construction separates the invariant trace-space principle from the selectable radial approximation.

An outward coordinate $\xi_\infty\in[0,\infty)$ is introduced on each segment of the artificial boundary. The unbounded direction is represented by a radial function
\begin{equation}
P_q=P_q(\xi_\infty;k,\omega,\boldsymbol{\theta}_q),
\qquad
P_q(0;k,\omega,\boldsymbol{\theta}_q)=1,
\label{eq:IE_propagation_IGA}
\end{equation}
where $q$ identifies a boundary segment and $\boldsymbol{\theta}_q$ collects the radial parameters assigned to that segment. Equation~\eqref{eq:IE_propagation_IGA} deliberately does not prescribe a particular exponential, decay coefficient, or spatial partition. These choices determine the radiation accuracy of a realization of the method, but they do not alter its trace-space construction.

The admissible radial family is defined by three requirements. First, the normalization $P_q(0)=1$ preserves the finite-dimensional trace at the near--far-field interface. Second, $P_q$ must represent an outgoing or evanescent contribution and contain no component that grows toward infinity. Third, $P_q$ and its radial derivative must make the far-field weak forms finite. In a mapped radial coordinate, the last requirement can be expressed abstractly as
\begin{equation}
\int_0^\infty
\left(
|P_q(\xi_\infty)|^2
+
|\partial_{\xi_\infty}P_q(\xi_\infty)|^2
\right)
w_q(\xi_\infty)\,\mathrm{d}\xi_\infty
<\infty,
\label{eq:radial_admissibility}
\end{equation}
where $w_q$ accounts for the algebraic growth introduced by the far-field mapping. For an undamped propagating contribution, the outgoing solution is understood through the limiting-absorption principle. Consequently, a vanishing explicit decay coefficient is admissible only when the complete radial factor remains outgoing and the associated weak forms are well defined.

For an isotropic exterior medium, the candidate transverse wavenumbers are obtained from the bulk or surface-wave dispersion relations using $k$ from Eq.~\eqref{eq:k_def}:
\begin{equation}
\kappa_i^2=k_i^2-k^2,
\qquad
k_i=\frac{\omega}{c_i}.
\label{eq:kappa_def_infinite}
\end{equation}
The physically relevant branch is selected by the outgoing radiation condition, equivalently by continuation from a weakly dissipative medium. This principle covers propagating, evanescent, and damped components without tying the formulation to a code-dependent algebraic sign convention. Candidate wave-informed functions may be used to construct $P_q$, while their adequacy for a finite truncation boundary is evaluated by the controlled numerical assessment in Section~\ref{VerificationExamples}.

The numerical study uses an all-S realization, in which the S-wave-informed radial family is assigned to every artificial-boundary segment. Its explicit radial factor is
\begin{equation}
P_q(\xi_\infty)
=
\exp\!\left[-\left(\alpha_{S,q}-\mathrm{i}\kappa_S\right)\xi_\infty\right],
\qquad
\kappa_S
=
\left[\left(\frac{\omega}{c_{S,\infty}}\right)^2-k^2\right]_{\mathrm{out}}^{1/2},
\label{eq:allS_radial_factor}
\end{equation}
where the subscript ``out'' denotes the outgoing branch. Its reference decay is
\begin{equation}
\alpha_{S,q}^{\mathrm{ref}}
=
\frac{1}{2L_q}
\left(
1+\frac{k^2}{k^2+\kappa_S^2}
\right).
\label{eq:alphaS_reference}
\end{equation}
Here $L_q$ is the characteristic outward distance assigned to segment $q$ by the radial map. For the rectangular domains used in the verification and engineering examples, $L_q=R$ on the lateral boundary and $L_q=H$ on the lower boundary. Thus, $L_q$ is a declared geometric input rather than a fitted material parameter; a non-rectangular radial map must analogously store its segment-wise characteristic distance. In damped calculations, $c_{S,\infty}=\sqrt{G_\infty/\rho_\infty}$ is the reference speed computed from the storage modulus $G_\infty=E_\infty/[2(1+\nu_\infty)]$. Material attenuation remains in the complex tensor $\mathbf{D}^{\ast}$ of Eq.~\eqref{eq:D_tensor_iso}; a complex wave speed is not substituted into Eqs.~\eqref{eq:kappa_def_infinite} and~\eqref{eq:alphaS_reference}.

\begin{figure}[t]
  \centering
  \includegraphics[width=0.65\linewidth]{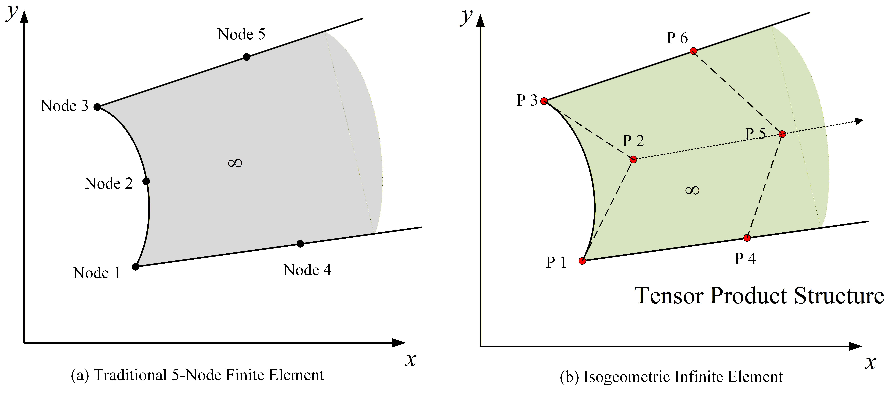}
  \caption{Illustration of semi-infinite domain modeling. (a) Conventional finite-element representation using nodal interpolation in a truncated exterior domain. (b) NURBS-based infinite element constructed from the boundary NURBS representation and an outward propagation function for the unbounded direction.}
  \label{iem}
\end{figure}

The far-field geometry preserves the NURBS representation of the truncation boundary and extends it in the outward direction through a radial mapping. Let $s\in[0,1]$ be the parametric coordinate along the boundary segment $\Gamma_b^q$, with NURBS basis functions $\{R_{A,q}^{b}(s)\}_{A=1}^{n_{b,q}}$ and boundary control points $\{\mathbf{P}_{A,q}^{b}\}_{A=1}^{n_{b,q}}$. By restricting the general NURBS-curve representation in Eq.~\eqref{eq:NURBS_curve} to segment $q$, its geometry is denoted by $\mathbf{x}_{b,q}(s)=(x_{b,q}(s),y_{b,q}(s))$.

The discrete compatibility on each segment can now be stated precisely. Let $\mathcal{V}_h^n$ be the vector-valued near-field IGA trial space, let $\gamma_{b,q}$ denote its trace on $\Gamma_b^q$, and let $\{\mathbf{e}_\ell\}_{\ell=1}^{3}$ be the Cartesian basis. The near-field trace space associated with segment $q$ is
\begin{equation}
\mathcal{T}_{h,q}
:=
\gamma_{b,q}\mathcal{V}_h^n
=
\operatorname{span}
\left\{
R_{A,q}^b(s)\mathbf{e}_\ell:
A=1,\ldots,n_{b,q},\ \ell=1,2,3
\right\}.
\label{eq:nurbs_trace_space}
\end{equation}
For a far-field segment $q$, the proposed trial space is the tensor product
\begin{equation}
\mathcal{V}_{h,q}^f
=
\operatorname{span}
\left\{
R_{A,q}^b(s)P_q(\xi_\infty)\mathbf{e}_\ell:
A=1,\ldots,n_{b,q},\ \ell=1,2,3
\right\}.
\label{eq:far_field_trial_space}
\end{equation}
Because $P_q(0)=1$, its boundary trace satisfies
\begin{equation}
\left.\mathcal{V}_{h,q}^f\right|_{\xi_\infty=0}
=
\mathcal{T}_{h,q}
=
\gamma_{b,q}\mathcal{V}_h^n.
\label{eq:trace_space_conformity}
\end{equation}
Equation~\eqref{eq:trace_space_conformity} is the precise meaning of \emph{NURBS-trace consistency} in this work: the two discrete traces coincide on each artificial-boundary segment and share their coefficients. The global far-field space $\mathcal{V}_h^f$ is assembled from the segment spaces $\mathcal{V}_{h,q}^f$, with common trace coefficients identified at shared boundary control points. This construction does not assert that the unbounded radial direction is itself represented by NURBS; that direction is represented by the analytical function $P_q$.

The geometry of an infinite segment is described by a regular mapping
\begin{equation}
\mathbf{x}_\perp
=
\mathcal{F}_q(s,\xi_\infty),
\qquad
\mathcal{F}_q(s,0)=\mathbf{x}_{b,q}(s),
\qquad
\|\mathcal{F}_q(s,\xi_\infty)\|\rightarrow\infty
\quad\text{as }\xi_\infty\rightarrow\infty.
\label{eq:IE_geom_tensor}
\end{equation}
Only the interface condition and regularity of $\mathcal{F}_q$ are essential to the trace coupling. Pole-based, normal-extrusion, or other admissible radial mappings may therefore be used without changing the discrete interface space, provided that the mapping remains one-to-one and its Jacobian is nonsingular.

In the numerical implementation, the far-field matrices are assembled from the infinite-element contributions associated with the artificial-boundary segments. The segment mapping $\mathcal{F}_q$ supplies the geometric Jacobian in the boundary-coordinate integrals, while the radial dependence is carried by $P_q$ and its derivative.

The coefficient realization of Eqs.~\eqref{eq:far_field_trial_space}--\eqref{eq:trace_space_conformity} is obtained as follows. The segment coefficient $\mathbf{d}_{A,q}(k,\omega)$ has the same $[d_x,d_y,d_z]^T$ ordering as the near-field coefficient $\mathbf{d}_A$ in Eq.~\eqref{eq:IGA_u_interp} and is identified with the corresponding global trace unknown. For a far-field segment $q$, the displacement amplitude is approximated as
\begin{equation}
\hat{\mathbf{u}}^{\,f}(s,\xi_\infty;k,\omega)
=
\sum_{A=1}^{n_{b,q}}
R_{A,q}^{b}(s)\,P_q(\xi_\infty)\,\mathbf{d}_{A,q}(k,\omega).
\label{eq:IE_u_tensor}
\end{equation}
At $\xi_\infty=0$, Eq.~\eqref{eq:IE_u_tensor} reduces to the near-field boundary interpolation. Displacement continuity is therefore imposed by coefficient identification, without projection, mortar variables, or an additional interface transfer operation.

Derivatives with respect to $(x,y)$ follow from the Jacobian of the mapping in Eq.~\eqref{eq:IE_geom_tensor}. Define
\begin{equation}
\mathbf{J}_f(s,\xi_\infty)
=
\frac{\partial\mathbf{x}_\perp}{\partial(s,\xi_\infty)},
\qquad
J_f=\det\mathbf{J}_f .
\label{eq:IE_Jacobian}
\end{equation}
Define the tensor-product basis $\Phi_{A,q}(s,\xi_\infty)=R_{A,q}^b(s)P_q(\xi_\infty)$. Its cross-sectional derivatives follow from the chain rule through $\mathbf{J}_f^{-T}$, while the longitudinal derivative uses the $-\mathrm{i}k$ operator defined in Eq.~\eqref{eq:2p5d_map}. Consequently, the far-field strain retains all six components of the 2.5D linear viscoelastic vector wave problem and can be written as
\begin{equation}
\boldsymbol{\varepsilon}^{\,f}
=
\mathbf{B}_f(k)\mathbf{d}_b ,
\label{eq:IE_B_IGA}
\end{equation}
where $\mathbf{d}_b$ stacks the boundary control-point degrees of freedom. The operator $\mathbf{B}_f(k)$ is obtained by applying the 2.5D strain construction of Eqs.~\eqref{eq:2p5d_strains} and~\eqref{eq:B_block} to $\Phi_{A,q}$; its entries therefore follow the same Fourier convention and coordinate orientation as the near-field operator. The radial model therefore changes the exterior approximation through $P_q$ and $\partial_{\xi_\infty}P_q$, without reducing the vector problem to a scalar or plane-strain model.

For each $(k,\omega)$, the near- and far-field approximations form a conforming coupled space, where $\gamma_b$ denotes the assembled trace operator on $\Gamma_b$:
\begin{equation}
\mathcal{V}_h
=
\left\{
(\mathbf{v}_h^n,\mathbf{v}_h^f)
\in\mathcal{V}_h^n\times\mathcal{V}_h^f:
\gamma_b\mathbf{v}_h^n
=
\left.\mathbf{v}_h^f\right|_{\xi_\infty=0}
\right\}.
\label{eq:coupled_trace_space}
\end{equation}
The same frequency-domain weak form in both subdomains gives the discrete problem: find $\mathbf{u}_h\in\mathcal{V}_h$ such that
\begin{equation}
a_n(\mathbf{v}_h,\mathbf{u}_h)
+a_f(\mathbf{v}_h,\mathbf{u}_h)
-\omega^2
\left[
m_n(\mathbf{v}_h,\mathbf{u}_h)
+m_f(\mathbf{v}_h,\mathbf{u}_h)
\right]
=
\ell(\mathbf{v}_h)
\label{eq:coupled_weak_form}
\end{equation}
for all admissible test functions $\mathbf{v}_h$. The far-field forms are
\begin{equation}
\begin{aligned}
a_f(\mathbf{v}_h,\mathbf{u}_h)
&=
\int_{\Omega_f}
\boldsymbol{\varepsilon}(\mathbf{v}_h)^{H}
\mathbf{D}^{\ast}
\boldsymbol{\varepsilon}(\mathbf{u}_h)
\,\mathrm{d}\Omega,\\
m_f(\mathbf{v}_h,\mathbf{u}_h)
&=
\int_{\Omega_f}
\mathbf{v}_h^{H}\rho\mathbf{u}_h
\,\mathrm{d}\Omega,
\end{aligned}
\label{eq:far_field_forms}
\end{equation}
The corresponding near-field element contributions are defined in Eq.~\eqref{eq:Ke_Me_param}. The superscript $H$ denotes the conjugate transpose under the adopted complex test-space convention.

Because the interface coefficients are shared, assembly of Eq.~\eqref{eq:coupled_weak_form} gives the global system
\begin{equation}
\left[
\mathbf{K}_n(k)+\mathbf{K}_f(k)
-\omega^2\left(\mathbf{M}_n+\mathbf{M}_f\right)
\right]\mathbf{d}
=
\mathbf{f}(k,\omega).
\label{eq:global_coupled_IGA_IE}
\end{equation}
Equation~\eqref{eq:global_coupled_IGA_IE} is the bounded-system relation in Eq.~\eqref{eq:discrete_system_kw} augmented by the exterior matrices $\mathbf{K}_f$ and $\mathbf{M}_f$ on the shared trace.
No projection, penalty, mortar variable, or separately interpolated interface traction is required. For Eq.~\eqref{eq:allS_radial_factor}, let $\gamma_q=\alpha_{S,q}-\mathrm{i}\kappa_S$. The radial moments required by both far-field matrices are evaluated analytically; in particular,
\begin{equation}
I_{0,q}
=\int_0^\infty P_q\overline{P_q}\,\mathrm{d}\xi_\infty
=\frac{1}{\gamma_q+\overline{\gamma_q}},
\qquad
\Re(\gamma_q)>0,
\label{eq:radial_moment_exact}
\end{equation}
and derivative moments follow by multiplication with $-\gamma_q$ and $-\overline{\gamma_q}$. In the optional calibrated comparison, a zero multiplier only removes the additional reference decay; the corresponding basis is retained only when the outgoing or evanescent part still gives $\Re(\gamma_q)>0$. An undamped propagating factor with both zero explicit decay and $\Re(\gamma_q)=0$ is rejected because Eq.~\eqref{eq:radial_moment_exact} is then divergent. Hence no finite radial cutoff or radial quadrature is introduced in the production matrices. The remaining boundary-coordinate integrals use $n_g=\max(3,p_b+1)$ Gauss--Legendre points on every nonzero NURBS knot span, where $p_b$ is the degree of the boundary trace. Section~\ref{VerificationExamples} independently verifies the analytical radial moments by direct adaptive complex integration on the native interval $[0,\infty)$. This audit uses no user-defined finite cutoff or variable transformation and reports the applied tolerances and matrix norm. The boundary quadrature is separately checked by repeating the cubic-trace response with three, four, and five points per knot span.

This construction separates the invariant contribution of the proposed NBIEM from the selectable exterior model. The invariant part is the tensor product of an admissible radial function with the NURBS trace of the bounded discretization, together with coefficient-level continuity in Eq.~\eqref{eq:trace_space_conformity}. The radial family, its parameters, and the truncation distance influence approximation quality and may be revised without changing this coupling principle. Section~\ref{VerificationExamples} therefore assesses these choices through response sensitivity and boundary enlargement instead of treating a particular empirical parameter formula as part of the mathematical definition.

\begin{figure}[H]
  \centering
  \includegraphics[width=0.96\linewidth]{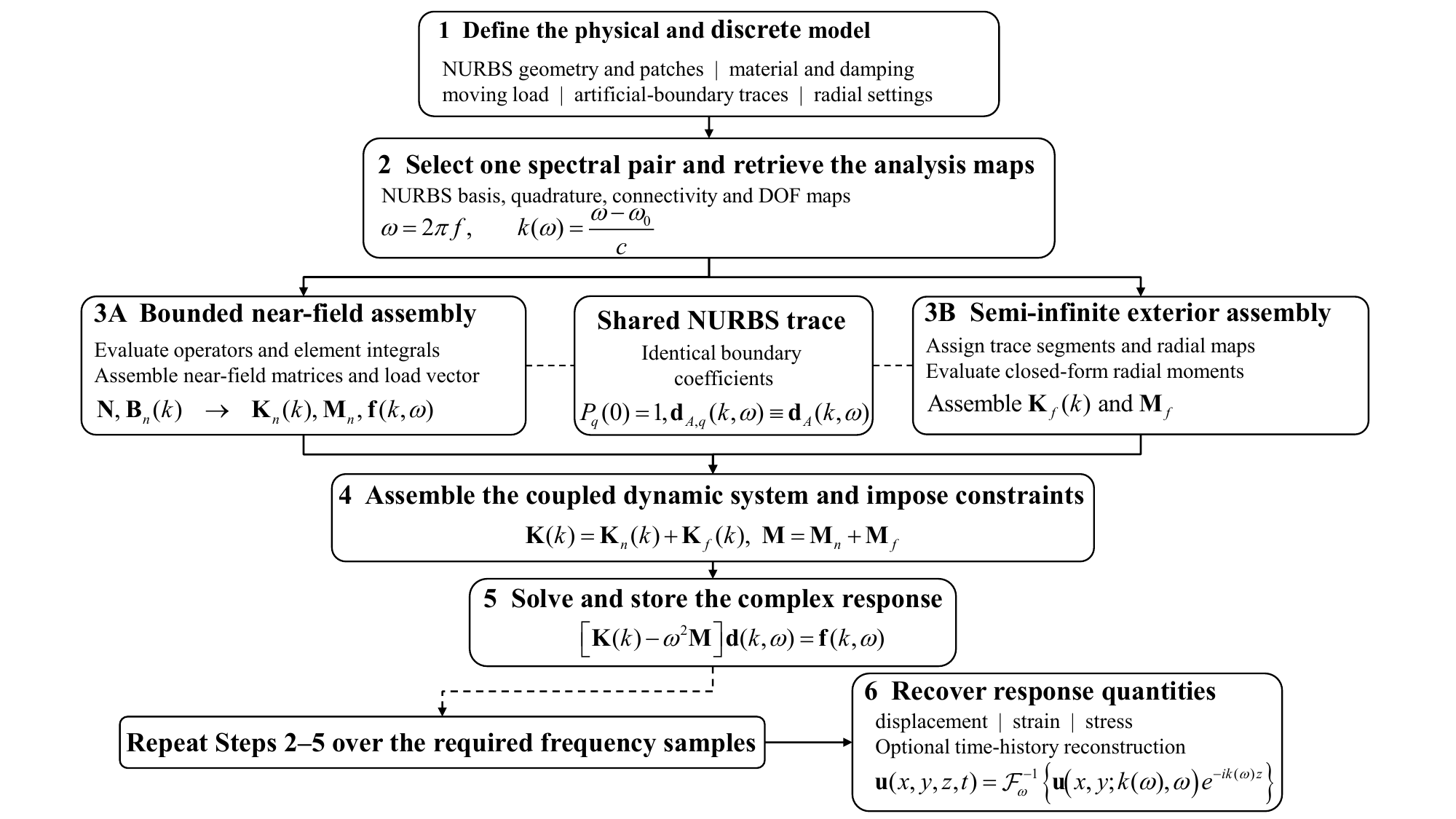}
  \caption{Implementation workflow of the proposed 2.5D NURBS-trace finite/infinite-element method. The bounded near-field and semi-infinite exterior matrices are assembled at each frequency--wavenumber pair through the shared NURBS trace, coupled after application of the prescribed constraints, and solved before response recovery and optional inverse Fourier reconstruction.}
  \label{fig:nbiem-implementation-flowchart}
\end{figure}

{
For implementation, the principal computational sequence is summarized in \fref{fig:nbiem-implementation-flowchart}. After defining the physical and discrete model, each frequency--wavenumber pair is used to assemble the bounded near-field and semi-infinite exterior contributions through their shared NURBS trace. The constrained complex dynamic system is then solved over the required spectral grid, followed by recovery of the displacement, strain, and stress responses and, when required, inverse Fourier reconstruction.
}

\section{Numerical verification and methodological assessment}
\label{numerexamp}

This section evaluates the proposed NBIEM through a series of numerical studies with increasing complexity. The studies assess three main aspects of the formulation: (i) the interior-field accuracy of the NURBS-based discretization for displacement and stress responses; (ii) the radiation performance of the NURBS-based infinite elements, including boundary-sensitivity checks, low-frequency radial sensitivity, and comparison with existing boundary-treatment strategies; and (iii) the applicability of the method to heterogeneous and engineering-oriented geotechnical configurations involving material interfaces, geometric transitions, and embedded structures.

The first group of tests uses closed-form half-space solutions to verify displacement and stress frequency-response functions under different moving-load speed regimes. The second group focuses on the artificial-boundary treatment at low frequencies, where boundary effects are more pronounced, and uses the same setting to assess the default radial workflow, examine an optional calibrated comparison, and compare NBIEM with FEM/IEM and IGA/SBIGA in terms of accuracy, phase response, and computational cost. The remaining examples apply the formulation to layered ground, ballastless-track sections, and buried-structure configurations, using engineering-oriented geotechnical models with multi-material domains and locally refined multi-patch discretizations. Unless otherwise stated, the baseline configuration is a homogeneous linear viscoelastic half-space, optionally containing a buried structure, subjected to a time-harmonic load traveling at a constant speed $c$ along the longitudinal direction, with load position $z(t)=c\,t$. The gravitational acceleration is fixed at $g=9.81~\mathrm{m/s^2}$. The numerical settings are chosen to isolate wave-propagation, boundary-treatment, discretization, and load-speed effects.

\subsection{Verification examples}\label{VerificationExamples}
\subsubsection{Interior-field verification at relatively high frequency}

The first verification study considers benchmark half-space problems with closed-form solutions to assess the interior-field accuracy of the proposed formulation in a controlled setting. 
A relatively high-frequency case is selected because it provides a stringent test of the spatial discretization, while the influence of the truncation boundary remains limited. 
Accordingly, the present results are interpreted primarily as a verification of the interior displacement and stress responses, whereas the boundary treatment and the selection of the infinite-element parameter $\alpha$ are examined separately in the following subsection under low-frequency conditions. Owing to symmetry with respect to the vertical plane containing the load trajectory, only one half of the domain is discretized using an isogeometric analysis (IGA) mesh, and the far-field radiation condition is imposed through isogeometric infinite elements. For the half-space considered in this verification study, the compressional, shear, and Rayleigh wave speeds are $c_P=173.2~\mathrm{m/s}$, $c_S=100~\mathrm{m/s}$, and $c_R=92.1~\mathrm{m/s}$, respectively. To control discretization-induced dispersion and maintain phase accuracy at the excitation frequency, the minimum near-field range $R$ and the maximum element size $L$ are selected using a fixed wavelength-based resolution criterion. The selected near-field range and element size satisfy
\begin{equation}\label{eq:mesh_rule}
L \le \frac{\lambda_s'}{6},
\qquad
R \ge \frac{\lambda_s'}{2},
\qquad
\lambda_s'=\frac{2\pi}{k_s'} ,
\end{equation}
where $\lambda_s'$ is the effective shear wavelength and $k_s'=|\kappa_S|$, with $\kappa_S$ obtained from Eq.~\eqref{eq:kappa_def_infinite} for $i=S$ and $k$ from Eq.~\eqref{eq:k_def}.
For the present benchmark, the mesh is selected according to the most restrictive requirement among the tested cases and is then kept unchanged for the different load speeds.

{All results in this benchmark are reported in normalized form. For a load acting in the $y$-direction, the normalized $x$-, $y$-, and $z$-directed displacement responses are denoted by $U_y$, $V_y$, and $W_y$, respectively; the uppercase letter identifies the response direction, whereas the subscript identifies the load direction. They are defined by}
\begin{equation}\label{eq:norm_freq_disp}
U_y=\frac{2\pi G}{c}\,\hat u_y(k,\omega),
\qquad
V_y=\frac{2\pi G}{c}\,\hat v_y(k,\omega),
\qquad
W_y=\frac{2\pi G}{c}\,\hat w_y(k,\omega),
\end{equation}
where $G$ is the shear modulus.

The first interior-field benchmark uses a homogeneous half-space without a buried structure and a mass density of $\rho=2000~\mathrm{kg/m^3}$.
Three representative speed regimes are examined: a sub-Rayleigh case ($c=90~\mathrm{m/s}$), a super-shear but sub-compressional case ($c=120~\mathrm{m/s}$), and a super-compressional case ($c=200~\mathrm{m/s}$).
The material damping ratio of the half-space is set to $\zeta=0.05$.

The analytical displacement frequency-response functions follow \citet{YangLiu-56}:
\begin{equation}\label{eq:yang_frf_disp}
\hat u_y^{\mathrm{ana}}(k,\omega)
=\frac{1}{2\pi\mu}\int_{-\infty}^{\infty}
\frac{-\mathrm{i}k_x}{2Q}
\left[
\left(k^2+k_x^2-\frac{1}{2}k_s^2\right)\exp(-m_1 y)
- m_1 m_2 \exp(-m_2 y)
\right]
\exp(\mathrm{i}k_x x)\,\mathrm{d}k_x ,
\end{equation}
\begin{equation}\label{eq:yang_frf_disp_v}
\hat v_y^{\mathrm{ana}}(k,\omega)
=\frac{1}{2\pi\mu}\int_{-\infty}^{\infty}
\frac{-m_1}{2Q}
\left[
\left(k^2+k_x^2-\frac{1}{2}k_s^2\right)\exp(-m_1 y)
-\left(k^2+k_x^2\right)\exp(-m_2 y)
\right]
\exp(\mathrm{i}k_x x)\,\mathrm{d}k_x ,
\end{equation}
\begin{equation}\label{eq:yang_frf_disp_w}
\hat w_y^{\mathrm{ana}}(k,\omega)
=\frac{1}{2\pi\mu}\int_{-\infty}^{\infty}
\frac{\mathrm{i}k}{2Q}
\left[
\left(k^2+k_x^2-\frac{1}{2}k_s^2\right)\exp(-m_1 y)
- m_1 m_2 \exp(-m_2 y)
\right]
\exp(\mathrm{i}k_x x)\,\mathrm{d}k_x .
\end{equation}
\noindent Here $\mu$ denotes the shear modulus, $k_x$ is the Fourier integration variable, and $k$ is the longitudinal wavenumber defined in Eq.~\eqref{eq:k_def} for the selected moving-load harmonic.
The compressional and shear wavenumbers are $k_p=\omega/c_P$ and $k_s=\omega/c_S$.
The auxiliary quantities are
\begin{equation}\label{eq:Q_m1_m2_def}
Q=\left(k^2+k_x^2-\frac{1}{2}k_s^2\right)^2
- m_1 m_2\left(k^2+k_x^2\right),\qquad
m_1=\left(k^2+k_x^2-k_p^2\right)^{1/2},\qquad
m_2=\left(k^2+k_x^2-k_s^2\right)^{1/2}.
\end{equation}

\begin{figure}[!ht]
  \centering
  \begin{subfigure}[t]{0.36\textwidth}
    \centering
    \includegraphics[width=\linewidth]{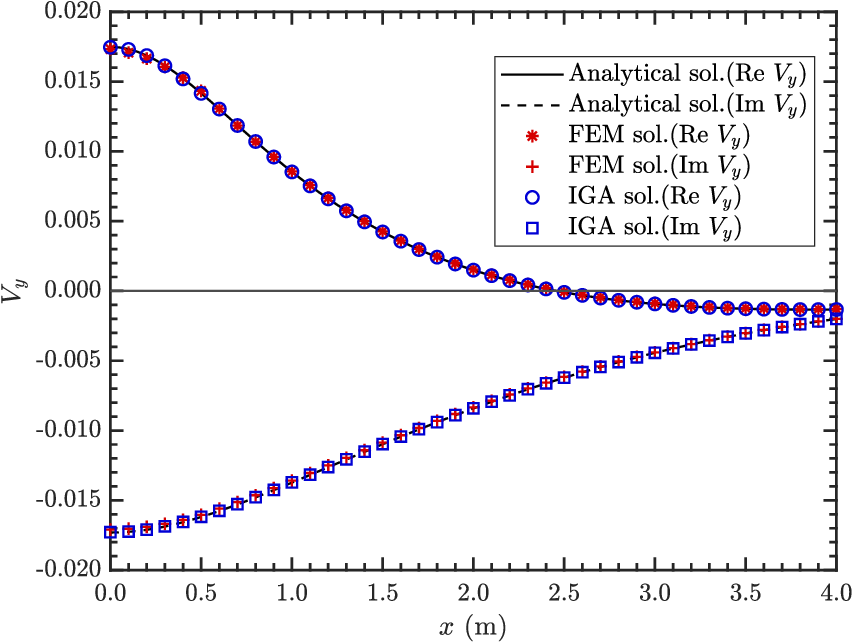}
    \caption{}
  \end{subfigure}\hspace{5mm}
  \begin{subfigure}[t]{0.36\textwidth}
    \centering
    \includegraphics[width=\linewidth]{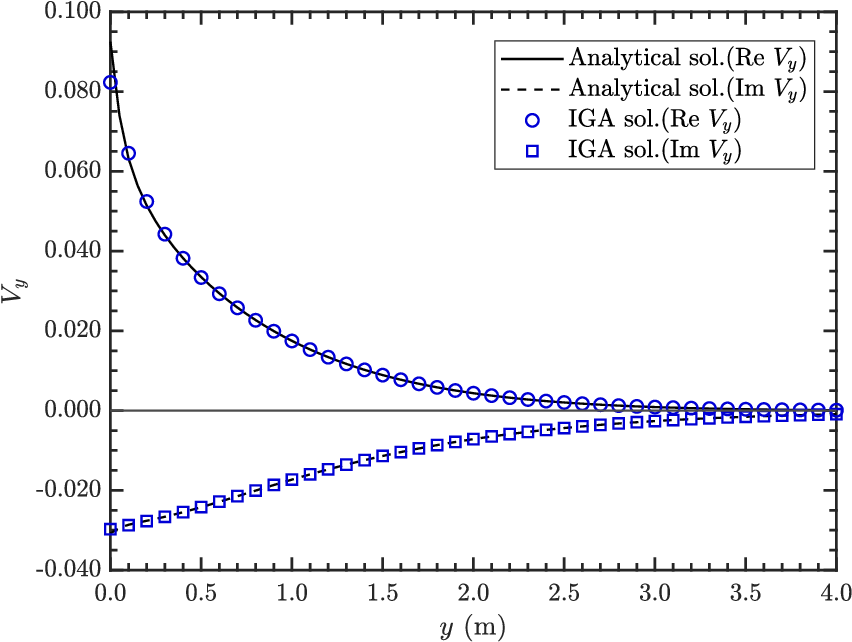}
    \caption{}
  \end{subfigure}
  \begin{subfigure}[t]{0.36\textwidth}
    \centering
    \includegraphics[width=\linewidth]{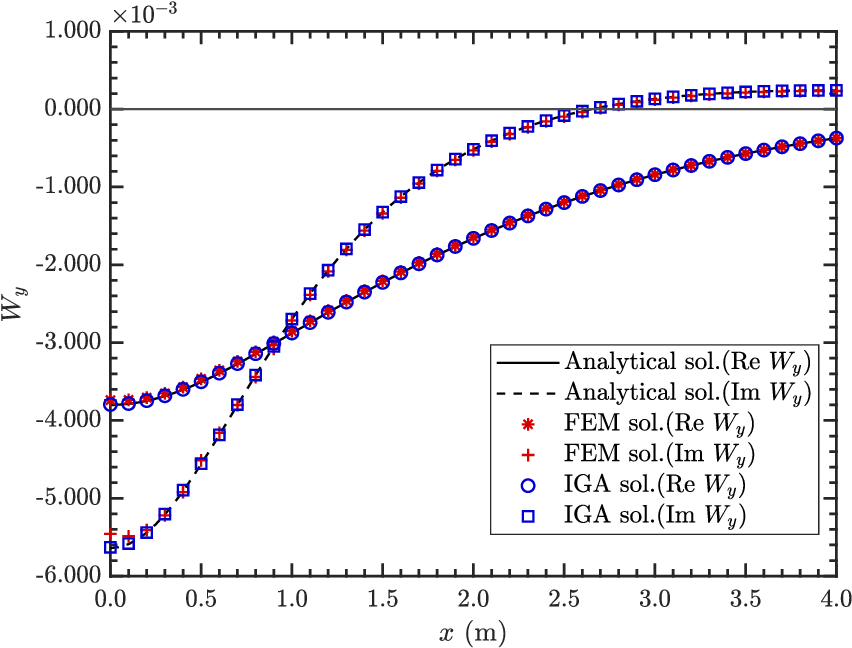}
    \caption{}
  \end{subfigure}\hspace{5mm}
  \begin{subfigure}[t]{0.36\textwidth}
    \centering
    \includegraphics[width=\linewidth]{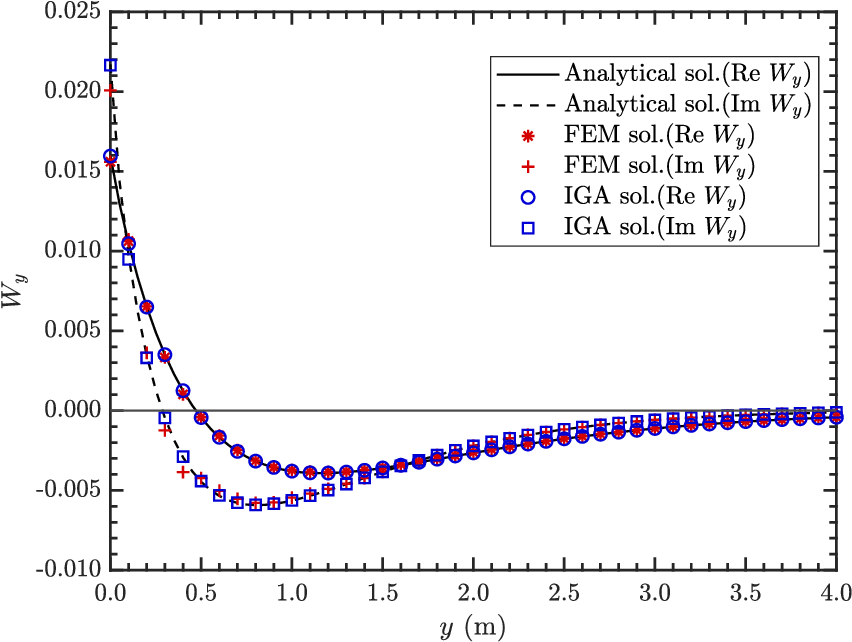}
    \caption{}
  \end{subfigure}
\caption{Normalized frequency-domain displacement components $V_y$ (vertical) and $W_y$ (longitudinal) for $f=32~\mathrm{Hz}$, $f_0=0~\mathrm{Hz}$, $c=90~\mathrm{m/s}$ (sub-Rayleigh, $c<c_R$), and $\zeta=0.05$. Panels (a) and (c): $x$-profiles at $y=1~\mathrm{m}$. Panels (b) and (d): depth-wise profiles at $x=0~\mathrm{m}$. IGA and FEM are compared under matched spatial resolution against the analytical solution.}
  \label{FRF_c90}
\end{figure}

\begin{figure}[!ht]
  \centering
  \begin{subfigure}[t]{0.36\textwidth}
    \centering
    \includegraphics[width=\linewidth]{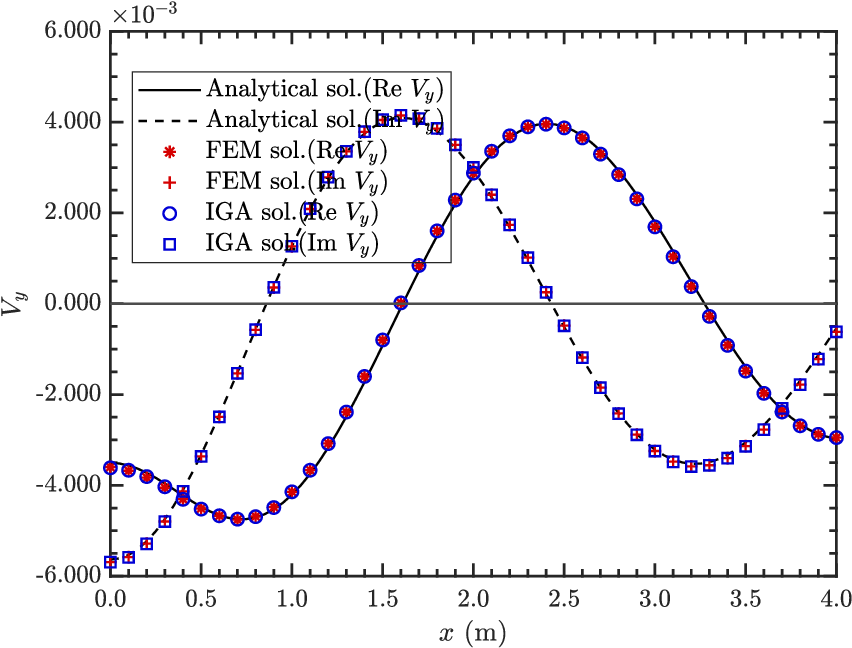}
    \caption{}
  \end{subfigure}\hspace{5mm}
  \begin{subfigure}[t]{0.36\textwidth}
    \centering
    \includegraphics[width=\linewidth]{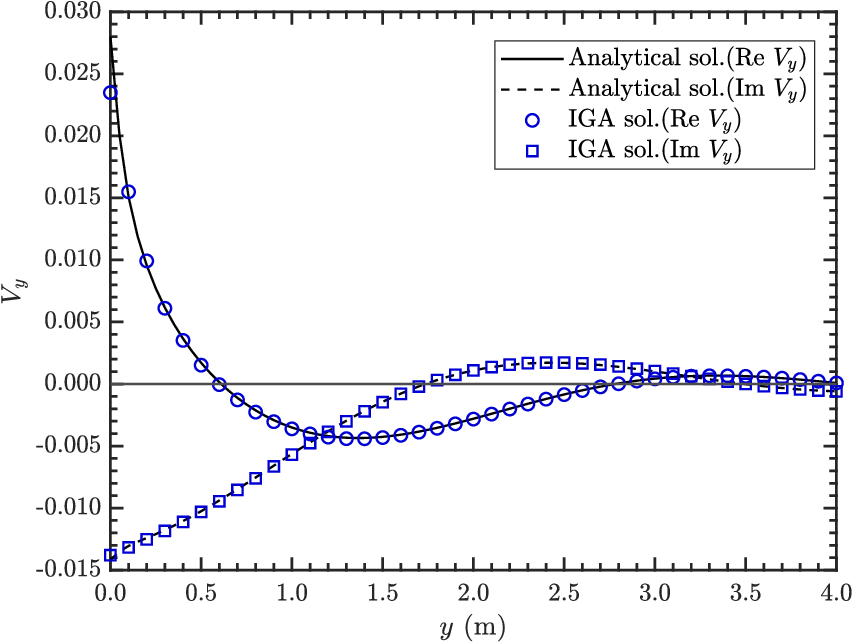}
    \caption{}
  \end{subfigure}
  \begin{subfigure}[t]{0.36\textwidth}
    \centering
    \includegraphics[width=\linewidth]{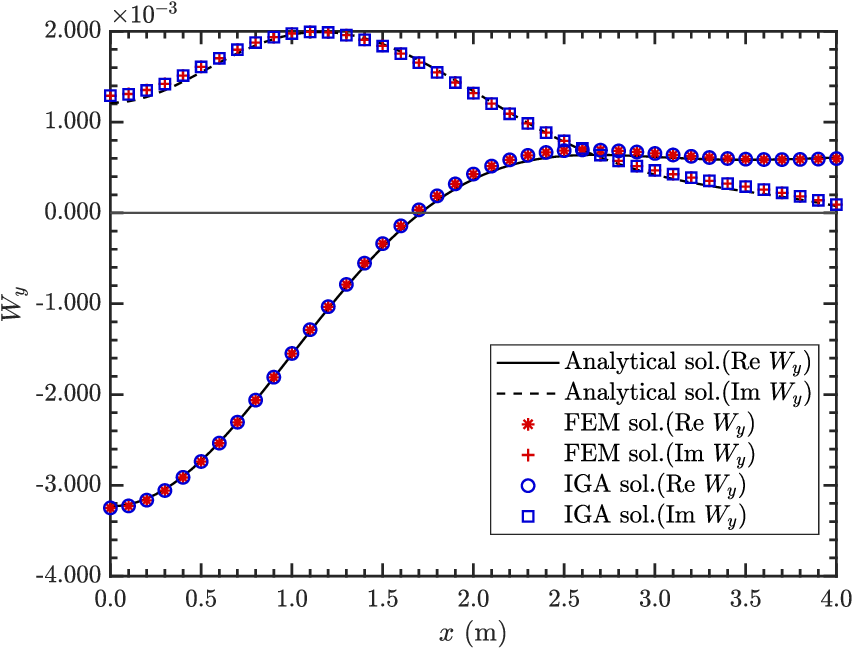}
    \caption{}
  \end{subfigure}\hspace{5mm}
  \begin{subfigure}[t]{0.36\textwidth}
    \centering
    \includegraphics[width=\linewidth]{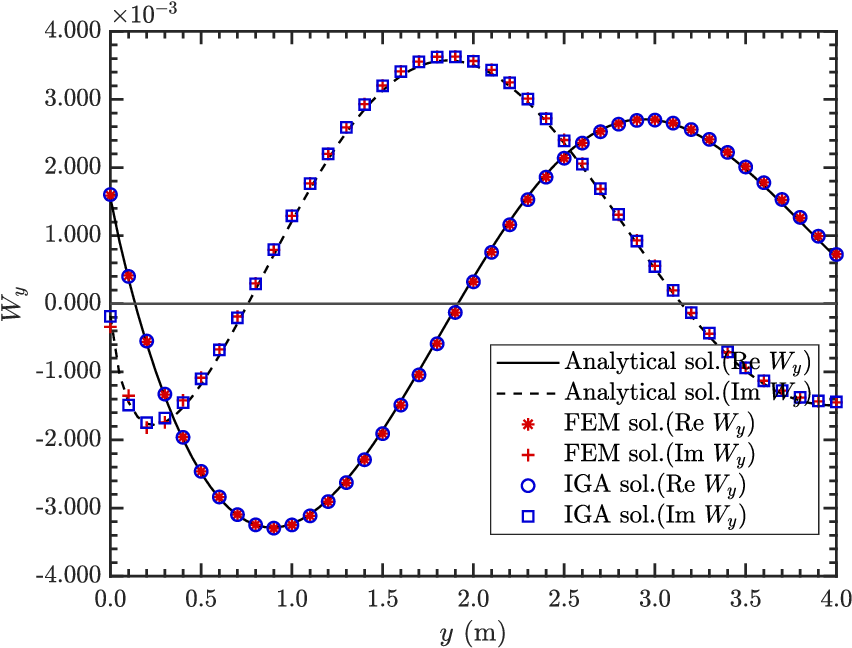}
    \caption{}
  \end{subfigure}
  \caption{Normalized frequency-domain displacement components $V_y$ and $W_y$ for $f=32~\mathrm{Hz}$, $f_0=0~\mathrm{Hz}$, $c=200~\mathrm{m/s}$ ($c>c_P$), and $\zeta=0.05$. Panels (a) and (c): $x$-profiles at $y=1~\mathrm{m}$. Panels (b) and (d): depth-wise profiles at $x=0~\mathrm{m}$.}
  \label{FRF_c200}
\end{figure}

The near-field truncation radius and the vertical extent of the computational domain are set to $R=10~\mathrm{m}$ and $H=10~\mathrm{m}$, respectively. In this frequency-domain verification, the analysis frequency is fixed at $f=\omega/(2\pi)=32~\mathrm{Hz}$. The complex normalized vertical and longitudinal displacement components, denoted by $V_y$ and $W_y$, are compared with the analytical solutions in Eqs.~\eqref{eq:yang_frf_disp_v} and~\eqref{eq:yang_frf_disp_w}. To keep the presentation compact while retaining the bounding response regimes, Figs.~\ref{FRF_c90} and~\ref{FRF_c200} show $c=90~\mathrm{m/s}$ (sub-Rayleigh) and $c=200~\mathrm{m/s}$ (super-compressional), respectively. The intermediate $c=120~\mathrm{m/s}$ regime is assessed through the more demanding stress comparison below.

Panels (a) and (c) show the $x$-profiles at $y=1~\mathrm{m}$, whereas panels (b) and (d) show the depth-wise profiles at $x=0~\mathrm{m}$. The response changes from predominantly non-oscillatory decay at $c=90~\mathrm{m/s}$ to an oscillatory pattern at $c=200~\mathrm{m/s}$. In both bounding regimes, the IGA and FEM real and imaginary parts are in good agreement with the analytical results.

Under matched spatial resolution, the depth-wise profiles show slightly smaller near-surface discrepancies for IGA in both $V_y$ and $W_y$, consistent with the higher-order continuity of the spline basis.

\begin{figure}[!ht]
  \centering
  \begin{subfigure}[t]{0.36\textwidth}
    \centering
    \includegraphics[width=\linewidth]{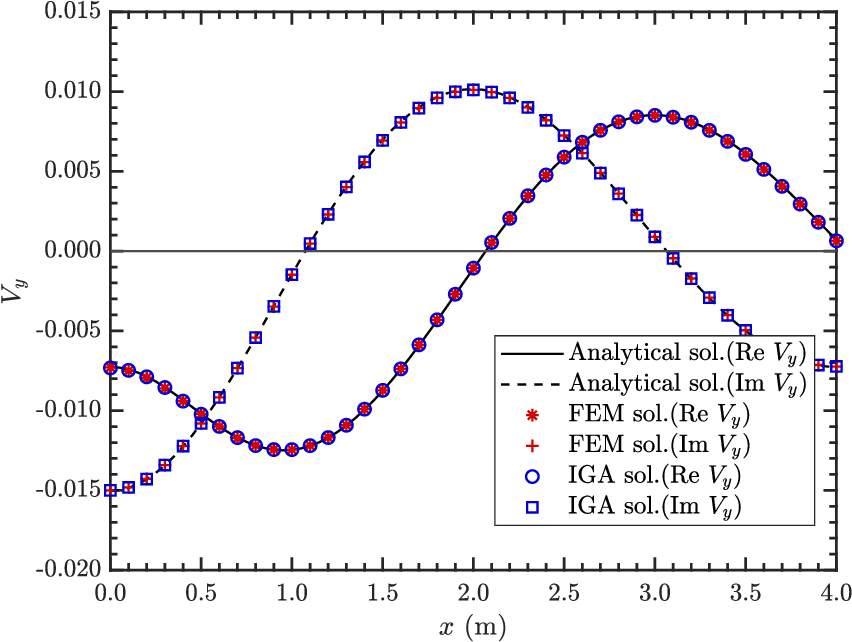}
    \caption{}
  \end{subfigure}\hspace{5mm}
  \begin{subfigure}[t]{0.36\textwidth}
    \centering
    \includegraphics[width=\linewidth]{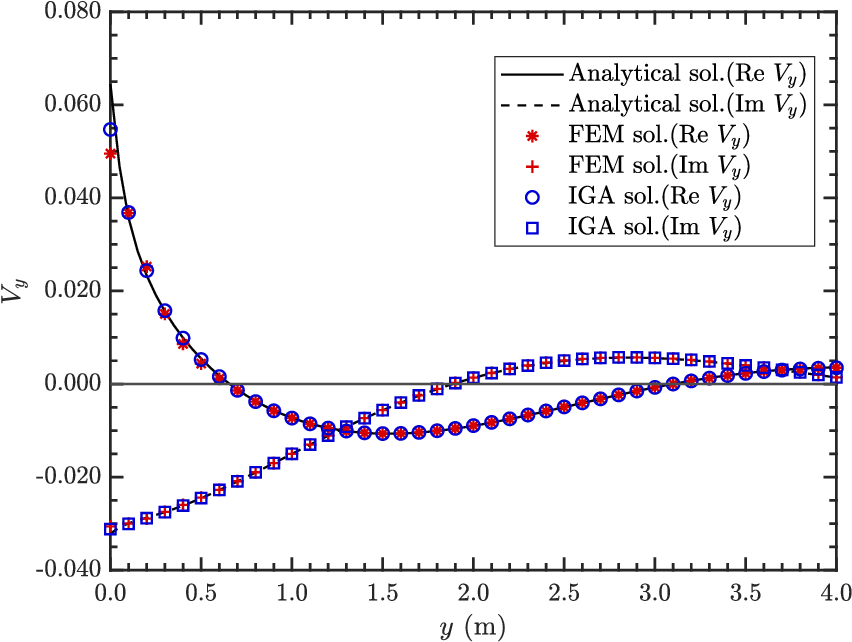}
    \caption{}
  \end{subfigure}
  \begin{subfigure}[t]{0.36\textwidth}
    \centering
    \includegraphics[width=\linewidth]{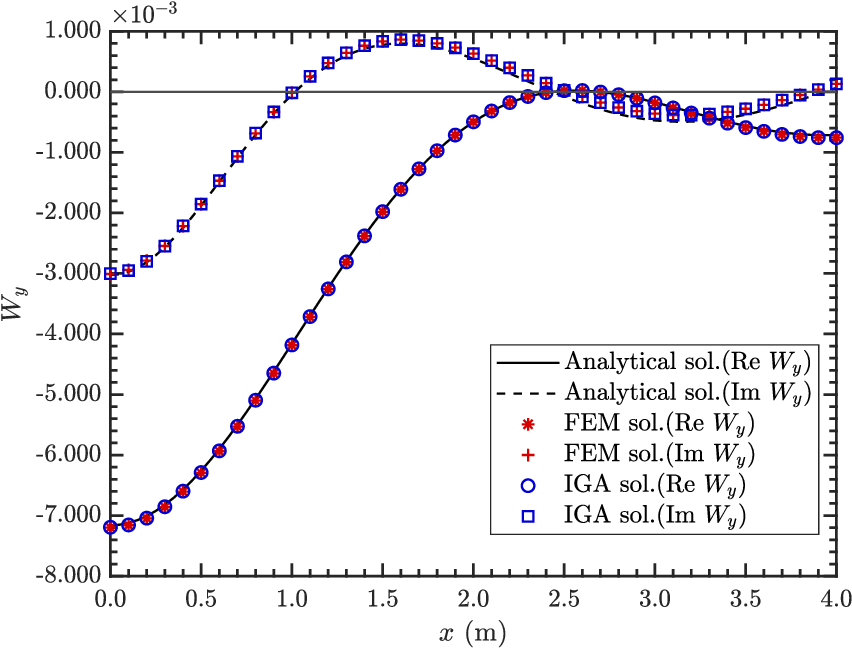}
    \caption{}
  \end{subfigure}\hspace{5mm}
  \begin{subfigure}[t]{0.36\textwidth}
    \centering
    \includegraphics[width=\linewidth]{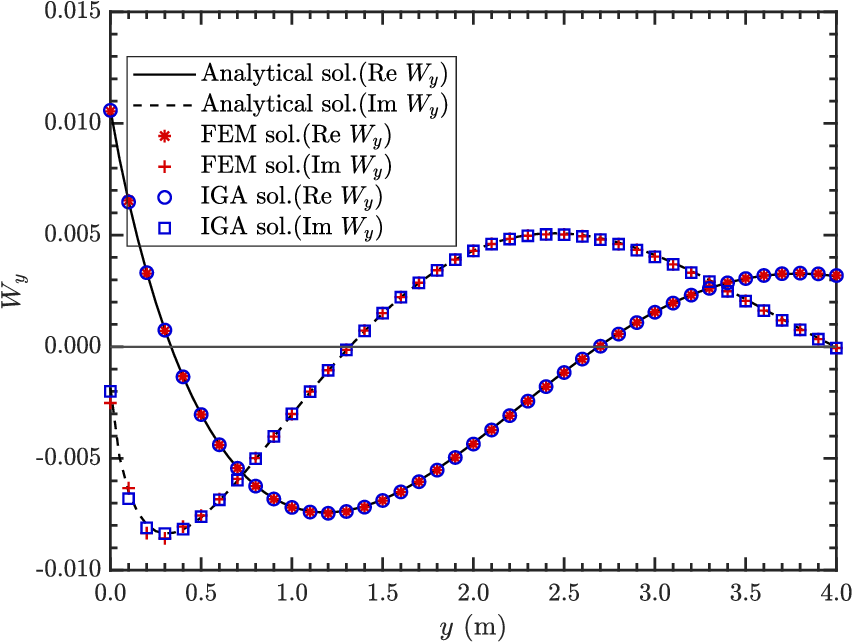}
    \caption{}
  \end{subfigure}
 \caption{Normalized frequency-domain displacement components $V_y$ and $W_y$ for $f=32~\mathrm{Hz}$, $f_0=10~\mathrm{Hz}$, $c=90~\mathrm{m/s}$, and $\zeta=0.05$. Panels (a) and (c): $x$-profiles at $y=1~\mathrm{m}$. Panels (b) and (d): depth-wise profiles at $x=0~\mathrm{m}$.}
  \label{FRF10}
\end{figure}

The effect of a nonzero load-oscillation frequency is examined with $f_0=10~\mathrm{Hz}$ while $f=32~\mathrm{Hz}$ is fixed. In \fref{FRF10}, IGA and FEM reproduce the analytical real and imaginary parts of $V_y$ and $W_y$. Relative to $f_0=0$, the $x$-profiles are more oscillatory and therefore impose a higher spatial-resolution demand; the IGA profiles retain slightly smaller near-surface discrepancies.

In addition to the displacement FRFs, stress FRFs are computed and validated. The analytical expressions for the stress components are given in Eqs.~\eqref{eq:yang_stress_xx}--\eqref{eq:yang_stress_xz}.

\begin{equation}
\begin{aligned}
\tilde{\sigma}_{xx}(k,\omega)
&= \frac{1}{2\pi\mu}\int_{-\infty}^{\infty}\frac{1}{Q}
\left[
\begin{aligned}
&\Big(\frac{\nu}{1-2\nu}k_p^2+k_x^2\Big)\Big(k^2+k_x^2-\frac{1}{2}k_s^2\Big)e^{-m_1 y}\\
&\quad - k_x^2 m_1 m_2 e^{-m_2 y}
\end{aligned}
\right]\cos(k_x x)\,\mathrm{d}k_x
\end{aligned}
\label{eq:yang_stress_xx}
\end{equation}

\begin{equation}
\begin{aligned}
\tilde{\sigma}_{yy}(k,\omega)
&= \frac{1}{2\pi\mu}\int_{-\infty}^{\infty}\frac{1}{Q}
\left[
\begin{aligned}
&\Big(\frac{\nu}{1-2\nu}k_p^2-m_1^2\Big)\Big(k^2+k_x^2-\frac{1}{2}k_s^2\Big)e^{-m_1 y}\\
&\quad + (k^2+k_x^2)m_1 m_2 e^{-m_2 y}
\end{aligned}
\right]\cos(k_x x)\,\mathrm{d}k_x
\end{aligned}
\label{eq:yang_stress_yy}
\end{equation}

\begin{equation}
\begin{aligned}
\tilde{\sigma}_{zz}(k,\omega)
&= \frac{1}{2\pi\mu}\int_{-\infty}^{\infty}\frac{1}{Q}
\left[
\begin{aligned}
&\Big(\frac{\nu}{1-2\nu}k_p^2+k^2\Big)\Big(k^2+k_x^2-\frac{1}{2}k_s^2\Big)e^{-m_1 y}\\
&\quad - k^2 m_1 m_2 e^{-m_2 y}
\end{aligned}
\right]\cos(k_x x)\,\mathrm{d}k_x
\end{aligned}
\label{eq:yang_stress_zz}
\end{equation}

\begin{equation}
\begin{aligned}
\tilde{\tau}_{xy}(k,\omega)
&= \frac{1}{2\pi\mu}\int_{-\infty}^{\infty}
\left[
\begin{aligned}
&-\frac{k_x m_1}{2Q}\Bigg(
2\Big(k^2+k_x^2-\frac{1}{2}k_s^2\Big)e^{-m_1 y}\\
&\quad - \Big(k^2+k_x^2+m_2^2\Big)e^{-m_2 y}
\Bigg)
\end{aligned}
\right]\sin(k_x x)\,\mathrm{d}k_x
\end{aligned}
\label{eq:yang_stress_xy}
\end{equation}

\begin{equation}
\begin{aligned}
\tilde{\tau}_{yz}(k,\omega)
&= \frac{1}{2\pi\mu}\int_{-\infty}^{\infty}
\left[
\begin{aligned}
&\frac{\mathrm{i} k m_1}{2Q}\Bigg(
-2\Big(k^2+k_x^2-\frac{1}{2}k_s^2\Big)e^{-m_1 y}\\
&\quad + \Big(k^2+k_x^2+m_2^2\Big)e^{-m_2 y}
\Bigg)
\end{aligned}
\right]\cos(k_x x)\,\mathrm{d}k_x
\end{aligned}
\label{eq:yang_stress_yz}
\end{equation}

\begin{equation}
\begin{aligned}
\tilde{\tau}_{xz}(k,\omega)
&= \frac{1}{2\pi\mu}\int_{-\infty}^{\infty}\frac{-\mathrm{i} k k_x}{Q}
\left[
\begin{aligned}
&\Big(k^2+k_x^2-\frac{1}{2}k_s^2\Big)e^{-m_1 y}\\
&\quad - m_1 m_2 e^{-m_2 y}
\end{aligned}
\right]\sin(k_x x)\,\mathrm{d}k_x
\end{aligned}
\label{eq:yang_stress_xz}
\end{equation}
In Eqs.~\eqref{eq:yang_stress_xx}--\eqref{eq:yang_stress_xz}, all symbols follow the definitions used for the displacement FRFs; $\mu$ is the shear modulus and $\nu$ denotes Poisson's ratio. The same single-harmonic relation $k=(\omega-\omega_0)/c$ from Eq.~\eqref{eq:k_def} is used, with $\omega=2\pi f$ and $\omega_0=2\pi f_0$.

\begin{figure}[!ht]
  \centering
  \begin{subfigure}[t]{0.36\textwidth}
    \centering
    \includegraphics[width=\linewidth]{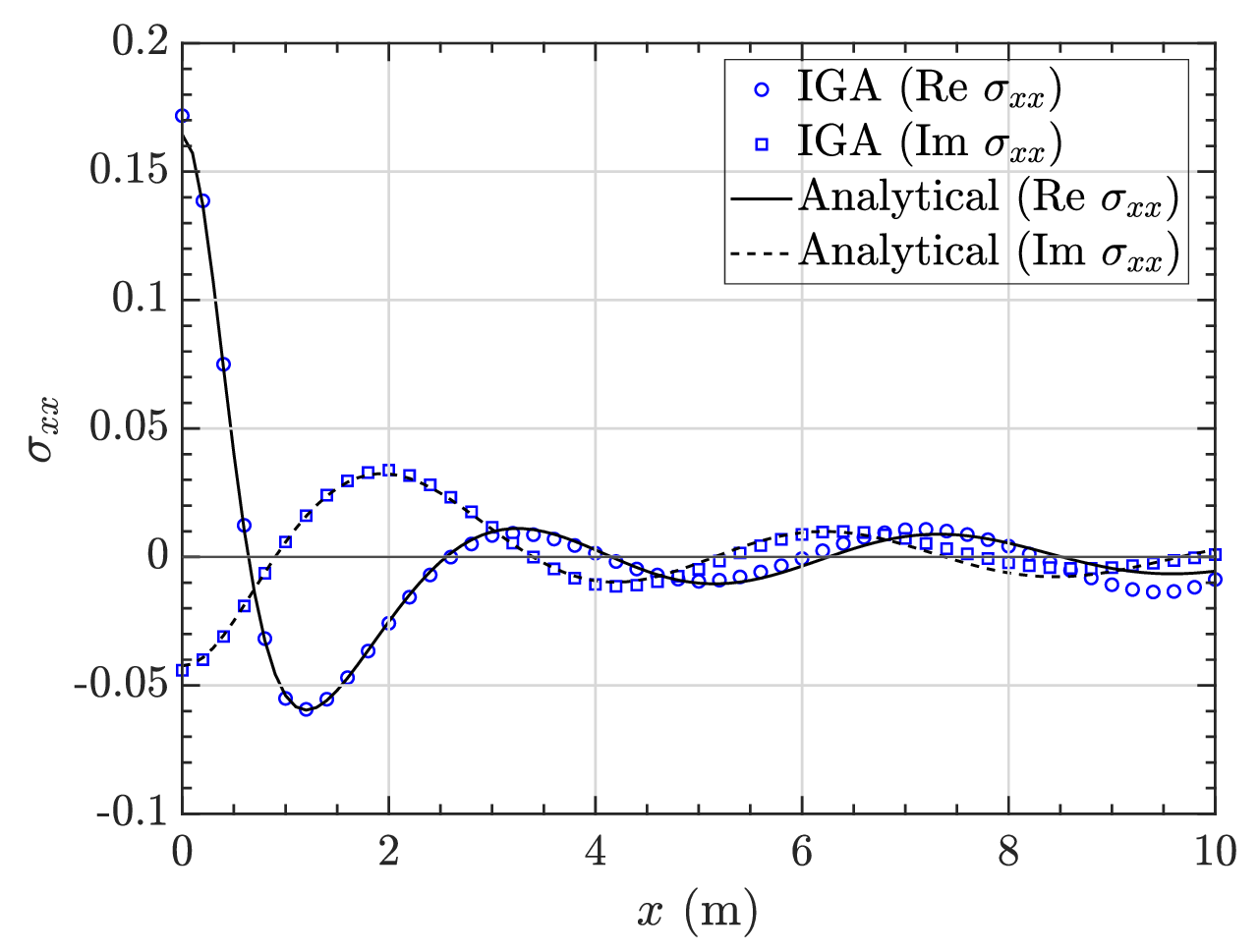}
    \caption{}
  \end{subfigure}\hspace{5mm}
  \begin{subfigure}[t]{0.36\textwidth}
    \centering
    \includegraphics[width=\linewidth]{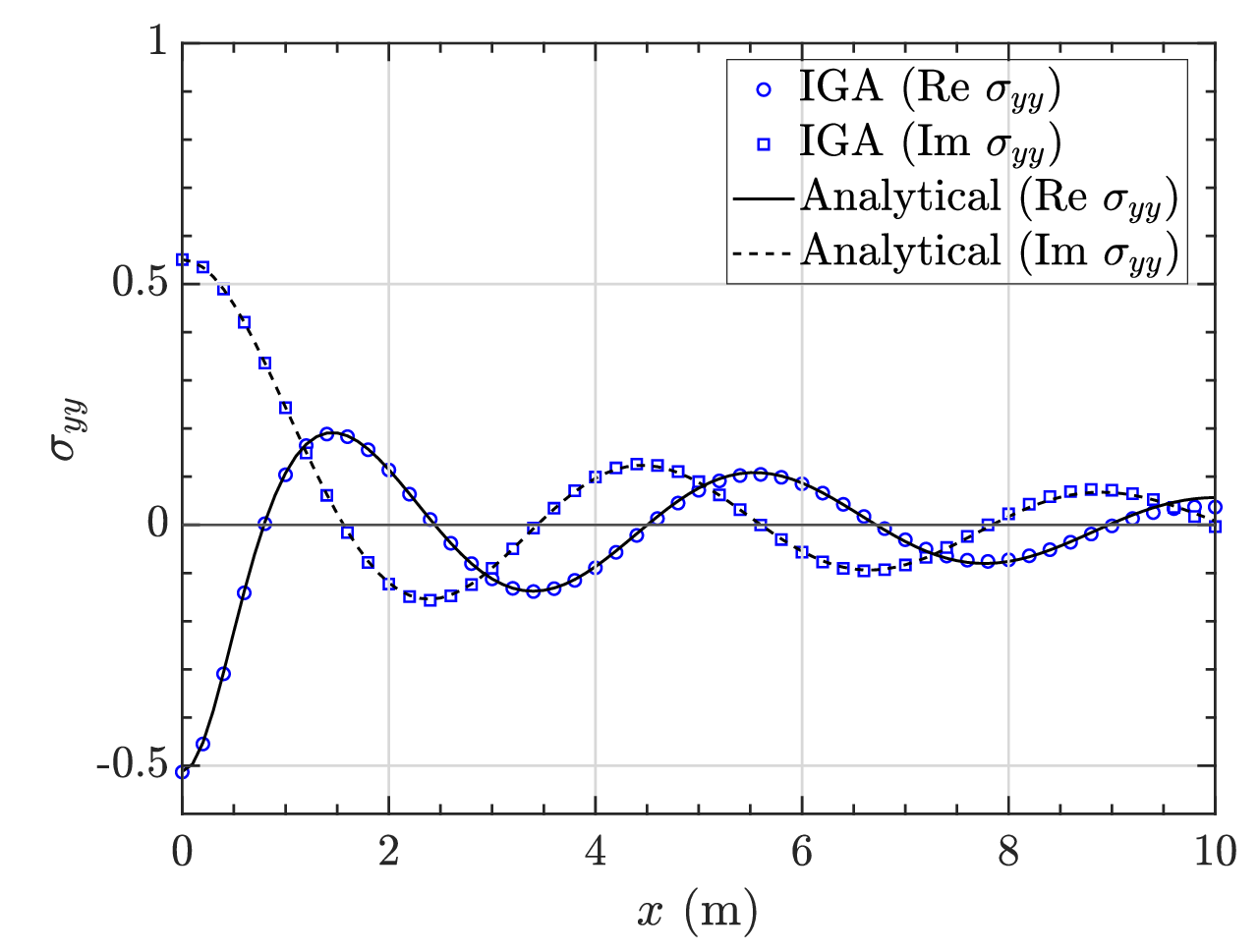}
    \caption{}
  \end{subfigure}
  \begin{subfigure}[t]{0.36\textwidth}
    \centering
    \includegraphics[width=\linewidth]{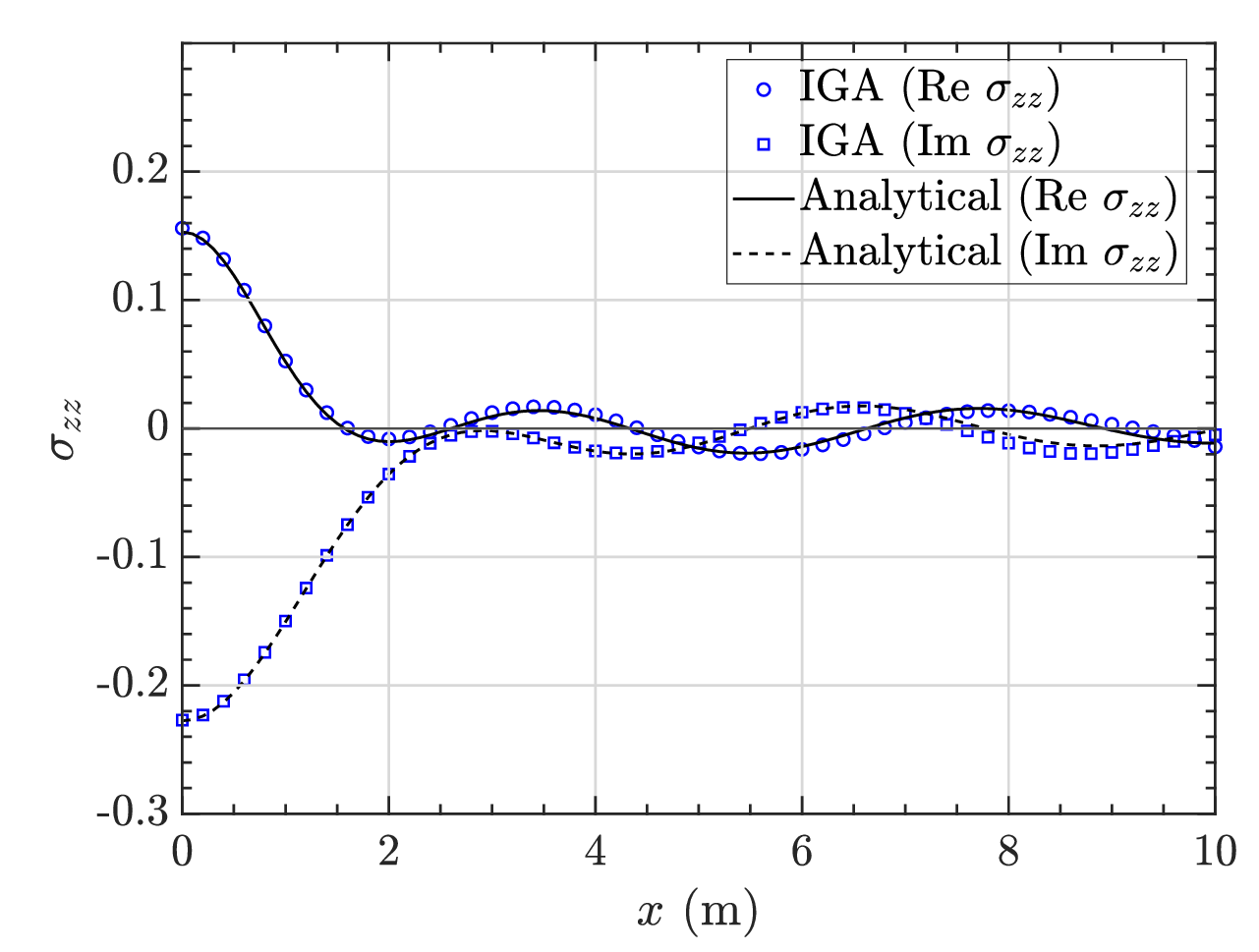}
    \caption{}
  \end{subfigure}\hspace{5mm}
  \begin{subfigure}[t]{0.36\textwidth}
    \centering
    \includegraphics[width=\linewidth]{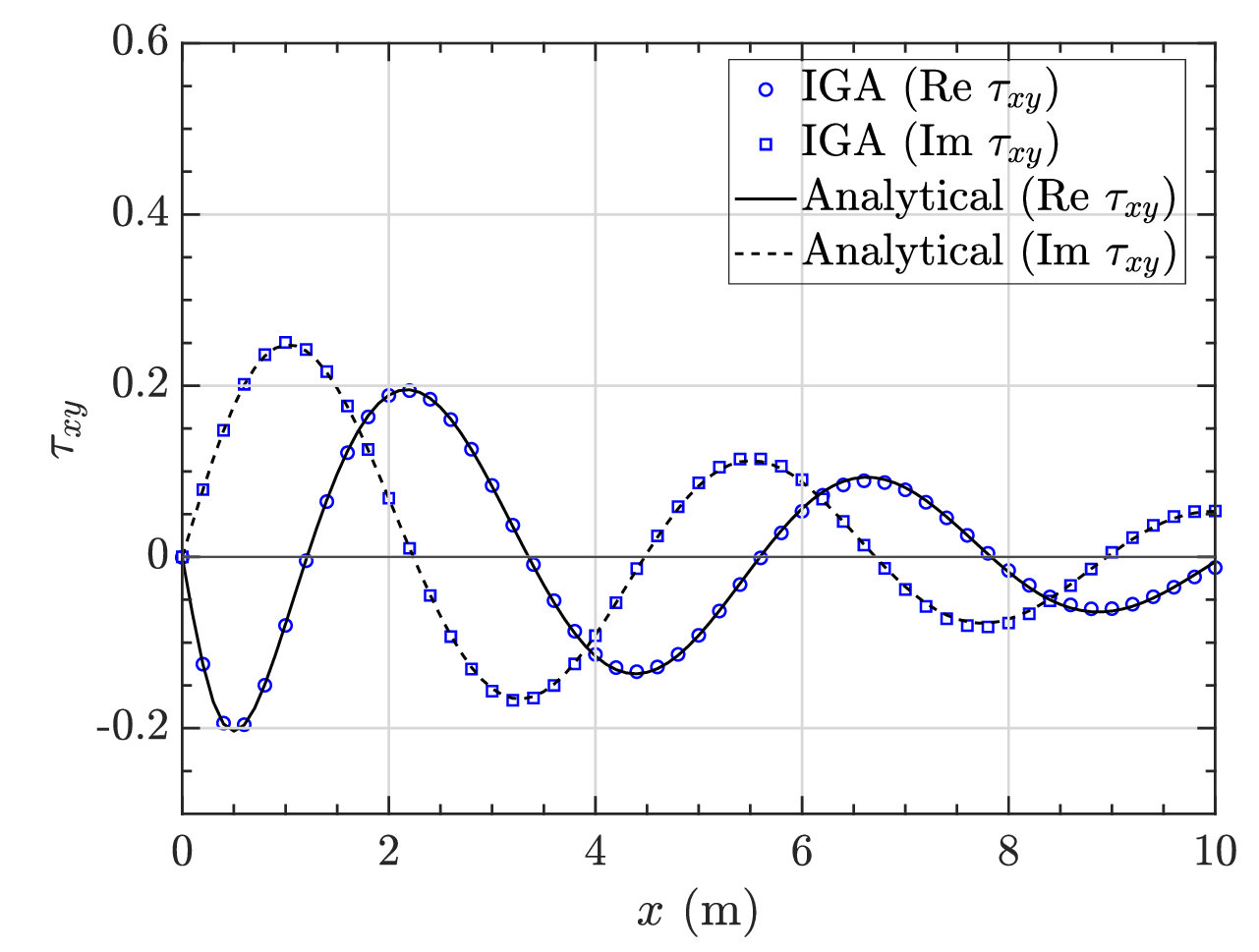}
    \caption{}
  \end{subfigure}
  \begin{subfigure}[t]{0.36\textwidth}
    \centering
    \includegraphics[width=\linewidth]{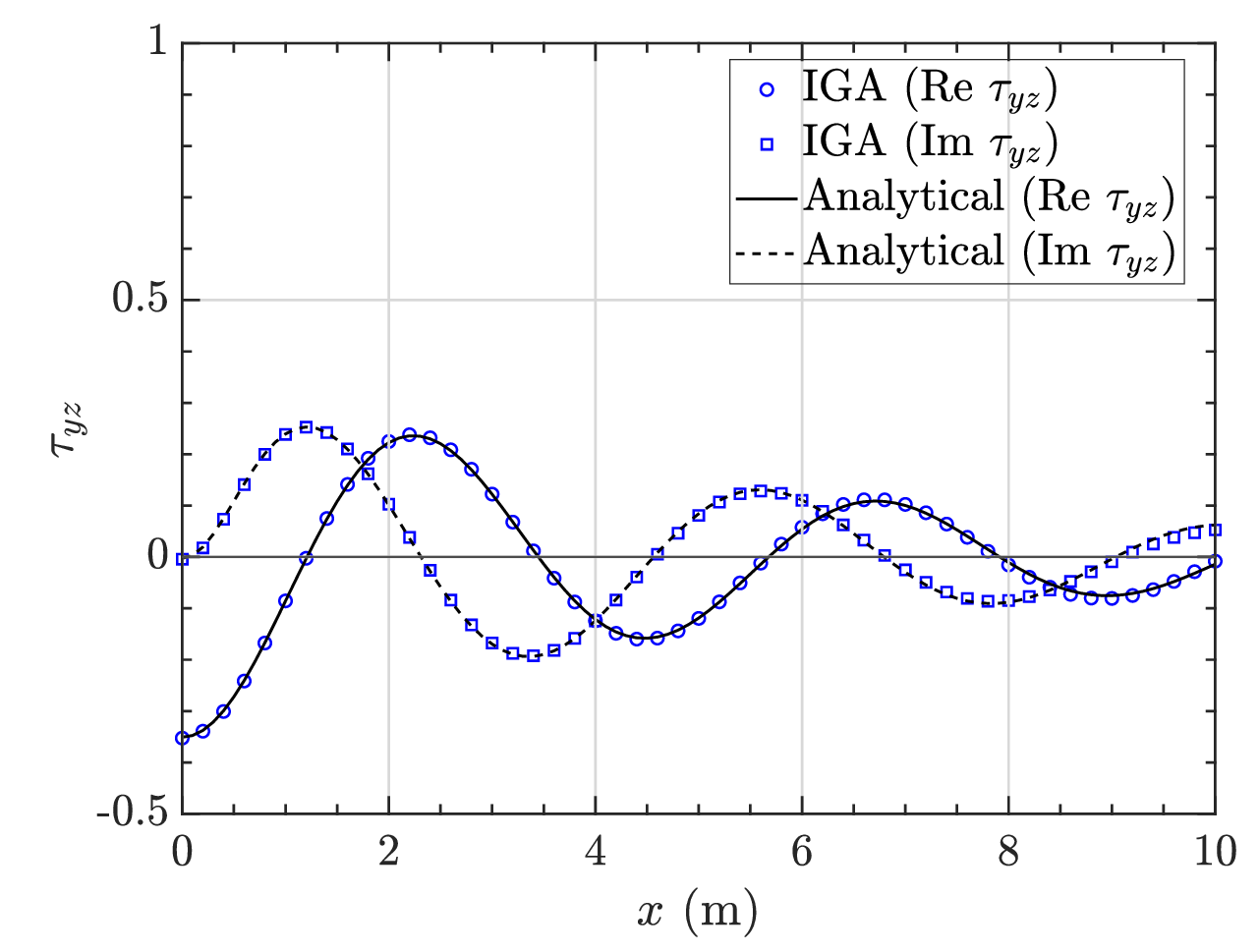}
    \caption{}
  \end{subfigure}\hspace{5mm}
  \begin{subfigure}[t]{0.36\textwidth}
    \centering
    \includegraphics[width=\linewidth]{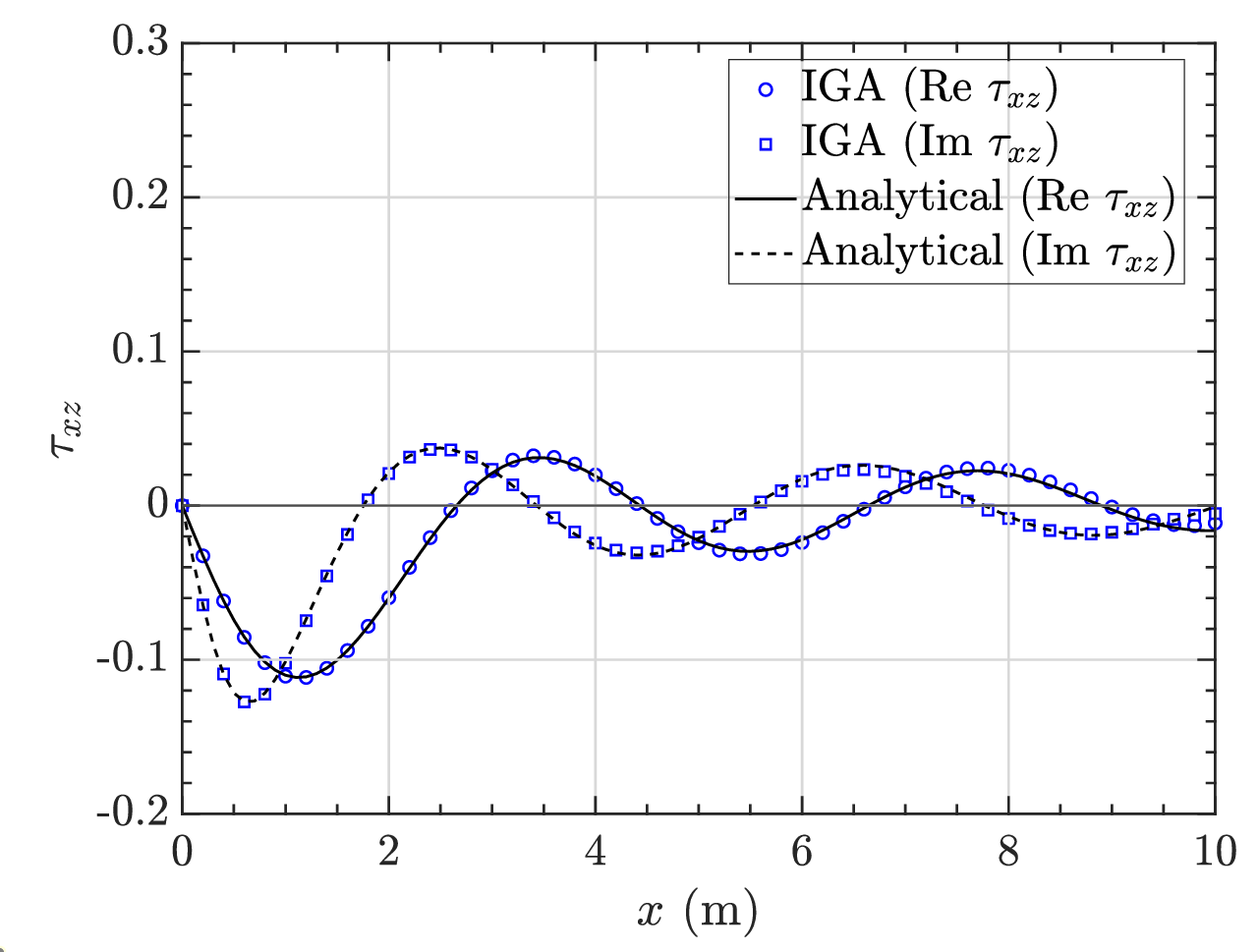}
    \caption{}
  \end{subfigure}
  \caption{Frequency-domain stresses for $f=32~\mathrm{Hz}$, $f_0=0~\mathrm{Hz}$, $c=120~\mathrm{m/s}$, and $\zeta=0.05$, plotted along the $x$-axis at $y=1~\mathrm{m}$. (a) $\sigma_{xx}$; (b) $\sigma_{yy}$; (c) $\sigma_{zz}$; (d) $\tau_{xy}$; (e) $\tau_{yz}$; (f) $\tau_{xz}$.}
  \label{stress_c120}
\end{figure}

The six stress components are compared along the $x$-axis at $y=1~\mathrm{m}$. Because stresses depend on displacement gradients, this provides a more stringent interior-field check than the displacement profiles alone. The retained $c=120~\mathrm{m/s}$ case is the more oscillatory of the evaluated stress regimes. As shown in \fref{stress_c120}, all six IGA components agree well with the analytical real and imaginary responses; the comparison therefore also checks stress recovery from the spline displacement field.

\begin{figure}[!ht]
  \centering
  \includegraphics[width=0.6\linewidth]{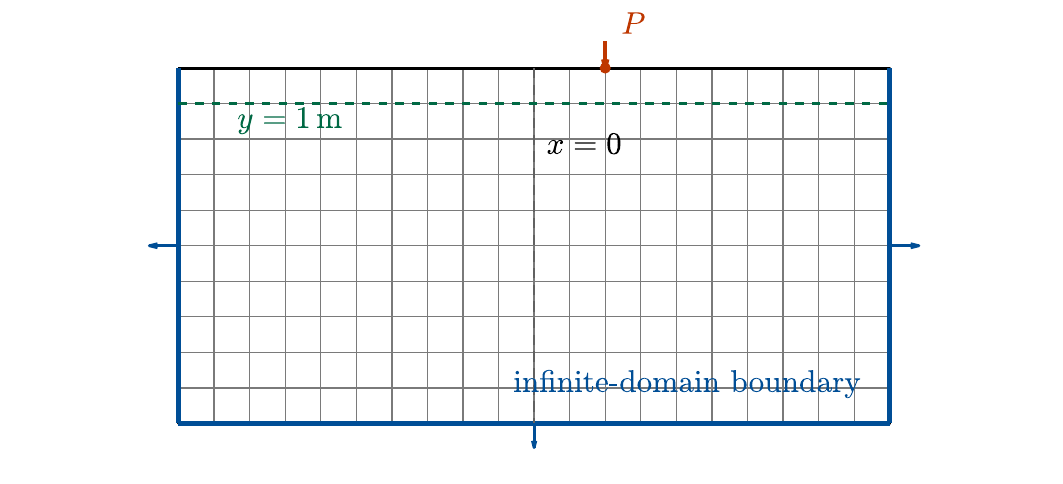}
  \caption{Full-domain configuration for the off-center moving-load verification. The bounded cross-section is coupled to fixed all-S NURBS infinite-domain boundaries along the left, right, and bottom traces. The vertical load is applied at $x_{\mathrm{load}}=2~\mathrm{m}$, and no symmetry constraint is imposed at $x=0$. The transverse receiver profile at $y=1~\mathrm{m}$ is indicated.}
  \label{fig:full-domain-offcenter-model}
\end{figure}

{
As a complementary implementation check, the preceding half-domain benchmark is repeated on the complete transverse cross-section without imposing symmetry at $x=0$. The bounded computational domain is $x\in[-R,R]$ and $y\in[0,H]$, with $R=H=10~\mathrm{m}$. The left, right, and bottom traces are all coupled to the same fixed all-S NURBS infinite-element realization used elsewhere in the paper, while the free surface remains traction-loaded. The formal plotting case uses the $20\times10$ near-field mesh, corresponding to 792 displacement degrees of freedom. The principal off-center case places the vertical load at $x_{\mathrm{load}}=2~\mathrm{m}$, and the analytical half-space response is evaluated after the transverse coordinate shift $x-x_{\mathrm{load}}$. The transverse receiver profile is located at $y=1~\mathrm{m}$, and the five representative receivers are defined relative to the load as $(-4,1)$, $(-2,1)$, $(2,1)$, $(4,1)$, and $(0,3)$ in the $(x-x_{\mathrm{load}},y)$ plane. The configuration is summarized in \fref{fig:full-domain-offcenter-model}.}

\begin{figure}[!ht]
  \centering
  \begin{subfigure}[t]{0.32\textwidth}
    \centering
    \includegraphics[width=\linewidth]{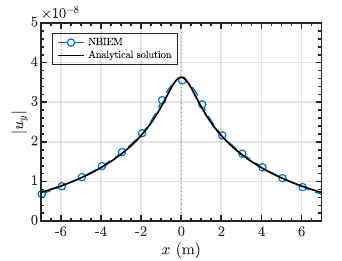}
    \caption{$x_{\mathrm{load}}=0~\mathrm{m}$}
  \end{subfigure}\hfill
  \begin{subfigure}[t]{0.32\textwidth}
    \centering
    \includegraphics[width=\linewidth]{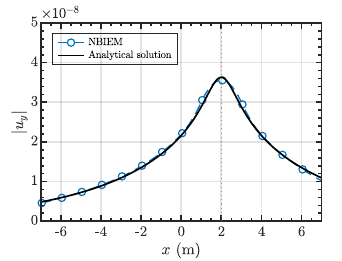}
    \caption{$x_{\mathrm{load}}=2~\mathrm{m}$}
  \end{subfigure}\hfill
  \begin{subfigure}[t]{0.32\textwidth}
    \centering
    \includegraphics[width=\linewidth]{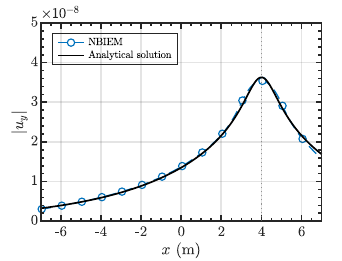}
    \caption{$x_{\mathrm{load}}=4~\mathrm{m}$}
  \end{subfigure}
  \caption{Comparison of the vertical-displacement amplitude along $y=1~\mathrm{m}$ for different transverse load positions at $f=10~\mathrm{Hz}$. NBIEM results are compared with the translated analytical half-space solution.}
  \label{fig:full-domain-load-position}
\end{figure}

{
A transverse load-position consistency check is carried out at $f=10~\mathrm{Hz}$ for $x_{\mathrm{load}}=0$, 2, and $4~\mathrm{m}$, as shown in \fref{fig:full-domain-load-position}. It tests coordinate translation, left--right boundary assembly, full-domain load application, and the behavior when the load is closer to one artificial boundary; it is not interpreted as a physical eccentricity parametric study. A local frequency-robustness check is also evaluated at $x_{\mathrm{load}}=2~\mathrm{m}$ for $f=8$, 10, and $12~\mathrm{Hz}$ and is quantified below. The retained curves report vertical-displacement amplitude, whereas the complex $L_2$ and amplitude-weighted phase errors are evaluated from the original complex responses.
}

{
For the representative case with $x_{\mathrm{load}}=2~\mathrm{m}$, $f=10~\mathrm{Hz}$, and a $20\times10$ mesh, the common-physical-point mesh audit gives a transverse complex $L_2$ error of 4.86\% for the vertical displacement, an amplitude $L_2$ error of 2.76\%, and an amplitude-weighted phase error of $2.33^\circ$. For the final global profiles used in the plots, the corresponding load-position check gives vertical complex errors of 6.36\%, 7.25\%, and 9.97\% for $x_{\mathrm{load}}=0$, 2, and $4~\mathrm{m}$, respectively, with amplitude errors of 4.46\%, 5.03\%, and 6.70\%. For the frequency check at $x_{\mathrm{load}}=2~\mathrm{m}$, the global-profile complex errors are 9.76\%, 7.25\%, and 5.50\% for 8, 10, and $12~\mathrm{Hz}$, and the corresponding phase errors are $4.50^\circ$, $3.22^\circ$, and $2.39^\circ$. The mesh-comparison audit gives M16--M20 transverse vertical-response differences of 1.38\% at $10~\mathrm{Hz}$ and 1.56\% at $12~\mathrm{Hz}$, indicating that the response is mesh-stabilized from the $16\times8$ discretization onward. The 25\% domain-enlargement check changes the 10-Hz vertical response by 3.19\% on the transverse profile, 1.35\% on the depth profile, and 3.08\% at the representative points. These results provide an additional full-domain implementation check without reliance on transverse symmetry.
}

\begin{table}[H]
\centering
\footnotesize
\setlength{\tabcolsep}{4pt}
\renewcommand{\arraystretch}{0.95}
\caption{Representative-point comparison for the full-domain off-center verification at $x_{\mathrm{load}}=2~\mathrm{m}$ and $f=10~\mathrm{Hz}$.}
\label{tab:full-domain-representative-points}
{
\begin{tabular}{lcccccc}
\toprule
Point & $x-x_{\mathrm{load}}$ (m) & $y$ (m) & $|u_y^{\mathrm{NBIEM}}|$ ($10^{-8}$ m) & $|u_y^{\mathrm{ana}}|$ ($10^{-8}$ m) & $|\Delta u_y|$ ($10^{-8}$ m) & Phase difference (deg) \\
\midrule
$P_1$ & $-4$ & 1 & 1.386 & 1.352 & 0.0528 & 1.67 \\
$P_2$ & $-2$ & 1 & 2.192 & 2.161 & 0.0381 & 0.60 \\
$P_3$ & 2 & 1 & 2.178 & 2.161 & 0.0733 & 1.88 \\
$P_4$ & 4 & 1 & 1.324 & 1.352 & 0.1301 & 5.45 \\
$P_5$ & 0 & 3 & 1.910 & 1.881 & 0.0455 & 1.06 \\
\bottomrule
\end{tabular}
}
\end{table}

\subsubsection{Assessment of the radiation boundary and radial decay}

The high-frequency verification in the previous subsection primarily assesses the bounded-domain discretization because the artificial boundary is comparatively remote in terms of wavelength. The present subsection instead defines a default realization of the radiation treatment and a reproducible screening procedure for detecting sensitivity to the selected exterior setting without access to an analytical solution. A controlled mixed-wave calibration is then considered separately to quantify the additional accuracy attainable within its verified parameter envelope.

\paragraph{Reference-free all-S boundary assessment}
The assessment uses a fixed all-S exterior model as the default radial realization of the trace-space formulation in Section~\ref{subsec:IGA_infinite}. The same S-wave-informed radial family is applied on all artificial-boundary segments, and its reference decay coefficient is scaled as
\begin{equation}
\alpha_{S,q}=\beta_S\alpha_{S,q}^{\mathrm{ref}},
\qquad
\beta_S=1\quad\text{by default},
\label{eq:allS_default}
\end{equation}
where $q$ identifies a boundary segment and $\alpha_{S,q}^{\mathrm{ref}}$ is given explicitly by Eq.~\eqref{eq:alphaS_reference}. In the present rectangular domains, $L_q=R$ for a lateral segment and $L_q=H$ for a lower segment. The dimensionless exterior frequency is
\begin{equation}
\eta_S=\frac{\omega R}{c_{S,\infty}},
\label{eq:etaS_alpha}
\end{equation}
where $R$ is the truncation radius and $c_{S,\infty}$ is the shear-wave speed of the exterior material. This definition depends on the far field rather than on local inclusions inside the bounded domain. The all-S realization removes the need to prescribe an a priori Rayleigh/S-wave partition of the artificial boundary and leaves only one radial family to be checked.

For problems without an analytical reference, boundary adequacy is tested by repeating the solution for the fixed perturbation set
\begin{equation}
\mathcal{B}=\{0.75,1.00,1.25\}.
\label{eq:beta_perturbation_set}
\end{equation}
Let $x_r=0,0.2,\ldots,4~\mathrm{m}$ and $y_r=1~\mathrm{m}$ denote the fixed physical receivers used in this audit. The response vector stacks the vertical component and the sign-aligned horizontal component at these receivers,
\begin{equation}
\mathbf{u}_{\beta}
=
\left[
\tilde{u}_{v}(x_1,y_1;\beta),\ldots,\tilde{u}_{v}(x_{N_r},y_{N_r};\beta),
\tilde{u}_{h}^{\mathrm{sa}}(x_1,y_1;\beta),\ldots,\tilde{u}_{h}^{\mathrm{sa}}(x_{N_r},y_{N_r};\beta)
\right]^T,
\label{eq:boundary_response_vector}
\end{equation}
where ``sa'' denotes the fixed sign transformation required to reconcile the analytical and numerical transverse-axis conventions. The dimensionless radial-sensitivity indicator is defined as
\begin{equation}
I_{\beta}
=
\max_{\substack{\beta_a,\beta_b\in\mathcal{B}\\\beta_a<\beta_b}}
\frac{\|\mathbf{u}_{\beta_a}-\mathbf{u}_{\beta_b}\|_2}
{\|\mathbf{u}_{\beta=1}\|_2}.
\label{eq:beta_spread}
\end{equation}
\begin{table}[H]

\centering
\footnotesize
\setlength{\tabcolsep}{3pt}
\renewcommand{\arraystretch}{0.95}
\caption{Parameter-robustness matrix for the fixed all-S workflow. This table is an $I_{\beta}$ sensitivity screen only; it neither establishes solution accuracy nor validates the calibrated mixed-wave rule in Eqs.~\eqref{eq:betaR_calibrated}--\eqref{eq:betaS_calibrated}. Boundary-placement assessment additionally requires $I_D$.}
\label{tab:allS_robustness_matrix}
\begin{tabular}{L{2.35cm}L{4.15cm}L{3.0cm}L{4.55cm}}
\toprule
Varied factor & Tested settings & $I_{\beta}$ & Principal observation \\
\midrule
Exterior frequency & $f=2,5,8$ Hz; $\eta_S=1.257,3.142,5.027$ & $0.2853,0.0825,0.0118$ & Sensitivity decreases with exterior distance measured in wavelengths; the 2-Hz case triggers enlargement. \\
Load-speed ratio & $M_S=0.6,0.9,1.2$ at $\eta_S=\pi$ & $0.0002,0.0825,0.3153$ & The default is insensitive in the sub-shear cases; the super-shear case triggers enlargement. \\
Poisson ratio & $\nu=0.15,0.25,0.35,0.45$ & $0.1191$--$0.0448$ & All $I_{\beta}$ values remain below the preliminary sensitivity screen; $I_D$ is still required. \\
Damping ratio & $\zeta=0.01,0.05,0.10$ & $0.1323$--$0.0392$ & Sensitivity decreases as material attenuation increases. \\
Boundary aspect ratio & $H/R=0.5,1,2$ & $0.2157,0.0825,0.0265$ & A shallow lower boundary is detected directly by the indicator. \\
Similarity scaling & $c_S=80,100,120$ m/s at fixed $\eta_S$ and $M_S$ & $0.082481$ in all cases & Dimensionless invariance is recovered to numerical precision. \\
Trace resolution & $8\times4$, $12\times6$, $16\times8$ & $0.0824$--$0.0825$ & The indicator is independent of trace refinement over the tested range. \\
NURBS degree & quadratic and cubic trace bases & $0.08248,0.08249$ & The diagnostic is unchanged by degree elevation. \\
Near-field contrast & homogeneous, stiff inclusion, local soft layer & $0.0825,0.0319,0.1852$ & The local soft layer triggers boundary enlargement; exterior material remains unchanged. \\
\bottomrule
\end{tabular}
\end{table}

The heterogeneous holdout uses the same $R=H=10~\mathrm{m}$ exterior and changes only the bounded material. The stiff inclusion occupies $2\leq x\leq5~\mathrm{m}$ and $3\leq y\leq6~\mathrm{m}$, with $E_{\mathrm{inc}}/E_{\infty}=5$, $\nu_{\mathrm{inc}}=0.20$, and $\rho_{\mathrm{inc}}/\rho_{\infty}=1.20$. The local soft layer occupies $0\leq x\leq7~\mathrm{m}$ and $0\leq y\leq3~\mathrm{m}$, with $E_{\mathrm{soft}}/E_{\infty}=0.35$, $\nu_{\mathrm{soft}}=0.35$, and $\rho_{\mathrm{soft}}/\rho_{\infty}=0.90$.

The subscript $\beta$ indicates that the sensitivity is evaluated by perturbing $\beta_S$; $I_{\beta}$ is not a material or infinite-element parameter. A small value shows that the response is stationary with respect to a finite perturbation of the radial decay. A second, independent check complements this decay-parameter test by comparing two successive near-field domains $D_j$ and $D_{j+1}$ while preserving the physical receiver locations and the resolution per unit length. Here $\mathbf{u}_{D_j}$ has exactly the component ordering and receiver coordinates in Eq.~\eqref{eq:boundary_response_vector} and is evaluated with $\beta_S=1$. Between successive evaluations, it differs only through the domain $D_j$:
\begin{equation}
I_D^{(j)}
=
\frac{\|\mathbf{u}_{D_{j+1}}-\mathbf{u}_{D_j}\|_2}
{\|\mathbf{u}_{D_{j+1}}\|_2}.
\label{eq:domain_change_indicator}
\end{equation}
Neither $I_{\beta}$ nor $I_D$ uses the analytical solution. The working value $0.15$ is applied to both indicators in the present screening study and is audited below against the true complex displacement error. The two tests answer different questions: $I_{\beta}$ detects decay-parameter sensitivity on a fixed domain, whereas $I_D$ detects residual dependence on boundary placement even when the response happens to be stationary with respect to $\beta_S$.

Table~\ref{tab:allS_robustness_matrix} establishes parameter robustness of the all-S model, but parameter robustness is not equivalent to accuracy. The indicator pair is therefore calibrated against the closed-form half-space solution using ordered domain-enlargement sequences at low frequency, at super-shear speed, and for the lower-boundary depth. The analytical response is used only in this audit and is not used to choose $\beta_S$, $I_{\beta}$, or $I_D$.

\begin{figure}[!ht]

\centering
\includegraphics[width=0.88\linewidth]{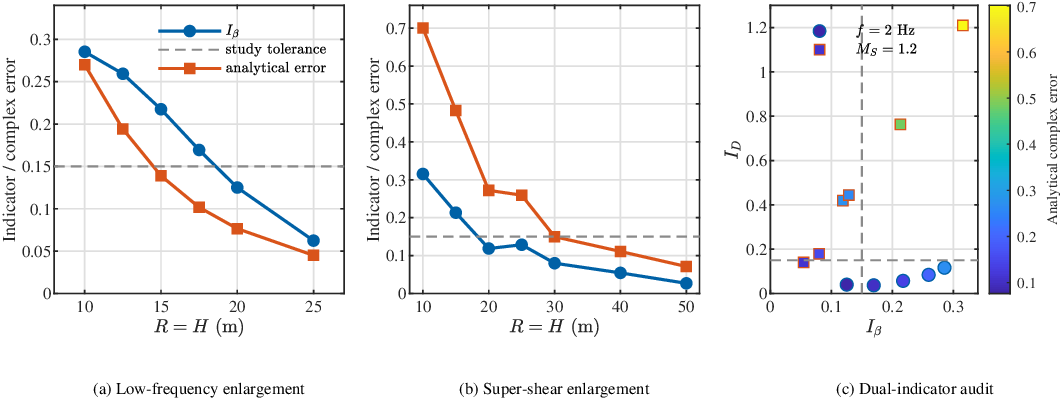}
\caption{Analytical audit of the two reference-free boundary indicators. (a) Simultaneous reduction of $I_{\beta}$ and the true sign-aligned complex displacement error for the $2$-Hz domain sweep; (b) Corresponding super-shear sweep; (c) $I_D$ versus $I_{\beta}$ for successive enlargements, colored by the analytical error on the current domain. The dashed lines show the study screening value of $0.15$.}
\label{fig:boundary_indicator_validation}
\end{figure}

\begin{table}[!htbp]

\centering
\footnotesize
\setlength{\tabcolsep}{4pt}
\caption{Selected analytical audit cases for the all-S boundary workflow. The indicators are computed without the analytical solution; $e_j$ and $e_{j+1}$ are the sign-aligned complex displacement errors on the current and enlarged domains.}
\label{tab:boundary_indicator_audit}
\begin{tabular}{lccccc}
\toprule
Case and successive domains & $I_{\beta}^{(j)}$ & $I_D^{(j)}$ & $e_j$ & $e_{j+1}$ & Indicator result at $D_j$ \\
\midrule
$f=2$ Hz, $R_j/R_{j+1}=10/12.5$ m & 0.2853 & 0.1164 & 0.2700 & 0.1941 & enlarge \\
$f=2$ Hz, $R_j/R_{j+1}=20/25$ m & 0.1251 & 0.0400 & 0.0765 & 0.0450 & dual screen met \\
$M_S=1.2$, $R_j/R_{j+1}=10/15$ m & 0.3153 & 1.2096 & 0.7007 & 0.4825 & enlarge \\
$M_S=1.2$, $R_j/R_{j+1}=20/25$ m & 0.1186 & 0.4191 & 0.2721 & 0.2594 & enlarge by $I_D$ \\
$M_S=1.2$, $R_j/R_{j+1}=40/50$ m & 0.0544 & 0.1400 & 0.1109 & 0.0712 & dual screen met \\
$H_j/H_{j+1}=10/12.5$ m & 0.0825 & 0.0109 & 0.0612 & 0.0541 & dual screen met \\
\bottomrule
\end{tabular}
\end{table}

Figure~\ref{fig:boundary_indicator_validation} and Table~\ref{tab:boundary_indicator_audit} demonstrate why both checks are retained. In the 2-Hz sequence, $I_{\beta}$ and the true error decrease together from $(0.2853,0.2700)$ at $R=10~\mathrm{m}$ to $(0.0624,0.0450)$ at $R=25~\mathrm{m}$. The super-shear case supplies the more demanding counter-test: at $R=20~\mathrm{m}$, $I_{\beta}=0.1186$ alone would indicate apparent stability although the true error remains $0.2721$; the large value $I_D=0.4191$ identifies this false acceptance. For the successive domains $R_j=40~\mathrm{m}$ and $R_{j+1}=50~\mathrm{m}$, the indicators are $I_{\beta}=0.0544$ and $I_D=0.1400$, while the corresponding analytical errors are $0.1109$ and $0.0712$. Across the tested enlargement sequences, cases satisfying both $I_{\beta}\leq0.15$ and $I_D\leq0.15$ have a current-domain analytical error no larger than $0.1109$, while the enlarged-domain error is no larger than $0.0712$. The value $0.15$ is therefore used as an empirically audited boundary-sensitivity screening level for this study, with the tested dual-pass cases reaching approximately $11\%$ or smaller analytical error on the current domain.

The all-S workflow is consequently staged but reference-free: compute the fixed $\beta_S=1$ response and its three-point perturbation, then repeat the final candidate on one enlarged domain. The boundary placement is classified as insensitive under the present screen only when both Eqs.~\eqref{eq:beta_spread} and \eqref{eq:domain_change_indicator} satisfy the study values; otherwise, the near field is enlarged without reference-based retuning. This dual check preserves the simplicity of the all-S realization while guarding against a radial family that is insensitive to $\beta_S$ but still affected by boundary placement. Accuracy is established separately by an analytical benchmark or conventional discretization refinement whenever such evidence is available.

\paragraph{Controlled mixed-wave calibration}
To quantify the additional accuracy that radial calibration can provide in a compact domain, the mixed-wave calibration developed in this work is retained as a controlled secondary experiment. It assigns Rayleigh- and S-wave-informed radial functions to the surface-adjacent and remaining exterior segments, respectively, and scales their reference decay coefficients by $\beta_R^{\mathrm{cal}}(\eta_S)$ and $\beta_S^{\mathrm{cal}}(\eta_S)$. Frequency-by-frequency minimization of the sign-aligned complex displacement error in the homogeneous reference model is summarized by
\begin{equation}
\beta_R^{\mathrm{cal}}(\eta_S)=
\begin{cases}
0, & 0.314\leq\eta_S<0.628,\\
\max\!\left(0,-0.0644\eta_S^2+0.8014\eta_S-0.5029\right),
& 0.628\leq\eta_S\leq5.027,
\end{cases}
\label{eq:betaR_calibrated}
\end{equation}
and
\begin{equation}
\beta_S^{\mathrm{cal}}(\eta_S)=
\begin{cases}
2.364\eta_S-0.0105, & 0.314\leq\eta_S<0.628,\\
0.0206\eta_S^3-0.2028\eta_S^2+0.9636\eta_S+0.9325,
& 0.628\leq\eta_S\leq5.027.
\end{cases}
\label{eq:betaS_calibrated}
\end{equation}
These curves compactly represent the calibration model and are not additional constitutive parameters of NBIEM. At the two diagnostic points with $\eta_S>5.027$, the polynomial branches are evaluated by direct continuation solely to inspect behavior beyond the calibration interval; their range of claimed validity is not extended.

\begin{figure}[!ht]

\centering
\includegraphics[width=0.48\linewidth]{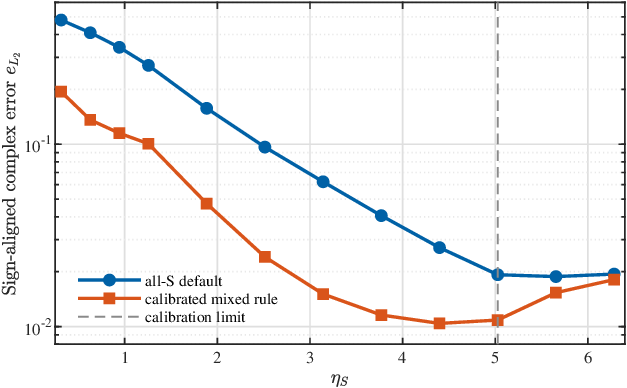}
\caption{Sign-aligned complex displacement errors of the all-S default and calibrated mixed-wave rule in the homogeneous calibration configuration ($R=H=10~\mathrm{m}$, $M_S=0.9$, $\nu=0.25$, and $\zeta=0.05$). The dashed line marks the upper limit of the fitted interval. Results beyond this line are shown only to assess continuation outside the calibration model.}
\label{fig:calibrated_rule_effect}
\end{figure}

Figure~\ref{fig:calibrated_rule_effect} compares the absolute complex errors of the two radial treatments. Both improve rapidly as $\eta_S$ increases and reach errors of order $10^{-2}$ near the upper part of the calibration interval. The calibrated rule provides an additional reduction in the compact-domain low-frequency cases, whereas the separation between the curves becomes small once the all-S response is already weakly dependent on radial decay. The zero-frequency endpoint is represented numerically by the lowest positive frequency sample and is used only to close the discrete spectrum. The calibration is therefore retained as an optional accuracy enhancement within its declared model, rather than as a replacement for the transferable all-S workflow.

\paragraph{Controlled hybrid time reconstruction}
{\strut}
{
Although the governing systems are solved frequency by frequency, the resulting complex spectra can be synthesized into linear time histories through inverse Fourier reconstruction.
}
The practical time-domain influence of the low-frequency radial treatment is assessed using controlled inverse reconstructions over $0$--$200~\mathrm{Hz}$ and $0$--$400~\mathrm{Hz}$. Within the nominal $0$--$8~\mathrm{Hz}$ interval, the computed all-S and calibrated mixed-wave NBIEM spectra are retained, whereas both reconstructions use the same analytical half-space spectrum at higher frequencies. At the zero-frequency endpoint, the lowest positive frequency sample supplies the numerical endpoint value. This construction directly isolates whether the low-frequency radial-rule difference remains significant in conventional broadband ground-response reconstructions. The common cutoff is increased from $20$ to $400~\mathrm{Hz}$ to examine convergence with reconstruction bandwidth.

\begin{figure}[!ht]

\centering
\includegraphics[width=0.88\linewidth]{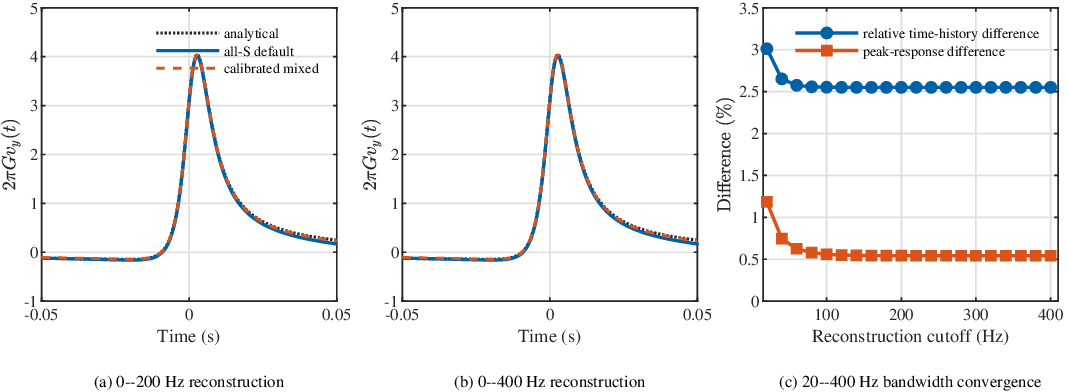}
\caption{Controlled broadband assessment of the all-S and calibrated mixed-wave radial treatments. (a) Analytical, all-S, and calibrated mixed-wave histories reconstructed over $0$--$200~\mathrm{Hz}$; (b) Corresponding histories over $0$--$400~\mathrm{Hz}$; (c) Relative time-history and peak-response differences as the common reconstruction cutoff is increased from $20$ to $400~\mathrm{Hz}$. The two numerical spectra differ only in the nominal low-frequency calibration interval and use identical analytical content at higher frequencies.}
\label{fig:time_reconstruction_comparison}
\end{figure}

Figure~\ref{fig:time_reconstruction_comparison}(a)--(b) shows that the all-S and calibrated mixed-wave histories are nearly coincident in both broadband reconstructions. For the reconstruction with a $200$-Hz cutoff, their relative time-history difference is $2.55\%$, the peak-response difference is $0.54\%$, the maximum pointwise difference is $1.60\%$, and the correlation coefficient is $0.9999$; the corresponding $400$-Hz-cutoff values are unchanged to the reported precision. Figure~\ref{fig:time_reconstruction_comparison}(c) further shows that the difference measures become essentially stationary once the reconstruction bandwidth exceeds approximately $80~\mathrm{Hz}$.

{
Relative to the analytical history on the same time grid, the $200$-Hz hybrid reconstruction gives time-history errors of $3.14\%$ and $0.843\%$, and peak-response errors of $0.729\%$ and $0.188\%$, for the all-S and calibrated mixed-wave spectra, respectively; the corresponding $400$-Hz values are unchanged to the reported precision. Because both numerical histories use the common analytical spectrum above $8~\mathrm{Hz}$, these measures quantify how the directly computed low-frequency differences propagate through the controlled reconstruction rather than constituting a standalone full-band transient validation.
}

A separate direct same-model Fourier-grid test holds each radial treatment fixed and compares its reconstructed history with the $\Delta f=0.05~\mathrm{Hz}$ result. Every NBIEM value on the requested grids within $0$--$8~\mathrm{Hz}$ is solved directly; no interpolated value is introduced as a new NBIEM sample. For $\Delta f=0.25$, $0.125$, and $0.10~\mathrm{Hz}$, the all-S grid errors are $1.03\%$, $0.39\%$, and $0.26\%$, while the calibrated mixed-wave grid errors are $1.75\%$, $0.65\%$, and $0.44\%$. These are same-model spectral-sampling errors of the controlled hybrid reconstruction, unlike the all-S/mixed-wave differences shown in Fig.~\ref{fig:time_reconstruction_comparison}. Thus, within the present controlled hybrid reconstructions, the calibrated rule improves the complex low-frequency response in Fig.~\ref{fig:calibrated_rule_effect}, but its influence on the broadband histories is limited. The fixed all-S realization preserves the principal waveform and peak response without calibration-dependent boundary partitioning.

The calibration and robustness results serve different purposes. Figure~\ref{fig:calibrated_rule_effect} establishes the accuracy gain attainable when a trusted reference is available, whereas Fig.~\ref{fig:boundary_indicator_validation} supports the complementary reference-free checks used when transferring the method to new configurations. The calibrated rule is therefore retained as a controlled enhancement within its declared envelope, while the all-S setting and Eqs.~\eqref{eq:beta_spread}--\eqref{eq:domain_change_indicator} define the transferable default workflow.

\paragraph{High-frequency phase and spectral audits}
Where an analytical response is available, accuracy is reported using the vertical displacement and a sign-aligned horizontal displacement. The analytical horizontal component uses a transverse-axis convention opposite to that of the numerical model; this deterministic coordinate transformation is applied before any combined norm or phase comparison. The resulting evidence is organized by error source in \tref{tab:error_budget}, rather than summarized by a single undifferentiated displacement error.

The amplitude-weighted phase error used here and in the following subsection is
\begin{equation}
e_{\phi}
=
\left[
\frac{
\displaystyle\sum_{m\in\mathcal{M}}\sum_{r\in\mathcal{S}}
w_{mr}\left(\phi^{\mathrm{num}}_m(\mathbf{x}_r)-\phi^{\mathrm{ana}}_m(\mathbf{x}_r)\right)^2
}{
\displaystyle\sum_{m\in\mathcal{M}}\sum_{r\in\mathcal{S}}w_{mr}
}
\right]^{1/2},
\qquad
w_{mr}=\left|\tilde{u}^{\mathrm{ana}}_m(\mathbf{x}_r)\right|^2,
\label{eq:method_phase_error}
\end{equation}
where $\phi=\operatorname{unwrap}(\arg\tilde{u})$ is evaluated separately for each displacement component along the ordered receiver profile before the component vectors are stacked. The weighting suppresses ill-conditioned phases near zeros of the analytical response.

\begin{figure}[!ht]

\centering
\includegraphics[width=0.76\linewidth]{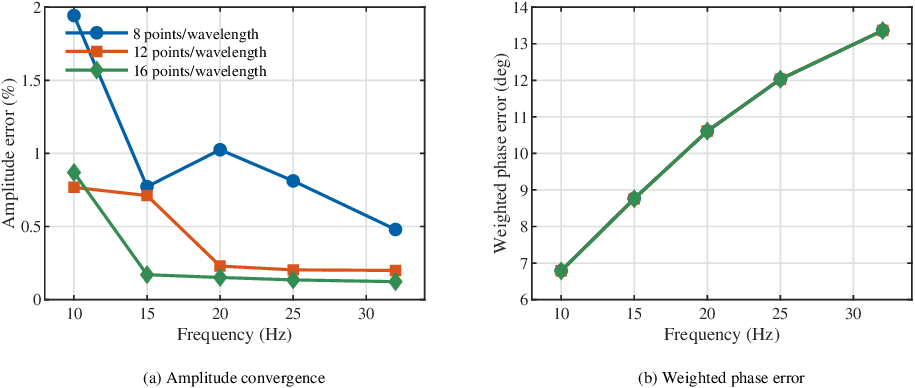}
\caption{Independent high-frequency mesh audit for NBIEM. (a) Amplitude error against the analytical half-space solution; (b) Amplitude-weighted phase error. Each curve maintains a fixed nominal number of points per S wavelength as frequency increases.}
\label{fig:high_frequency_audit}
\end{figure}

The high-frequency discretization is audited independently over $10$--$32~\mathrm{Hz}$ using five frequencies and three meshes that maintain approximately 8, 12, and 16 points per S wavelength. Figure~\ref{fig:high_frequency_audit} shows that the finest-mesh amplitude error is $0.87\%$ at $10~\mathrm{Hz}$ and below $0.2\%$ from $15$ to $32~\mathrm{Hz}$, whereas the amplitude-weighted phase error increases with frequency and changes little under proportional mesh refinement. This pattern is consistent with the acoustic IGA literature on pollution-type phase accumulation and high-frequency wave behavior \citep{KhajahAntoineBordas2019Acoustic}, but the audit does not uniquely separate that contribution from residual exterior-model error, radial approximation, systematic phase offset, or receiver-distance effects. It therefore provides a high-frequency phase diagnostic rather than a single-cause attribution. The audit also prevents the $2$--$8~\mathrm{Hz}$ method comparison from being used as the sole evidence for high-frequency behavior.

For inverse reconstruction, frequency and longitudinal-wavenumber sampling are not independent because Eq.~\eqref{eq:k_def} gives
\begin{equation}
\Delta k=\frac{\Delta\omega}{c}
=\frac{2\pi\Delta f}{c}.
\label{eq:wavenumber_sampling}
\end{equation}
Thus every frequency-grid refinement reported above simultaneously refines the moving-load wavenumber grid; a separate $k$ quadrature is not used for a single moving harmonic component.

\begin{table}[!htbp]

\centering
\scriptsize
\setlength{\tabcolsep}{3pt}
\renewcommand{\arraystretch}{0.92}
\caption{Numerical verification and sensitivity-assessment matrix. The reported displacement values are total errors under controlled refinement and are used to demonstrate stationarity; they are not interpreted as separately additive error-budget components.}
\label{tab:error_budget}
\begin{tabular}{L{2.45cm}L{4.15cm}L{4.05cm}L{3.05cm}}
\toprule
Numerical contribution & Isolation strategy & Numerical evidence & Assessment used in this work \\
\midrule
Cross-sectional discretization & Repeat the baseline case with coarse, reference, and fine NURBS meshes at fixed exterior model. & Total sign-aligned errors $0.0637$, $0.0612$, and $0.0613$ demonstrate mesh stationarity of the reported response measure. & Resolve the minimum wavelength and verify nested-mesh stationarity. \\
Spline degree and boundary quadrature & Elevate the quadratic trace to cubic and repeat the cubic result with 3, 4, and 5 Gauss points per knot span. & Degree-elevation errors are $0.0611907$ and $0.0612118$; the cubic response and $I_{\beta}$ change by less than $1.1\times10^{-9}$ from 3 to 5 points. & Use $n_g=\max(3,p_b+1)$ in production and confirm degree robustness. \\
Radiation boundary & Perturb $\beta_S$ and independently enlarge the bounded domain, then audit both indicators against the analytical error. & The $R=20$ m super-shear apparent pass by $I_{\beta}$ is rejected by $I_D=0.4191$; for $R_j/R_{j+1}=40/50$ m, the error falls from $0.1109$ to $0.0712$. & Apply both $I_{\beta}$ and $I_D$ for the sensitivity screen; verify accuracy separately. \\
Radial integration & Compare Eq.~\eqref{eq:radial_moment_exact} and its derivative-moment matrix with direct adaptive complex integration on $[0,\infty)$. & No finite cutoff or user variable transform; relative/absolute tolerances $10^{-12}/10^{-14}$ give maximum mass/stiffness residuals $1.58\times10^{-16}/1.42\times10^{-16}$. & Production radial moments are analytical; audit residuals use the relative Frobenius norm. \\
High-frequency phase behavior & Maintain 8, 12, and 16 points per wavelength over five frequencies from 10 to 32 Hz. & Finest-mesh amplitude error is $0.12$--$0.87\%$; phase growth is weakly affected by proportional mesh refinement. & Treat the result as consistent with, but not a unique identification of, accumulated phase pollution. \\
Spectral sampling & Directly solve every low-frequency NBIEM node on each fixed-model grid and map it to $k$ through Eq.~\eqref{eq:wavenumber_sampling}. & Against $\Delta f=0.05$ Hz, all-S errors at $0.25/0.125/0.10$ Hz are $1.03/0.39/0.26\%$; mixed-wave errors are $1.75/0.65/0.44\%$. & Verify direct same-model $\Delta f$ convergence; $\Delta k$ follows simultaneously. \\
\bottomrule
\end{tabular}
\end{table}

The assessment results support the following reproducible workflow. The fixed all-S setting is used without reference-based tuning, and candidate boundary placements are screened through both decay-parameter stability and domain-enlargement stability. This separates the transferable workflow from the calibrated mixed-wave accuracy enhancement while leaving the NURBS-trace coupling unchanged.

\subsubsection{Comparison with alternative boundary treatments}

To further assess the proposed NBIEM, its accuracy is compared with two boundary-treatment strategies that are closest to the present formulation. The conventional finite/infinite element formulation (FEM/IEM) is selected as the classical 2.5D moving-load reference because it uses the same finite/infinite-element radiation philosophy on a low-order finite-element trace \citep{YangLiu-56}. The scaled-boundary isogeometric analysis formulation (IGA/SBIGA) is selected as the spline-based reference because it represents the NURBS-based counterpart of the FEM--SBFEM route for 2.5D train-induced vibration problems \citep{BJie-72}. Perfectly matched layers provide another important family of artificial boundaries, and displacement-based 2.5D PML formulations have also been developed \citep{FrancoisSchevenelsPML-XX}. A PML baseline is not used here because its accuracy depends on absorbing-layer thickness, stretching or attenuation functions, and layer discretization, whereas the present comparison focuses on boundary-attached exterior discretizations coupled directly to the boundary trace. The selected cases sample the low-to-intermediate frequency band in which the artificial-boundary treatment remains visible; at higher frequencies, the dependence on the particular admissible boundary treatment decreases and the response becomes increasingly governed by the near-field discretization. The methods are compared under matched physical cross-sectional partitions. The resulting difference in global degrees of freedom is an intrinsic consequence of the respective approximation spaces and is therefore retained as part of the performance assessment.

This subsection compares both accuracy and the core cost of the unbounded-domain treatment. Accuracy is assessed over five matched refinement levels, whereas the timing sweep uses a dense sequence of meshes and supplemental large-mesh points so that the cost growth is represented directly rather than inferred from a few widely separated partitions. The reference-free all-S realization is adopted throughout the cross-method comparison and the subsequent engineering applications, without frequency-by-frequency optimization.

The displacement accuracy is first measured by the normalized complex $L_2$ error
\begin{equation}
e_{L_2}
=
\left[
\frac{
\displaystyle\sum_{m\in\mathcal{M}}\sum_{q\in\mathcal{S}}
\left|
\tilde{u}^{\mathrm{num}}_m(\mathbf{x}_q)
-
\tilde{u}^{\mathrm{ana}}_m(\mathbf{x}_q)
\right|^2
}{
\displaystyle\sum_{m\in\mathcal{M}}\sum_{q\in\mathcal{S}}
\left|
\tilde{u}^{\mathrm{ana}}_m(\mathbf{x}_q)
\right|^2
}
\right]^{1/2},
\label{eq:method_L2_error}
\end{equation}
where $\mathcal{M}$ denotes the selected displacement components and $\mathcal{S}$ denotes the sampling points along the verification profiles. Since the responses are complex-valued frequency-response functions, Eq.~\eqref{eq:method_L2_error} accounts for both real and imaginary parts. The metric is a discrete relative $\ell_2$ error over the receiver samples, denoted by $e_{L_2}$ for consistency with the figures; no quadrature weights are introduced in this profile-based comparison.

For wave-propagation problems, phase accuracy is assessed with the amplitude-weighted metric $e_{\phi}$ defined previously in Eq.~\eqref{eq:method_phase_error}.

\begin{figure}[!ht]
\centering
\includegraphics[width=0.65\linewidth]{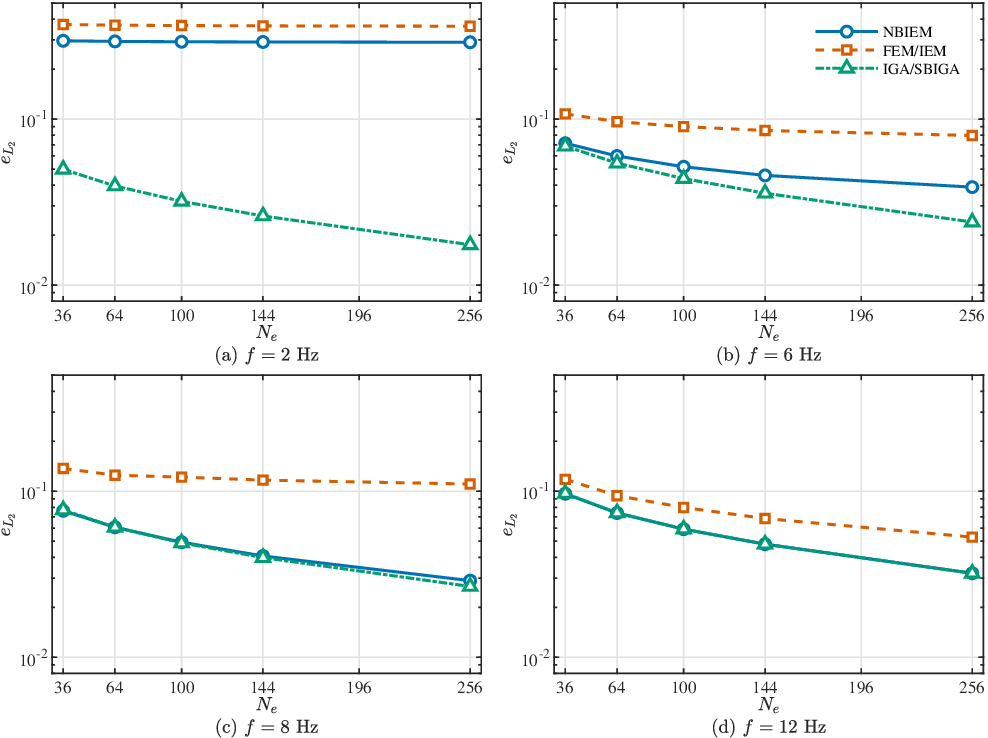}
\caption{Comparison of the normalized complex displacement error $e_{L_2}$ for FEM/IEM, NBIEM, and IGA/SBIGA under matched cross-sectional partitions. The error combines the vertical-displacement profiles along the horizontal and depth directions. The four subfigures correspond to $f=2$, $6$, $8$, and $12~\mathrm{Hz}$.}
\label{fig:method_L2_error}
\end{figure}

\fref{fig:method_L2_error} shows a smooth improvement of NBIEM as the dimensionless frequency increases. In this comparison, NBIEM uses the reference-free fixed all-S radial realization described in the preceding boundary-assessment study. At the representative matched partition $N_e=144$, NBIEM reduces the $2$-Hz displacement error from $0.3642$ for FEM/IEM to $0.2913$, while IGA/SBIGA gives $e_{L_2}=0.0261$. At $6~\mathrm{Hz}$, NBIEM further reduces the FEM/IEM error from $0.0854$ to $0.0458$ and approaches the IGA/SBIGA value of $0.0357$. The NBIEM and IGA/SBIGA curves then become nearly coincident: at $8~\mathrm{Hz}$ the errors are $0.0407$ for NBIEM, $0.0398$ for IGA/SBIGA, and $0.1167$ for FEM/IEM, while at $12~\mathrm{Hz}$ NBIEM and IGA/SBIGA both give approximately $0.0480$, compared with $0.0685$ for FEM/IEM. The refinement curves therefore show that NBIEM provides a favorable displacement accuracy--cost balance from the intermediate-frequency cases onward, while retaining a simpler far-field construction.

\begin{figure}[!ht]
\centering
\includegraphics[width=0.65\linewidth]{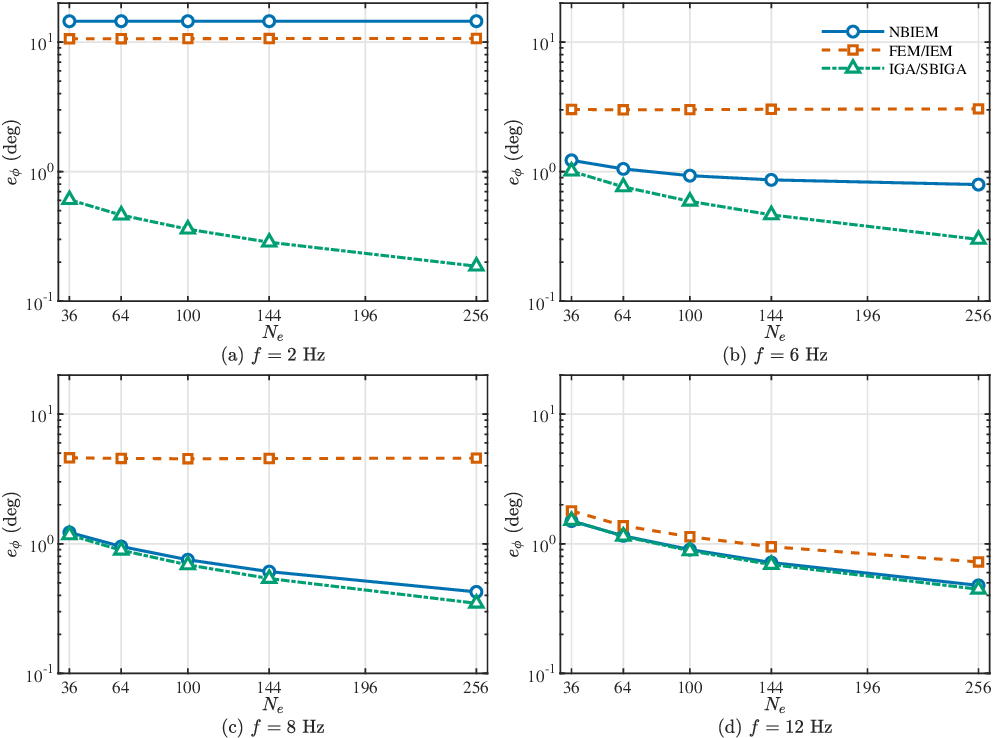}
\caption{Comparison of the amplitude-weighted phase error $e_{\phi}$ for FEM/IEM, NBIEM, and IGA/SBIGA under matched cross-sectional partitions. The phase error is reported in degrees and evaluated for $f=2$, $6$, $8$, and $12~\mathrm{Hz}$.}
\label{fig:method_phase_error}
\end{figure}

The phase comparison in \fref{fig:method_phase_error} shows the frequency-dependent improvement of NBIEM over the investigated range. At $N_e=144$, the $2$-Hz NBIEM phase error is $14.48^{\circ}$, while FEM/IEM and IGA/SBIGA give $10.63^{\circ}$ and $0.28^{\circ}$, respectively. As the frequency increases, NBIEM rapidly approaches the IGA/SBIGA result and improves on the conventional FEM/IEM exterior treatment. At $6~\mathrm{Hz}$, NBIEM reduces the FEM/IEM phase error from $3.03^{\circ}$ to $0.86^{\circ}$ and follows the IGA/SBIGA value of $0.46^{\circ}$. At $8~\mathrm{Hz}$, the corresponding errors are $0.61^{\circ}$, $4.55^{\circ}$, and $0.54^{\circ}$ for NBIEM, FEM/IEM, and IGA/SBIGA, respectively. At $12~\mathrm{Hz}$, all three methods show small phase errors, with $0.72^{\circ}$ for NBIEM, $0.95^{\circ}$ for FEM/IEM, and $0.69^{\circ}$ for IGA/SBIGA. The combined amplitude and phase results therefore support a frequency-aware accuracy--cost interpretation. At $2~\mathrm{Hz}$, NBIEM improves the combined displacement error relative to FEM/IEM but remains more phase-sensitive and less accurate than IGA/SBIGA. From $6~\mathrm{Hz}$ onward, its phase response approaches the IGA/SBIGA result while retaining a substantially simpler and less expensive far-field construction.

\tref{tab:method_8hz_summary} gives a representative equal-partition comparison at $8~\mathrm{Hz}$. At the same physical partition $N_e=144$, NBIEM and IGA/SBIGA have the same number of global displacement degrees of freedom and nearly the same displacement and phase errors, while the NBIEM wall-clock core time is much lower because no scaled-boundary far-field dynamic stiffness is solved. FEM/IEM uses more degrees of freedom under the same partition and remains less accurate for this case.

\begin{table}[H]
\centering
\caption{Representative accuracy--cost comparison at $f=8~\mathrm{Hz}$ and $N_e=144$. The core wall-clock time follows the same definition as in \fref{fig:method_cpu_time}.}
\label{tab:method_8hz_summary}
\begin{tabular}{lrrrr}
\toprule
Method & $N_{\mathrm{dof}}$ & $e_{L_2}$ & $e_{\phi}$ (deg) & Core wall-clock time (s) \\
\midrule
FEM/IEM   & 1443 & 0.1167 & 4.55 & 0.140 \\
NBIEM     &  588 & 0.0407 & 0.61 & 0.023 \\
IGA/SBIGA &  588 & 0.0398 & 0.54 & 168.1 \\
\bottomrule
\end{tabular}
\end{table}

\begin{figure}[!ht]
\centering
\includegraphics[width=3.5in]{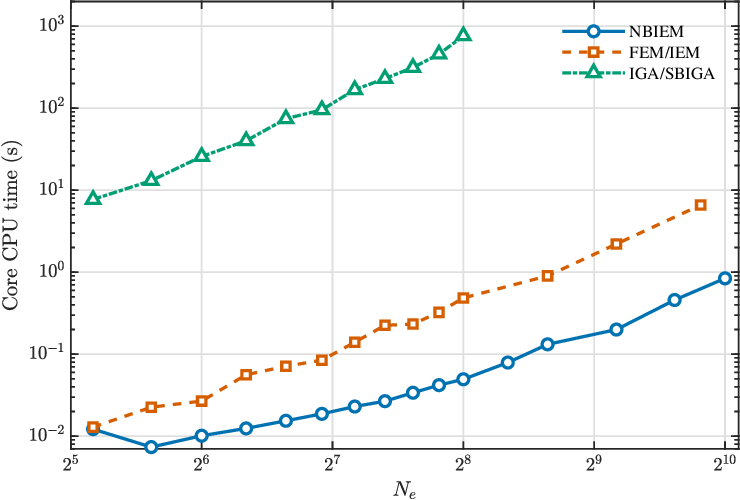}
\caption{Core wall-clock time for FEM/IEM, NBIEM, and IGA/SBIGA as a function of the number of cross-sectional elements $N_e$ for the method-comparison setting at $f=8~\mathrm{Hz}$. Supplemental large-mesh timing points are included for NBIEM and FEM/IEM to show the scaling trend. The reported time includes exterior-matrix construction, dynamic-system solution, and response extraction; common preprocessing, analytical-reference evaluation, and near-field assembly are excluded. The NBIEM values are medians of five repetitions at each mesh level.}
\label{fig:method_cpu_time}
\end{figure}

The computational cost of the three methods is further compared in \fref{fig:method_cpu_time}. The timings were obtained from the same in-house implementation using wall-clock measurements and sparse direct linear solves. To isolate the cost most directly affected by the unbounded-domain formulation, the reported core wall-clock time includes exterior-matrix construction, dynamic-system solution, and response extraction. Geometry and mesh generation, analytical-reference evaluation, and near-field matrix assembly are excluded from all three curves. All methods were evaluated in the same computational environment, and the reported wall-clock times are intended for relative performance comparison.

{
All wall-clock measurements reported in this comparison and in the tunnel assessment of Section~\ref{tunnel_burial_depth} were obtained on a workstation equipped with an Intel Core i5-14600KF processor (14 physical cores and 20 logical threads) and approximately 48~GB of RAM. The complex linear systems were solved using a sparse direct solver, and no application-level parallel processing was enabled.
}

As shown in \fref{fig:method_cpu_time}, FEM/IEM is inexpensive for relatively small meshes, but its computational cost increases rapidly as the mesh is refined. This is mainly due to the faster growth of the global degrees of freedom and the resulting increase in linear-solve cost. IGA/SBIGA provides a strong low-frequency comparison baseline, but it requires a much higher computational cost even for moderate mesh sizes because the scaled-boundary far-field dynamic stiffness has to be constructed and solved. In contrast, NBIEM maintains a much lower wall-clock time over the tested mesh range. In particular, at larger element counts, NBIEM remains substantially faster than FEM/IEM, while also avoiding the high boundary-assembly cost of IGA/SBIGA.

At $N_e=144$, NBIEM uses 588 degrees of freedom, approximately $40.7\%$ of the 1443 degrees of freedom in FEM/IEM, and its measured core wall-clock time is $0.023~\mathrm{s}$, compared with $0.140~\mathrm{s}$ for FEM/IEM and $168.1~\mathrm{s}$ for IGA/SBIGA. At $N_e=256$, the respective times are $0.049$, $0.485$, and $761.6~\mathrm{s}$. The supplemental large-mesh timing sweep gives the same growth trend: at matched partitions with $N_e=400$ and $576$, the NBIEM core wall-clock times remain $0.132$ and $0.200~\mathrm{s}$, whereas FEM/IEM increases to $0.901$ and $2.209~\mathrm{s}$; at the largest FEM/IEM timing point, $N_e=900$, the corresponding core wall-clock time reaches $6.588~\mathrm{s}$. On the present implementation and hardware, NBIEM is therefore already about one order of magnitude faster than FEM/IEM at moderate-to-large partitions, while also avoiding the iterative far-field dynamic-stiffness construction that dominates IGA/SBIGA. Together with the low-to-intermediate-frequency accuracy results, these measurements characterize the proposed formulation's frequency-dependent accuracy--cost balance for repeated frequency-domain simulations, with clear low-frequency sensitivity and behavior close to IGA/SBIGA from the representative medium-frequency cases onward.

This section verifies the proposed 2.5D NURBS-trace infinite-element formulation through comparisons with closed-form solutions over a range of moving-load speeds, including cases with a nonzero load-oscillation frequency. The displacement and stress results show that the formulation reproduces both primary response quantities and derivative-dependent fields in the tested settings. Under matched cross-sectional partitions, NBIEM gives a frequency-dependent accuracy--cost trade-off: it improves the displacement error relative to classical FEM/IEM and approaches IGA/SBIGA from the intermediate-frequency cases onward, while using substantially fewer globally independent degrees of freedom than the corresponding FEM/IEM discretization. Compared with IGA/SBIGA, it retains the NURBS-based near-field discretization but avoids the repeated construction of a scaled-boundary far-field dynamic stiffness, leading to a lower-cost exterior-domain treatment in the reported implementation. Unlike PML-based formulations, NBIEM does not require an additional absorbing layer or associated layer-parameter tuning, and unlike FE--BE coupling, it does not rely on problem-specific Green's functions or boundary-integral evaluation. The assessed all-S realization is therefore adopted in the subsequent single-frequency parametric and engineering-oriented investigations.

\subsection{\textit{Effect of treated depth \texorpdfstring{$h$}{h} in a soft-interlayer subgrade: comparison between the coupled pavement-subgrade model and the overall layered model}}
\label{sec:coupled-overall-model}

{
Before considering pavement-scale layered systems, a balanced two-layer half-space is first used as a compact cross-method consistency check for material-interface modeling. The computational domain has $R=H=12~\mathrm{m}$, with the material interface placed at the mid-depth $y=6~\mathrm{m}$. The upper and lower layers use the compacted-subgrade and foundation-soil material properties, respectively, and the same moving-load convention, displacement ordering, and exterior-boundary treatment are used in the NBIEM and FEM/IEM calculations. The comparison is made directly from the complex vertical displacement, rather than from amplitude-only postprocessing, along two horizontal receiver lines located $0.5~\mathrm{m}$ above and below the interface and along one depth profile at $x=2~\mathrm{m}$.
}

\begin{figure}[!ht]
  \centering
  \includegraphics[width=0.70\textwidth]{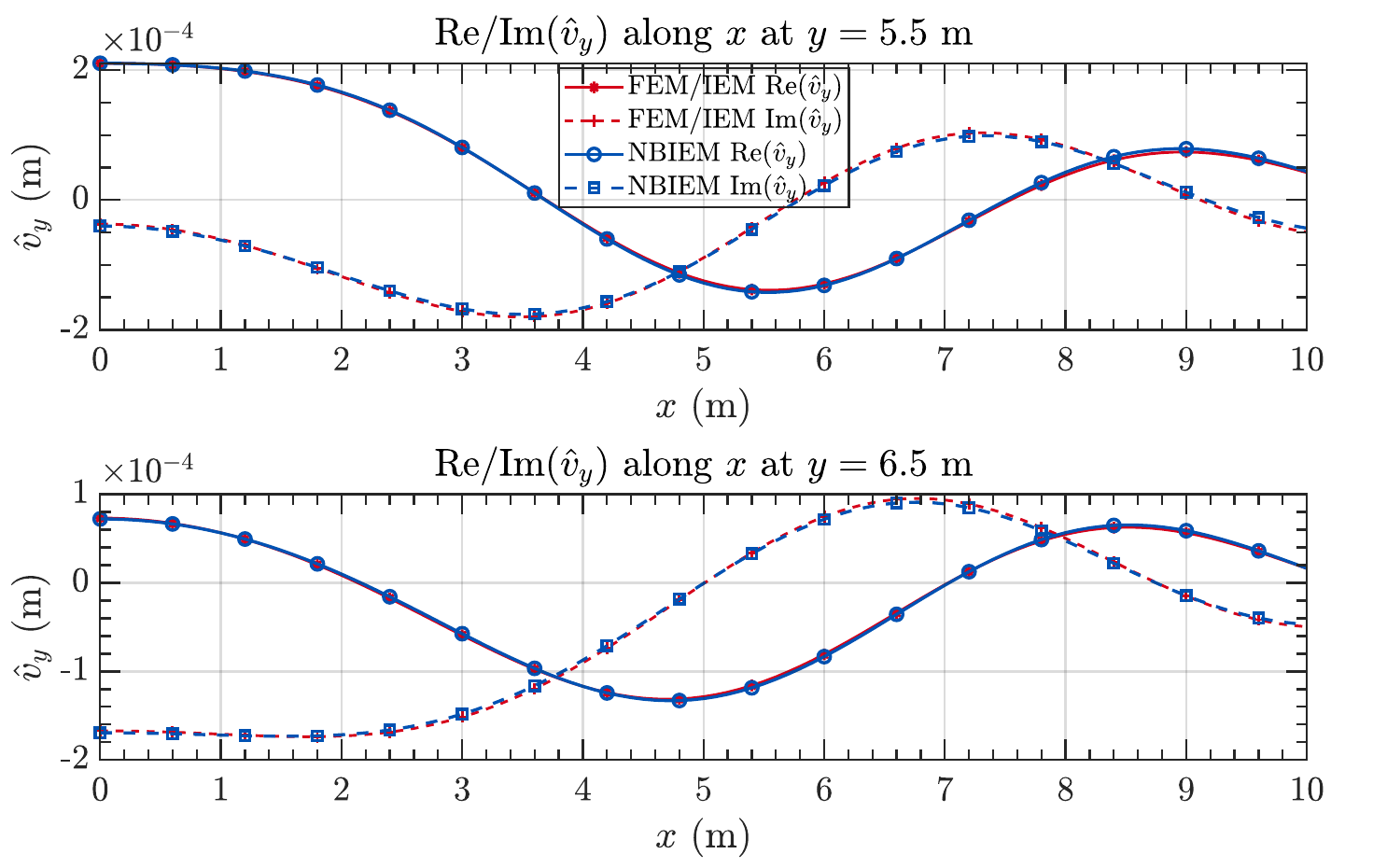}
  \caption{Real and imaginary parts of the complex vertical displacement $\hat v_y$ along horizontal profiles in the balanced two-layer half-space. The receiver lines are located at $y=5.5~\mathrm{m}$ and $y=6.5~\mathrm{m}$, i.e., $0.5~\mathrm{m}$ above and below the material interface at $y=6~\mathrm{m}$.}
  \label{fig:balanced-twolayer-horizontal-reim}
\end{figure}

\begin{figure}[!ht]
  \centering
  \includegraphics[width=0.60\textwidth]{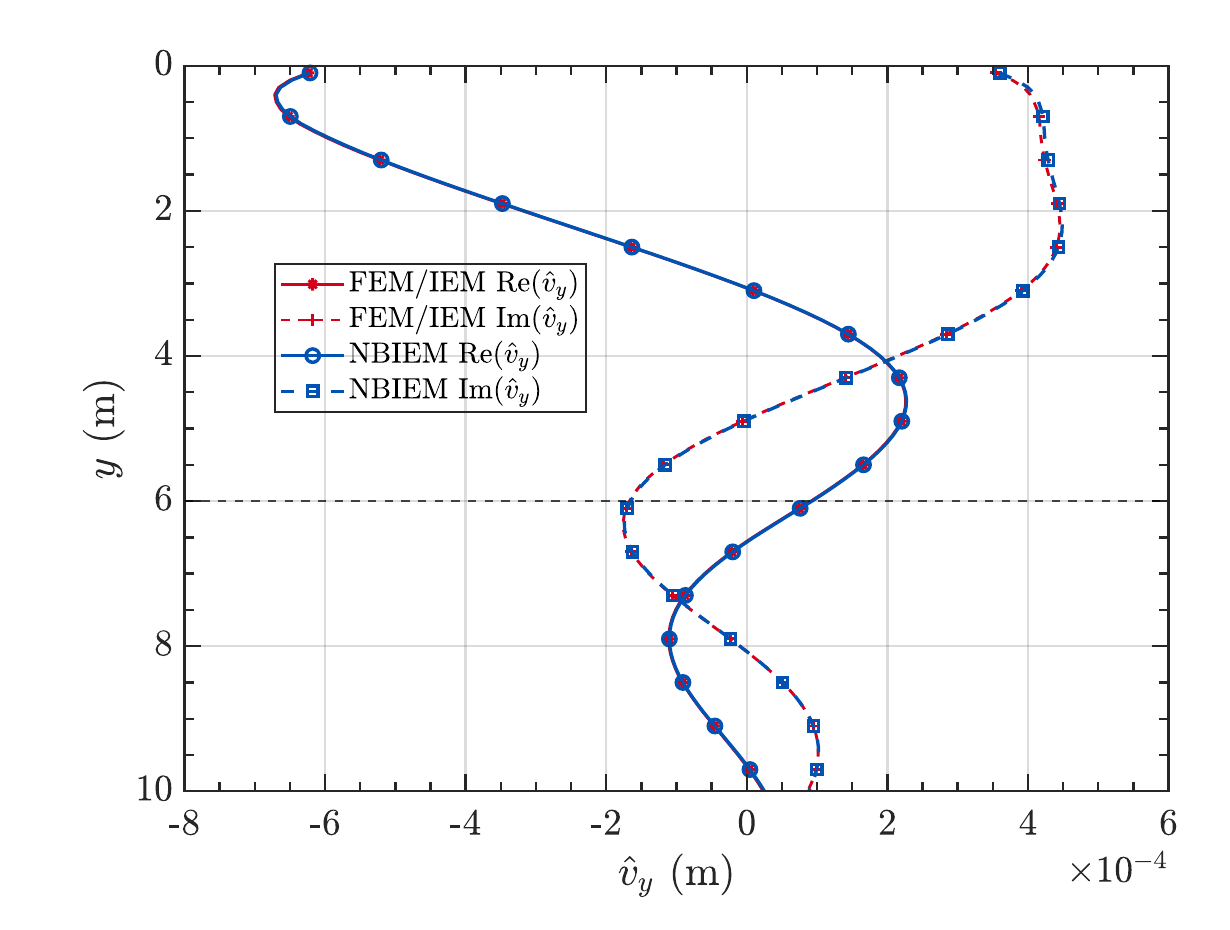}
  \caption{Real and imaginary parts of the complex vertical displacement $\hat v_y$ along the depth profile at $x=2~\mathrm{m}$ for the balanced two-layer half-space. The dashed horizontal line denotes the material interface at $y=6~\mathrm{m}$.}
  \label{fig:balanced-twolayer-depth-reim}
\end{figure}

{
As shown in \fref{fig:balanced-twolayer-horizontal-reim} and \fref{fig:balanced-twolayer-depth-reim}, the NBIEM and FEM/IEM responses agree closely in both real and imaginary parts, including the phase variation across the interface. Using the FEM/IEM profiles as the denominator, the complex relative differences are $2.80\%$ and $2.29\%$ for the two horizontal profiles and $0.82\%$ for the depth profile. This balanced layered-medium check provides a direct transition from the preceding homogeneous verification to the following engineering-oriented layered examples.
}

\begin{figure}[!ht]
  \centering
  \begin{subfigure}[t]{0.35\textwidth}
    \centering
    \includegraphics[width=\linewidth]{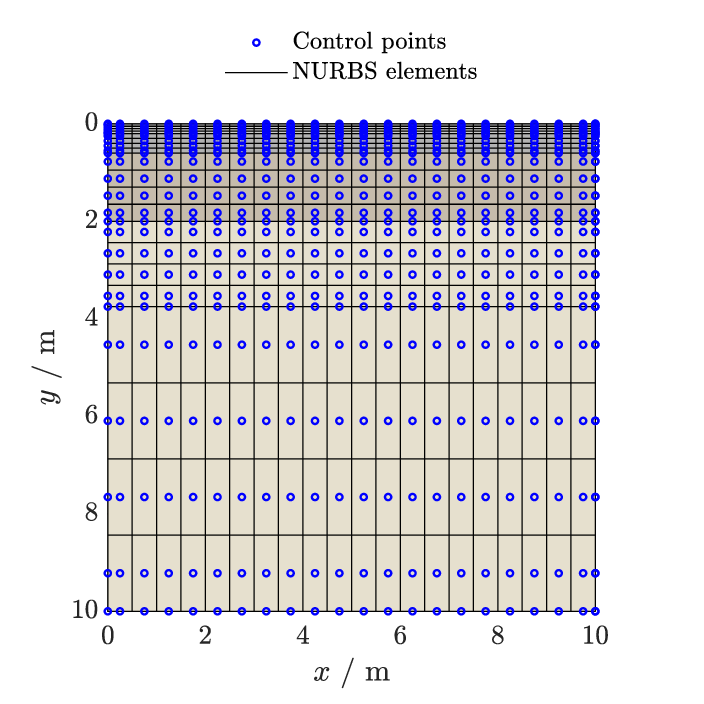}
    \caption{}
    \label{fig:overall-layered}
  \end{subfigure}\hspace{3mm}
  \begin{subfigure}[t]{0.35\textwidth}
    \centering
    \includegraphics[width=\linewidth]{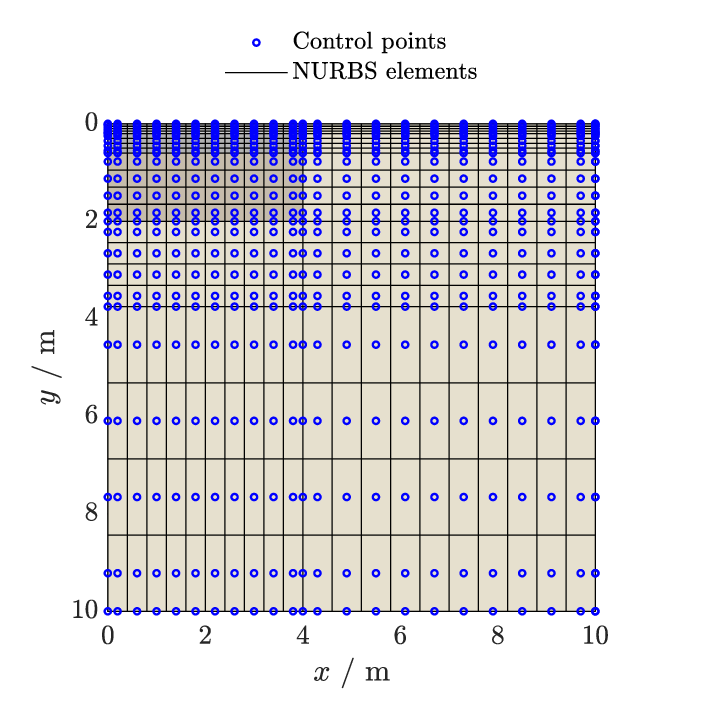}
    \caption{}
    \label{fig:coupled-model}
  \end{subfigure}
\caption{Isogeometric discretizations of the two modeling strategies.
(a) Overall layered model; (b) Coupled pavement--subgrade model.
Blue circles denote control points and black lines indicate NURBS elements.}
  \label{fig:coupled-overall}
\end{figure}

After this layered-medium consistency check, the proposed computational framework is applied to the dynamic analysis of layered pavement systems. For pavements constructed on a zero-fill foundation, many numerical studies adopt an overall layered model in which the pavement and supporting ground are treated as laterally unbounded. This representation is convenient to construct and solve, while the coupled model introduced below represents the finite pavement width explicitly. In the present study, two structural representations are considered and their applicability is distinguished by their geometric description. As illustrated in \fref{fig:coupled-overall}, the overall layered model (\fref{fig:overall-layered}) neglects the finite pavement width, whereas the coupled model (\fref{fig:coupled-model}) explicitly incorporates the actual pavement width and therefore provides a more faithful description of the cross-sectional geometry.

This section compares the dynamic responses of the two models for different transverse load locations across the pavement width. It also examines the influence of a soft interlayer embedded in the foundation in both modeling strategies to assess its effect on the coupled pavement--subgrade response. The material properties of each layer are summarized in \tref{tab:table1_material}.

\begin{table}[!htbp]
\centering
\caption{Material parameters of the layered ground with soft interlayer used in Example 4.2.}
\label{tab:table1_material}
\renewcommand{\arraystretch}{1.15}
\begin{tabular}{lcccc}
\hline
Layer & $E$ (MPa) & $\rho$ (kg/m$^{3}$) & $\nu$ (-) & $\zeta$ (-) \\
\hline
Surface layer               & 3450 & 2000 & 0.25 & 0.05 \\
Base layer                  & 1000 & 2000 & 0.25 & 0.05 \\
Compacted subgrade          & 60   & 1500 & 0.40 & 0.05 \\
Homogeneous foundation soil & 50   & 1400 & 0.30 & 0.05 \\
Soft interlayer             & 10   & 1200 & 0.35 & 0.05 \\
\hline
\end{tabular}
\end{table}

\begin{figure}[!ht]
  \centering
  \begin{subfigure}[t]{0.40\textwidth}
    \centering
    \includegraphics[width=\linewidth]{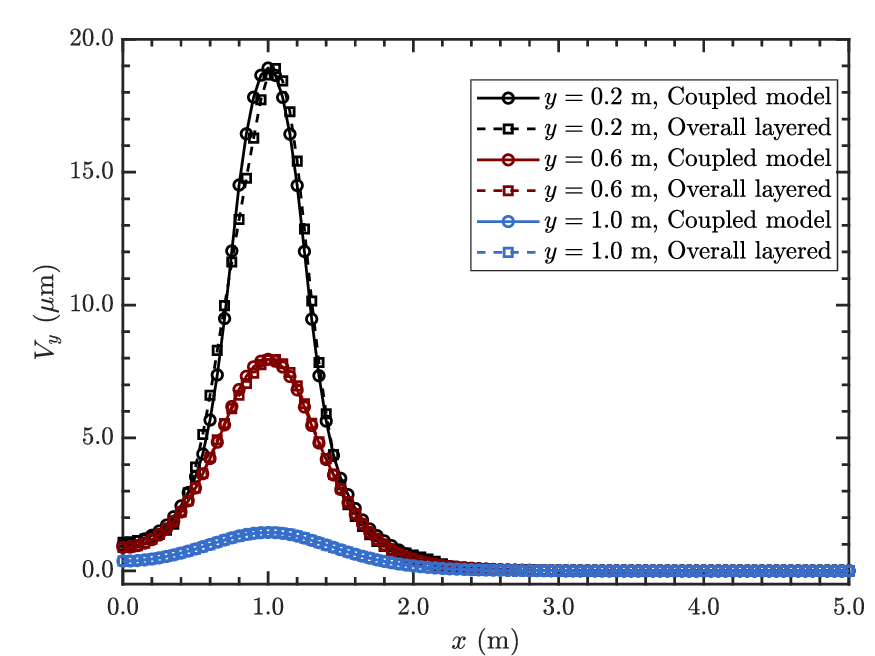}
    \caption{}    
  \end{subfigure}\hspace{3mm}
  \begin{subfigure}[t]{0.40\textwidth}
    \centering
    \includegraphics[width=\linewidth]{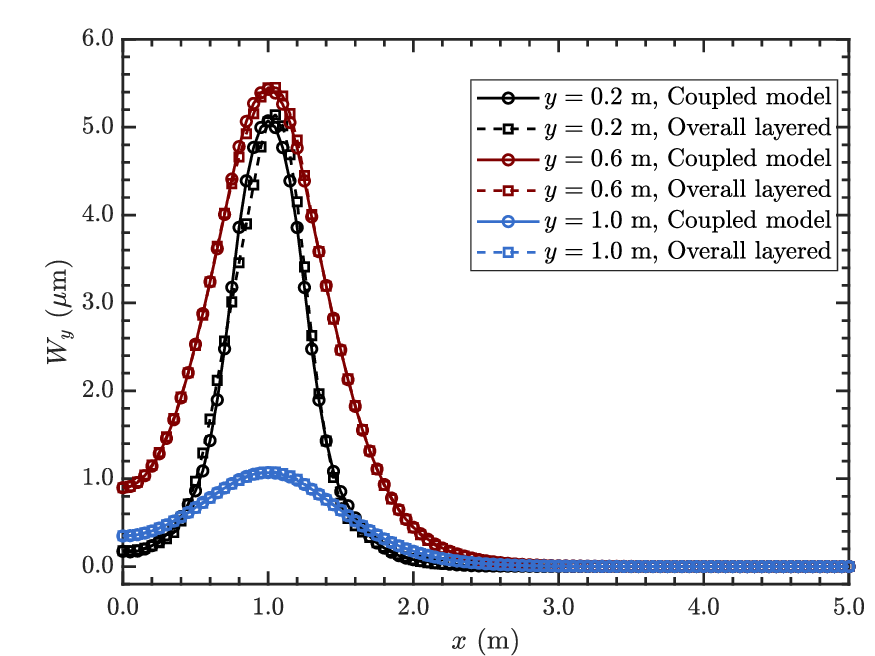}
    \caption{}  
  \end{subfigure}
\caption{Amplitude distributions of the vertical and longitudinal displacements under a single moving line load ($P=50~\mathrm{kN}$, $c=45~\mathrm{m/s}$, $f=32~\mathrm{Hz}$, $f_0=0$) with the load position at $x_0=1.0~\mathrm{m}$.
The horizontal axis represents the transverse coordinate $x$, and the vertical axes show the displacement magnitudes (a) $|V_y|$ and (b) $|W_y|$ extracted from the complex solution.
Results from the coupled pavement--subgrade model and the overall layered model are compared at $y=0.2$, $0.6$, and $1.0~\mathrm{m}$.}
\label{fig:resp_x1}
\end{figure}

\begin{figure}[!ht]
  \centering
  \begin{subfigure}[t]{0.40\textwidth}
    \centering
    \includegraphics[width=\linewidth]{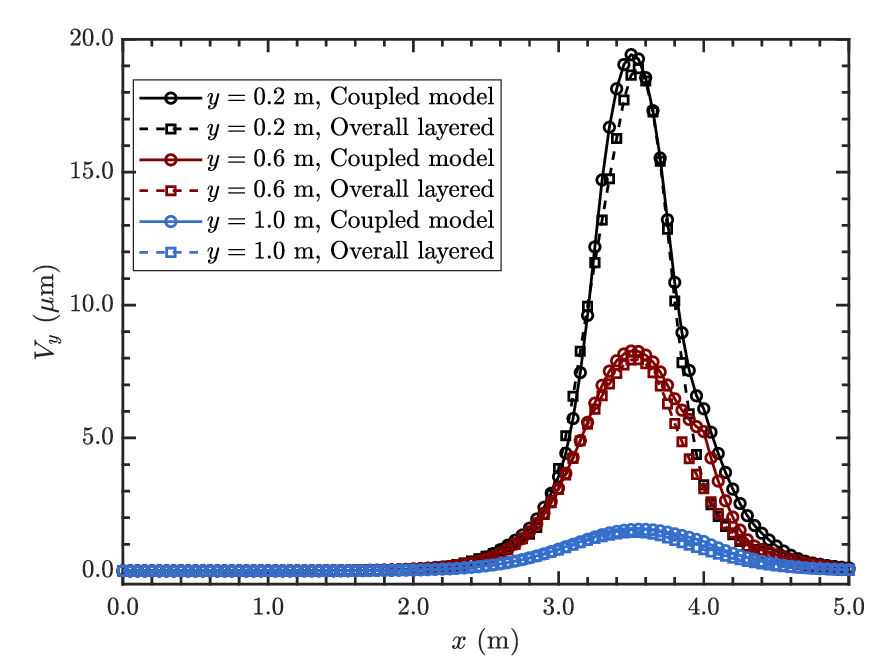}
    \caption{}
  \end{subfigure}\hspace{3mm}
  \begin{subfigure}[t]{0.40\textwidth}
    \centering
    \includegraphics[width=\linewidth]{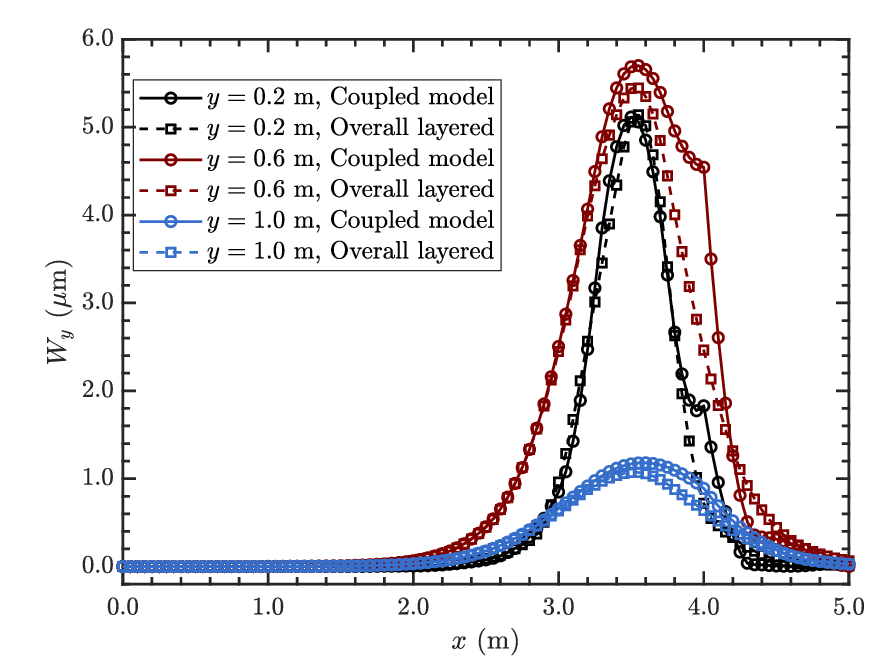}
    \caption{}
  \end{subfigure}
\caption{Amplitude distributions of (a) $|V_y|$ and (b) $|W_y|$ under a single moving line load ($P=50~\mathrm{kN}$, $c=45~\mathrm{m/s}$, $f=32~\mathrm{Hz}$, $f_0=0$) with the load position at $x_0=3.5~\mathrm{m}$.
The horizontal axis is $x$, and the vertical axis denotes the displacement magnitudes extracted from the complex solution.
Curves correspond to the coupled model and the overall layered model evaluated at $y=0.2$, $0.6$, and $1.0~\mathrm{m}$.
As the load approaches the coupling interface, small but visible differences between the two modeling strategies start to emerge, especially in the near-field region.}
\label{fig:resp_x35}
\end{figure}

\begin{figure}[!ht]
  \centering
  \begin{subfigure}[t]{0.40\textwidth}
    \centering
    \includegraphics[width=\linewidth]{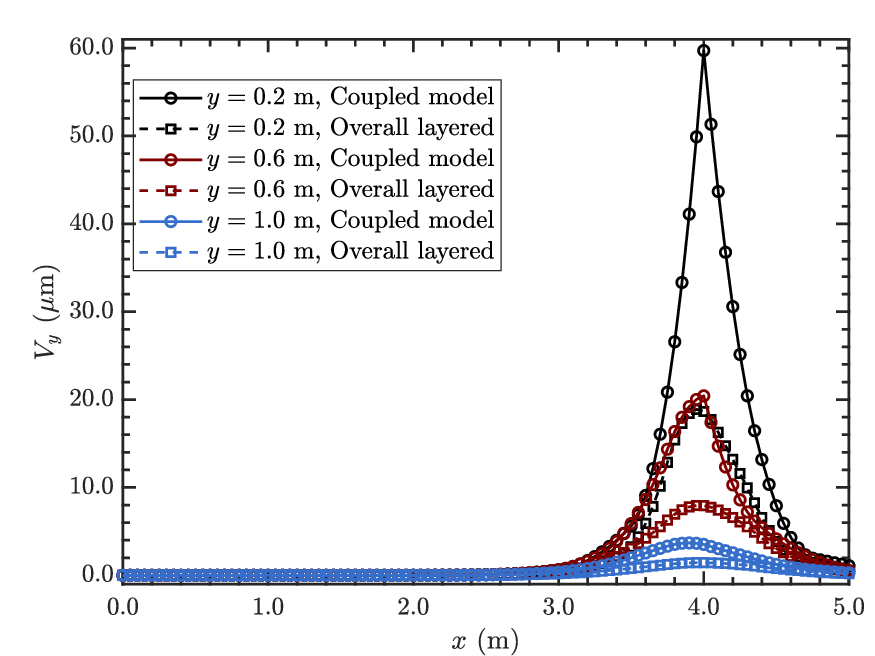}
    \caption{}
  \end{subfigure}\hspace{3mm}
  \begin{subfigure}[t]{0.40\textwidth}
    \centering
    \includegraphics[width=\linewidth]{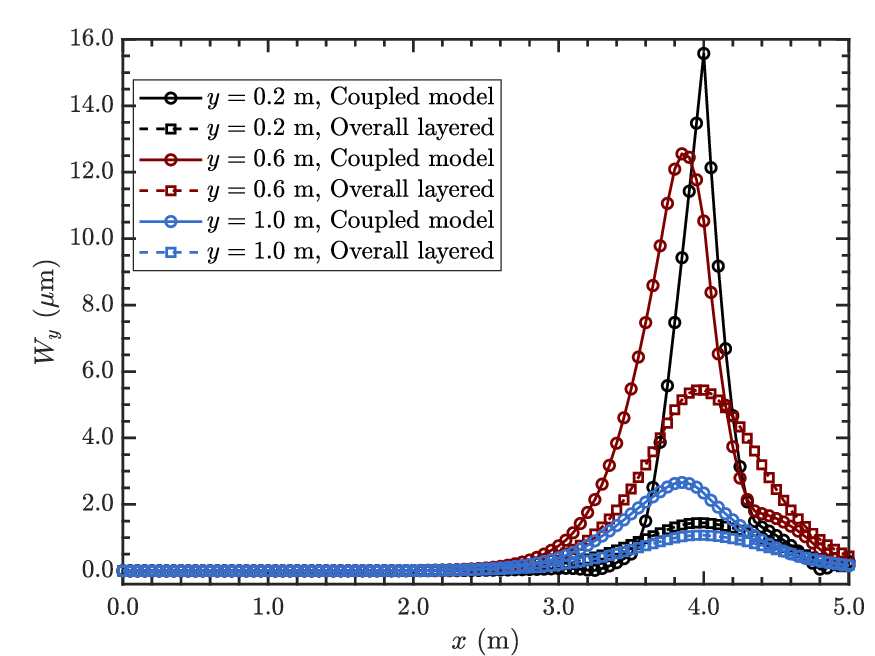}
    \caption{}
  \end{subfigure}
\caption{Amplitude distributions of (a) $|V_y|$ and (b) $|W_y|$ under a single moving line load ($P=50~\mathrm{kN}$, $c=45~\mathrm{m/s}$, $f=32~\mathrm{Hz}$, $f_0=0$) with the load position at the coupling interface ($x_0=4.0~\mathrm{m}$).
The displacement magnitudes are plotted along the $x$-direction for $y=0.2$, $0.6$, and $1.0~\mathrm{m}$.
Compared with the overall layered model, the coupled model predicts a noticeably different near-field response, indicating that the finite-width and coupling effects become important when the load is applied in the vicinity of the interface.}
\label{fig:resp_x4}
\end{figure}

A vertical concentrated load moving along a prescribed longitudinal line with total amplitude $P=50~\mathrm{kN}$ and speed $c=45~\mathrm{m/s}$ is applied on the pavement surface. This comparison is evaluated at a single analysis frequency \(f=32~\mathrm{Hz}\) with \(f_0=0\), so that \(\omega=2\pi f\) and the longitudinal wavenumber is \(k=(\omega-\omega_0)/c\). The pavement surface layer and base layer are assigned fixed thicknesses of $0.2~\mathrm{m}$ and $0.4~\mathrm{m}$, respectively. The compacted subgrade layer is taken as $1.4~\mathrm{m}$ thick, and the remaining depth is modeled as homogeneous foundation soil. The displacement magnitudes are extracted from the complex frequency-domain solution at the analysis angular frequency $\omega$ and are reported as
\[
U_y=\big|\hat u_y(k,\omega)\big|,\qquad
V_y=\big|\hat v_y(k,\omega)\big|,\qquad
W_y=\big|\hat w_y(k,\omega)\big|.
\]
Here the subscript $y$ denotes the $y$-directed moving load, while the uppercase letter identifies the $x$-, $y$-, or $z$-directed response. The amplitude profiles are evaluated along the transverse direction $x$ at three representative depths, $y=0.2$, $0.6$, and $1.0~\mathrm{m}$.
Figs.~\ref{fig:resp_x1}--\ref{fig:resp_x4} compare the coupled and overall layered models for $x_0=1.0$, 3.5, and $4.0~\mathrm{m}$. In every case, $V_y$ peaks near the load and decreases with distance and depth. The longitudinal response $W_y$ is non-monotonic and reaches its largest value at $y=0.6~\mathrm{m}$, coincident with the base--subgrade interface. This interface-adjacent amplification is consistent with wave redistribution across the stiffness and impedance contrast and motivates the strain-based assessment below.

\begin{figure}[!ht]
  \centering
  \begin{subfigure}[t]{0.35\textwidth}
    \centering
    \includegraphics[width=\linewidth]{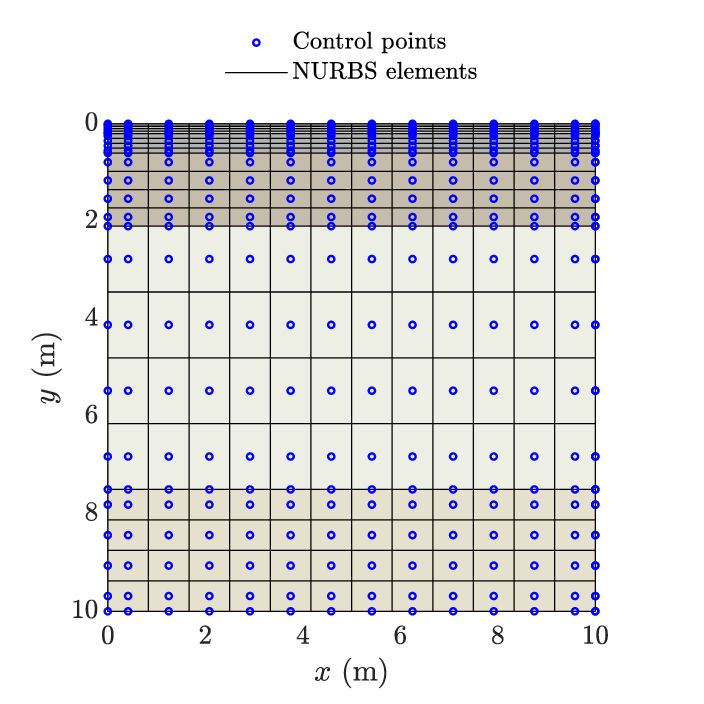}
    \caption{}
  \end{subfigure}\hspace{3mm}
  \begin{subfigure}[t]{0.35\textwidth}
    \centering
    \includegraphics[width=\linewidth]{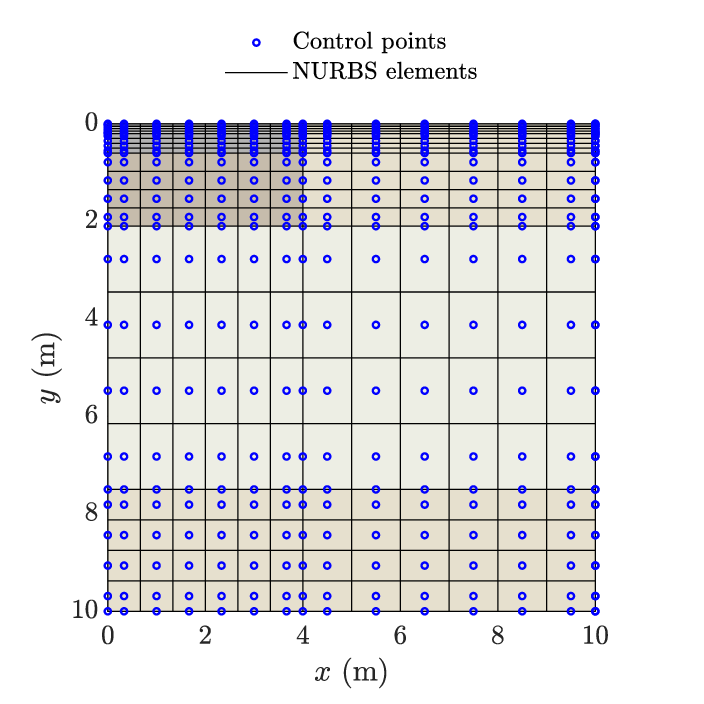}
    \caption{}
  \end{subfigure}
\caption{Isogeometric meshes for the two modeling strategies with a soft interlayer, shown for the treated depth $h=1.5~\mathrm{m}$. (a) Overall layered model; (b) Coupled pavement--subgrade model. Blue circles denote control points and black lines indicate NURBS elements; the light-gray region represents the soft interlayer.}
\label{fig:mesh_softlayer_h15}
\end{figure}

In both modeling strategies, a soft interlayer of identical depth is introduced beneath the pavement system. The foundation soil below $y=7.5~\mathrm{m}$ (measured from the free surface) is assumed to be normal soil. The analysis examines whether the influence of the underlying soft interlayer diminishes as the treated subgrade depth $h$ increases and identifies the range over which further increases in $h$ produce only diminishing response changes within the present linear model. \fref{fig:mesh_softlayer_h15} shows the isogeometric discretizations for the case $h=1.5~\mathrm{m}$, where the light-gray region denotes the soft interlayer.

\begin{figure}[!ht]
    \centering
    \includegraphics[width=3in]{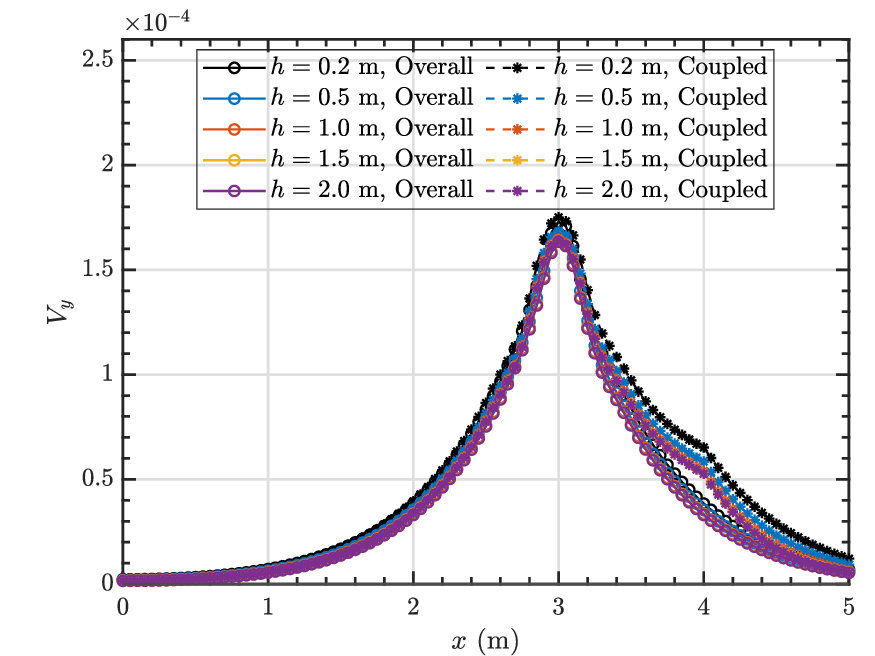}
\caption{Free-surface ($y=0$) distribution of the vertical displacement amplitude $|V_y|$ along the $x$-direction for different treated depths $h$.
Results from the overall layered model and the coupled pavement--subgrade model are compared.}
\label{fig:Uy_surface_x_h}
\end{figure}

\begin{figure}[!ht]
    \centering
    \includegraphics[width=3in]{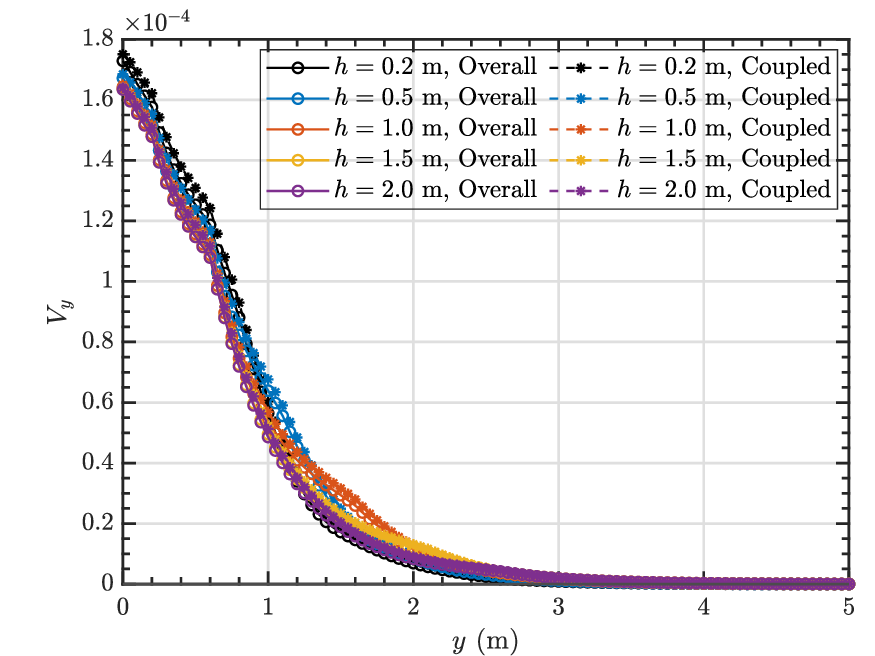}
\caption{Depth-wise variation of the vertical displacement amplitude $|V_y|$ along the $y$-direction directly beneath the load point ($x=x_0$) for different treated depths $h$.
Results from the overall layered model and the coupled pavement--subgrade model are compared.}
\label{fig:Uy_belowload_y_h}
\end{figure}

Figs.~\ref{fig:Uy_surface_x_h} and~\ref{fig:Uy_belowload_y_h} show the amplitude of the vertical displacement, denoted by $V_y$, for different treated depths $h$.
In this parametric study, a transversely concentrated vertical moving load with total amplitude $P=100~\mathrm{kN}$ is applied at $x_0=3~\mathrm{m}$. The load speed is set to $c=30~\mathrm{m/s}$, and the single-frequency response is evaluated at \(f=10~\mathrm{Hz}\) with \(f_0=0\). This loading configuration is adopted to approximate the effective action range of heavy-vehicle loading in typical service conditions, so that the evaluated sensitivity to $h$ is of practical relevance.
The top of the soft interlayer is located at
\begin{equation}\label{eq:y_int_def}
y_{\mathrm{int}} = 0.6 + h,
\end{equation}
where $0.6~\mathrm{m}$ is the total thickness of the surface and base layers (\tref{tab:table1_material}).

\fref{fig:Uy_surface_x_h} presents the free-surface response, i.e., the distribution of $V_y(x,y=0)$ along the transverse coordinate $x$. For each treated depth $h$, results from the overall layered model and the coupled pavement--subgrade model are compared. All curves exhibit a dominant peak near the loading position ($x\approx x_0$) and decay rapidly away from it, indicating that the response is mainly governed by a localized near-field zone. With increasing $h$, both the peak amplitude and the post-peak tail decrease, and the two-model predictions become closer. This behavior indicates that a deeper treated layer reduces the influence of the underlying soft interlayer on the surface motion and weakens the model discrepancy induced by the coupling effect.

\fref{fig:Uy_belowload_y_h} plots $V_y(x=x_0,y)$ versus depth $y$ beneath the load.
The amplitude decays quickly with depth, and the displacement becomes very small once $y$ exceeds approximately $2~\mathrm{m}$.
This indicates that, under a vehicle-representative moving load, the effective displacement response is mainly confined to the shallow subgrade.
Within this effective range, the influence of $h$ is most pronounced: the curves separate clearly in the shallow-to-intermediate depths and their relative ordering changes around $y\approx y_{\mathrm{int}}$, which shifts downward as $h$ increases.
This trend indicates that the impedance contrast across the soft-interlayer interface modifies the wave-field partition and the interference between downward-propagating and reflected components, thereby altering the near-field depth profile.

The interface-sensitive change of $V_y$ around $y\approx y_{\mathrm{int}}$ suggests that the interlayer stress response is most pronounced near the soft-interlayer top boundary.
Because stresses are governed by spatial derivatives of displacement, a material discontinuity can intensify the local gradients in the vicinity of the interface.
Therefore, increasing $h$ not only reduces the free-surface displacement level, but also shifts the stress-sensitive zone associated with the soft interlayer to a greater depth, weakening its impact on the upper pavement structure.

\begin{figure}[!ht]
  \centering
  \begin{subfigure}[t]{0.40\textwidth}
    \centering
    \includegraphics[width=\linewidth]{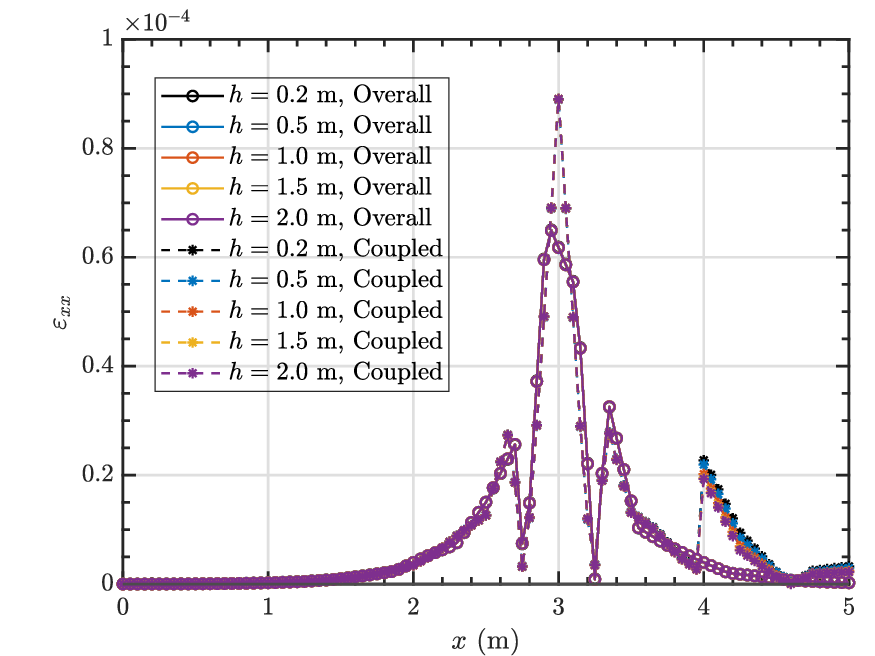}
    \caption{}
  \end{subfigure}\hspace{3mm}
  \begin{subfigure}[t]{0.40\textwidth}
    \centering
    \includegraphics[width=\linewidth]{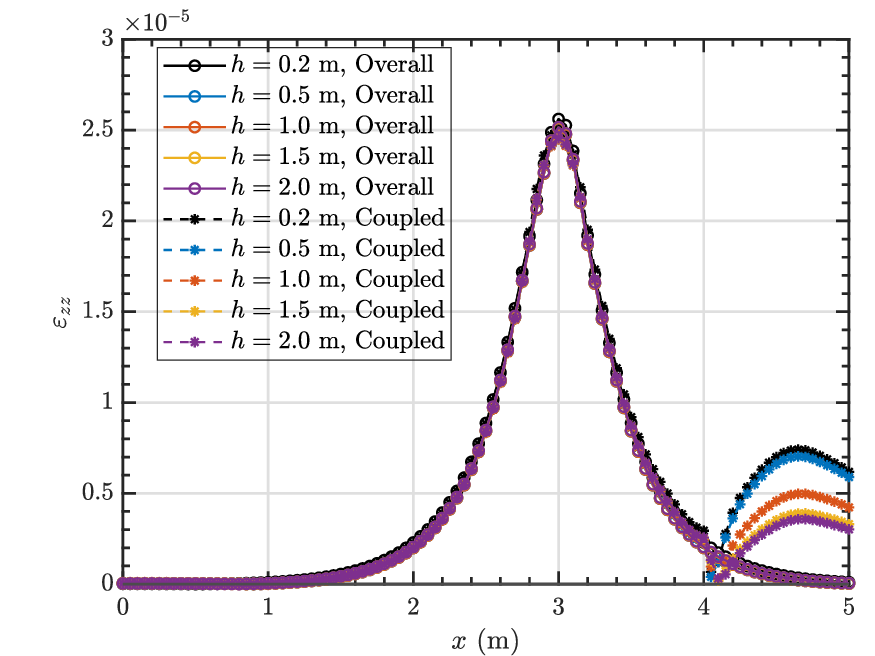}
    \caption{}
  \end{subfigure}
\caption{Tensile strain amplitudes at the bottom of the surface layer ($y=0.2~\mathrm{m}$) along the transverse coordinate $x$ under a transversely concentrated moving load ($P=100~\mathrm{kN}$, $c=30~\mathrm{m/s}$, $f=10~\mathrm{Hz}$, $f_0=0$) applied at $x_0=3~\mathrm{m}$. Results for different treated depths $h$ are compared between the overall layered model and the coupled pavement--subgrade model: (a) $\varepsilon_{xx}$; (b) $\varepsilon_{zz}$.}
  \label{fig:strain_surfbot_x}
\end{figure}

To further relate the displacement trends to cracking-related indicators, the tensile strain amplitudes at the bottom of the surface layer ($y=0.2~\mathrm{m}$) are examined.
\fref{fig:strain_surfbot_x} shows the distributions of $\varepsilon_{xx}$ and $\varepsilon_{zz}$ along the transverse coordinate $x$ for different treated depths $h$, with the overall layered model and the coupled model plotted together for each $h$.

For both components, the response is dominated by a pronounced near-field peak around the loading position ($x\approx x_0$), followed by a rapid decay away from the load.
Compared with $\varepsilon_{zz}$, $\varepsilon_{xx}$ exhibits a more concentrated peak and a stronger sensitivity to $h$ in the near-field region, indicating that the transverse tensile demand at the surface-layer bottom is more affected by the stiffness change induced by subgrade treatment. A second feature occurs near the coupling boundary at $x=4~\mathrm{m}$, where the two modeling strategies start to diverge.
In particular, $\varepsilon_{zz}$ from the overall layered model increases again for $x\gtrsim 4~\mathrm{m}$, whereas the coupled-model response remains much smaller in the same region.
By contrast, $\varepsilon_{xx}$ shows a sharper, more abrupt change near the coupling boundary and then decays rapidly, indicating a more localized sensitivity to the boundary-induced disturbance.
These different patterns suggest that the coupling boundary and the underlying soft interlayer affect the two tensile strain components through different mechanisms: $\varepsilon_{zz}$ is more influenced by reflected and redistributed wave components that can rebuild the tensile demand away from the load, while $\varepsilon_{xx}$ is dominated by localized curvature changes and interface-induced strain concentration.
Mechanically, these strain patterns imply that subgrade treatment not only reduces the displacement level, but also redistributes the deformation demand in the surface layer.
Increasing $h$ suppresses the interface-related response near the coupling boundary, while the enhanced near-field concentration of $\varepsilon_{xx}$ suggests a potential increase in local curvature of the surface layer.

\begin{figure}[!ht]
  \centering
  \begin{subfigure}[t]{0.40\textwidth}
    \centering
    \includegraphics[width=\linewidth]{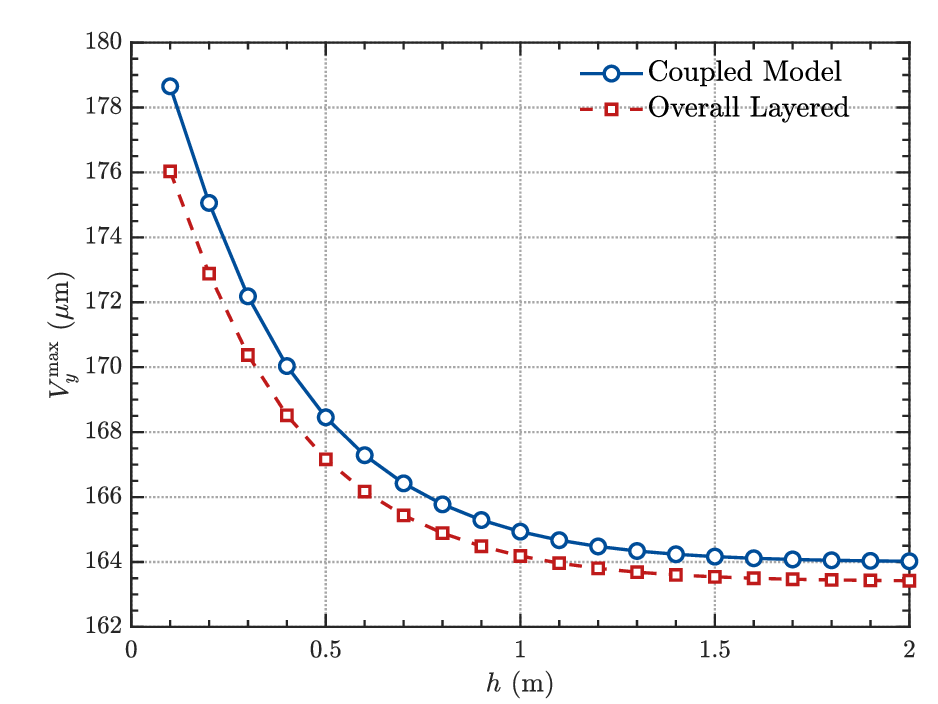}
    \caption{}
  \end{subfigure}\hspace{3mm}
  \begin{subfigure}[t]{0.40\textwidth}
    \centering
    \includegraphics[width=\linewidth]{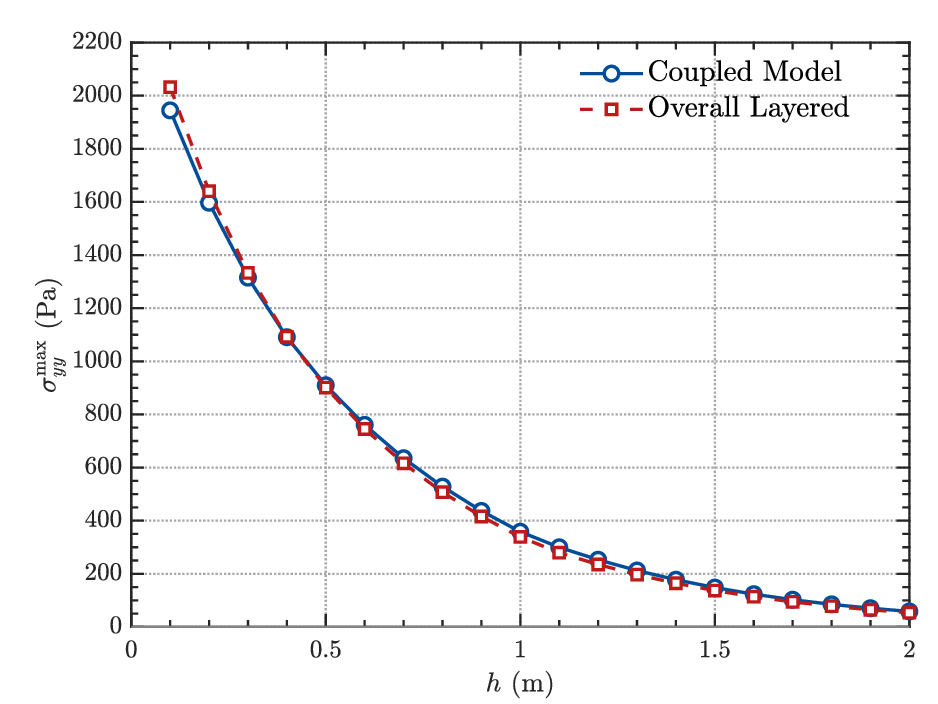}
    \caption{}
  \end{subfigure}
  \caption{Influence of treated depth $h$ on rutting-related response indicators. (a) Maximum amplitude of the surface vertical displacement $V_y^{\max}$; (b) Maximum amplitude of the vertical normal stress at the top of the soft interlayer, $\sigma_{yy}^{\max}(y=y_{\mathrm{int}})$, where $y_{\mathrm{int}}=0.6+h$. Results from the coupled model and the overall layered model are compared.}
  \label{fig:rutting_index_vs_h}
\end{figure}

\fref{fig:rutting_index_vs_h} summarizes two rutting-related indicators as functions of the treated depth $h$, namely the maximum free-surface vertical displacement amplitude $V_y^{\max}$ and the maximum vertical normal stress at the top of the soft interlayer, $\sigma_{yy}^{\max}(y=y_{\mathrm{int}})$ (in Pa).
Both indicators decrease monotonically as $h$ increases, and the reduction is most pronounced for shallow treatment depths.
When $h$ exceeds approximately $1.2$--$1.5~\mathrm{m}$, the curves gradually level off, indicating that further increases in the treated depth provide diminishing benefits.
In addition, the coupled model predicts consistently larger $V_y^{\max}$ than the overall layered model, especially for small $h$, whereas the two models give nearly identical predictions for $\sigma_{yy}^{\max}(y=y_{\mathrm{int}})$ over the full range of $h$.

Together with the tensile-strain results at the surface-layer bottom, these trends show that the treated depth $h$ affects several response measures simultaneously within the present linear frequency-domain model.
A larger $h$ reduces the surface displacement and the compressive stress transmitted to the soft interlayer, while also changing the strain concentration pattern in the surface layer.
The displacement, stress, and tensile-strain quantities should therefore be interpreted as response indicators rather than as direct rutting or cracking predictions.
For the present configuration, increasing $h$ up to about $1.2$--$1.5~\mathrm{m}$ provides substantial reductions in both $V_y^{\max}$ and $\sigma_{yy}^{\max}(y=y_{\mathrm{int}})$, while further deepening yields limited additional improvement.
\subsection{\textit{Speed-dependent responses of ballastless track models under moving line loads: embankment versus cutting}}
\label{ballastless_track_speed}

This example assesses the applicability of the proposed NBIEM to representative ballastless-track cross-sections involving multiple material regions, geometric transitions, and localized wave-field gradients. 
In contrast to the preceding verification examples, the analysis considers both the response induced by different moving-load speeds and the ability of the proposed NBIEM discretization to represent practical track-subgrade configurations through a multi-patch layout with compatible patch-wise refinement.

\begin{figure}[!ht]
  \centering
  \begin{subfigure}[t]{0.35\textwidth}
    \centering
    \includegraphics[height=1.2in,width=\linewidth,keepaspectratio]{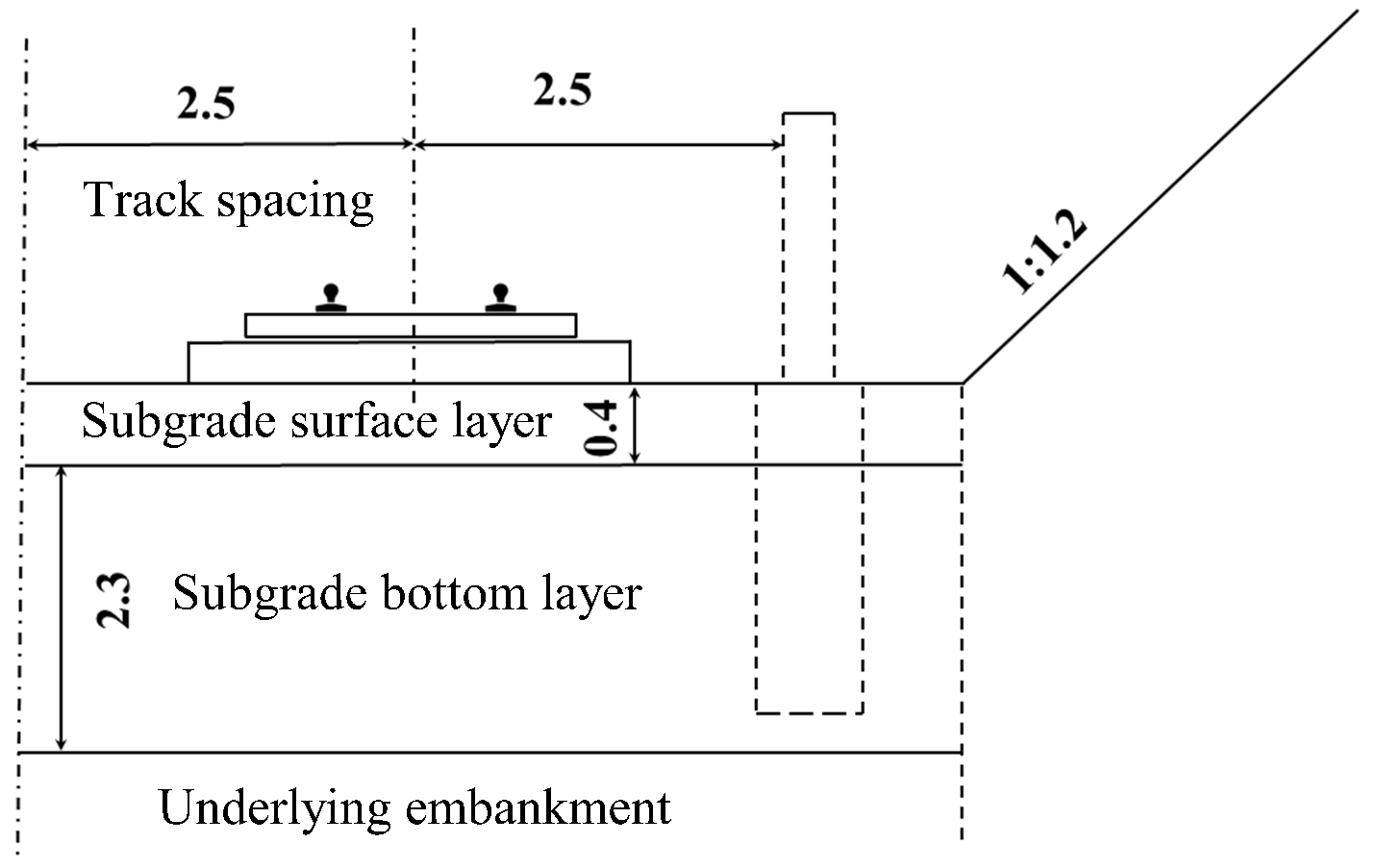}
    \caption{}
  \end{subfigure}\hspace{3mm}
  \begin{subfigure}[t]{0.45\textwidth}
    \centering
    \includegraphics[height=1.2in,width=\linewidth,keepaspectratio]{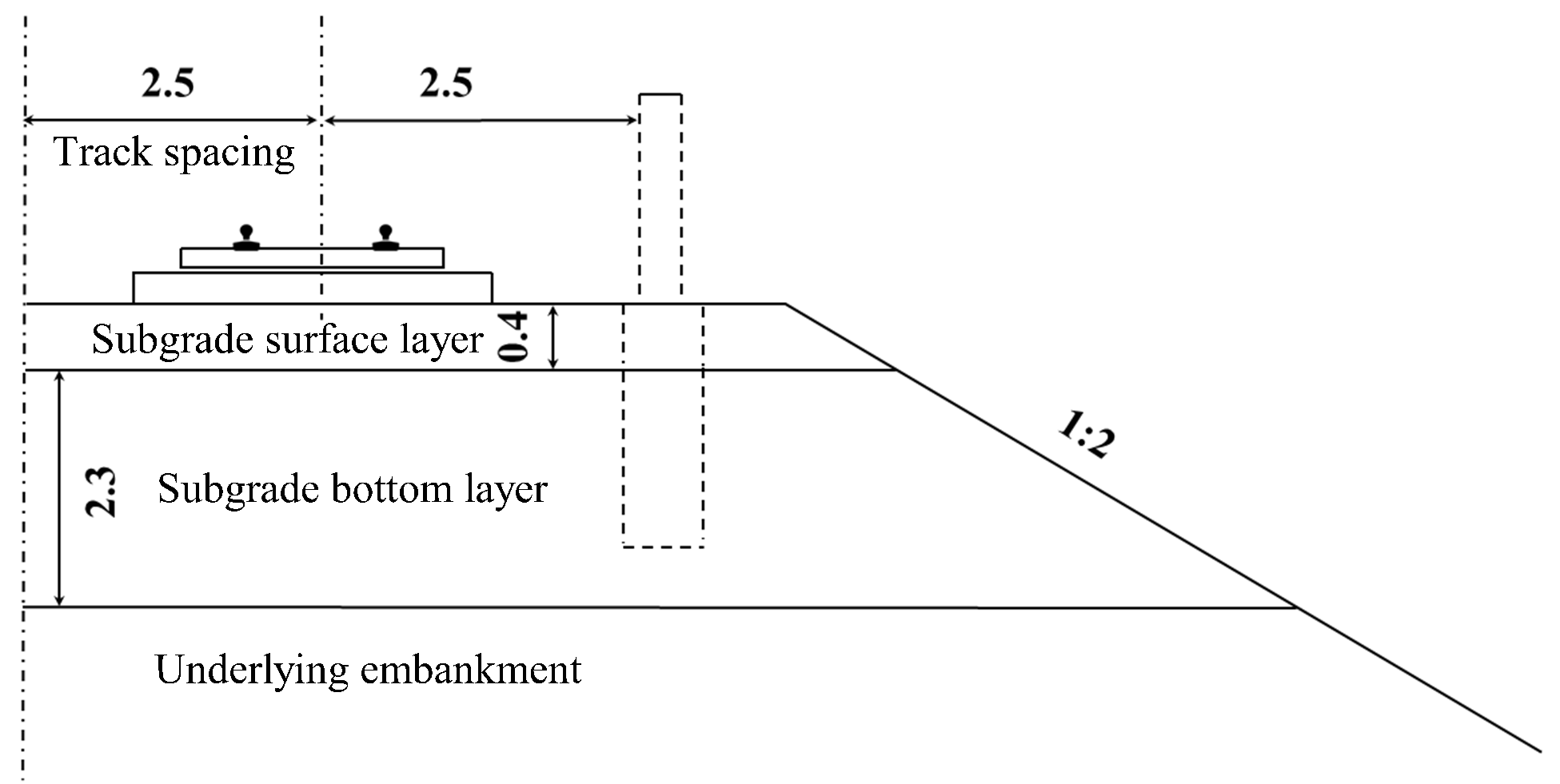}
    \caption{}
  \end{subfigure}
\caption{Schematic of the high-speed railway ballastless track cross-sections considered in this study: (a) Cutting section; (b) Embankment section. The model includes the track slab, subgrade surface layer, subgrade bottom layer, and the underlying foundation/ground (rails are not modeled).}
\label{fig:ballastless}
\end{figure}

Two typical ballastless-track configurations are considered, as illustrated in \fref{fig:ballastless}: an embankment-type section and a cutting-type section. 
The model is partitioned into four material regions: the track slab, the subgrade surface layer, the subgrade bottom layer, and the underlying embankment. The rails are omitted, and the excitation is represented as moving line loads applied to the track slab, allowing the comparison to focus on the influence of the ground configuration and load speed on the 2.5D wave field. The material properties are summarized in \tref{tab:mat_para_hsr}.

\begin{table}[H]
\centering
\caption{Material parameters for the ballastless track sections in Fig.~\ref{fig:ballastless}.}
\label{tab:mat_para_hsr}
\renewcommand{\arraystretch}{1.15}
\begin{tabular}{lcccc}
\hline
Material region & $E$ (GPa) & $\rho$ (kg/m$^{3}$) & $\nu$ (-) & $\zeta$ (-) \\
\hline
Track slab       & 32.5 & 2500 & 0.20 & 0.05 \\
Subgrade surface layer      & 0.12 & 2200 & 0.30 & 0.05 \\
Subgrade bottom layer       & 0.07 & 1900 & 0.30 & 0.05 \\
Underlying embankment/ground& 0.05 & 1400 & 0.35 & 0.05 \\
\hline
\end{tabular}
\end{table}

To represent the cutting- and embankment-type ballastless-track subgrades with sufficient geometric fidelity, the domain is constructed using a multi-patch isogeometric discretization. This strategy partitions the model into subregions with different materials and geometric features, while maintaining an analysis-suitable NURBS parameterization.

\begin{figure}[!ht]  
\centering  
\begin{subfigure}[t]{0.32\textwidth}    
\centering    
\includegraphics[width=\linewidth]{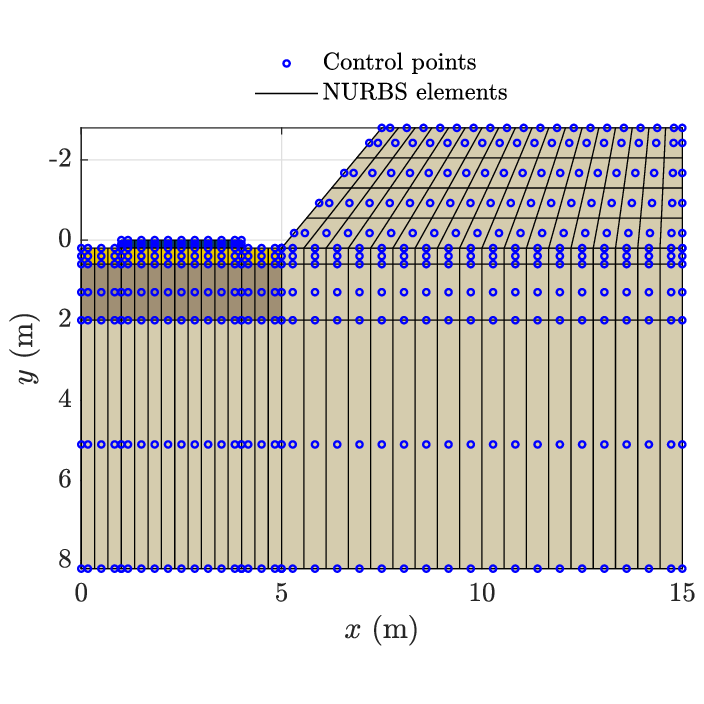}    
\caption{}  
\end{subfigure}\hspace{5mm}  
\begin{subfigure}[t]{0.32\textwidth}    
\centering    
\includegraphics[width=\linewidth]{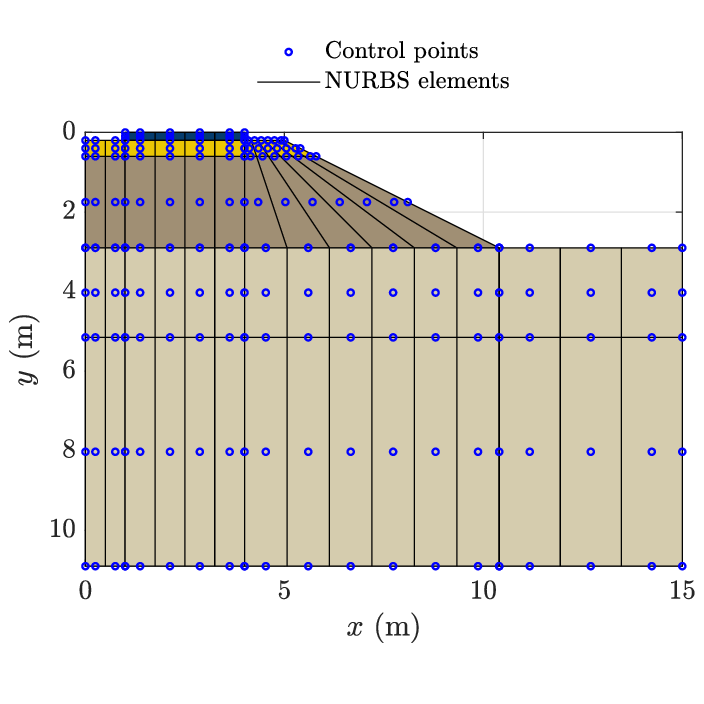}    
\caption{}  
\end{subfigure}  
\begin{subfigure}[t]{0.32\textwidth}    
\centering    
\includegraphics[width=\linewidth]{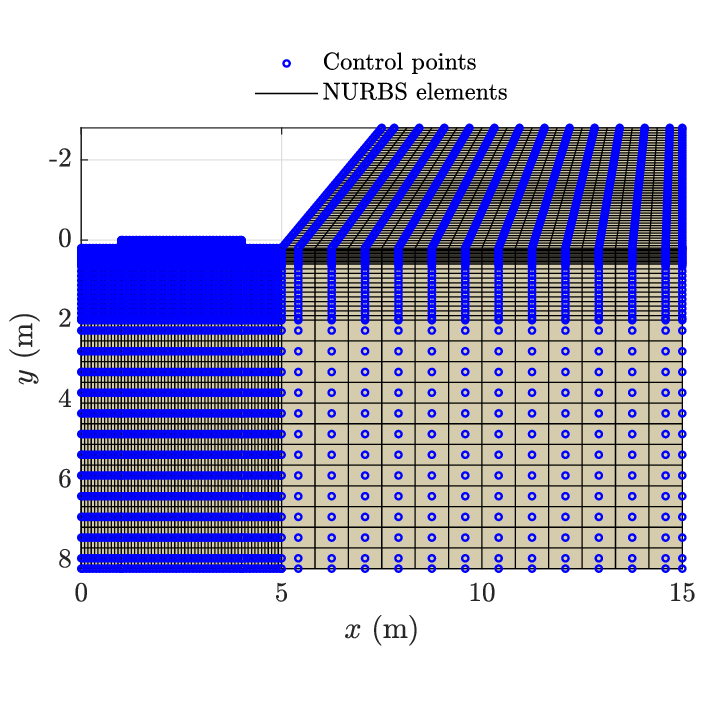}    
\caption{}  
\end{subfigure}\hspace{5mm}  
\begin{subfigure}[t]{0.32\textwidth}    
\centering    
\includegraphics[width=\linewidth]{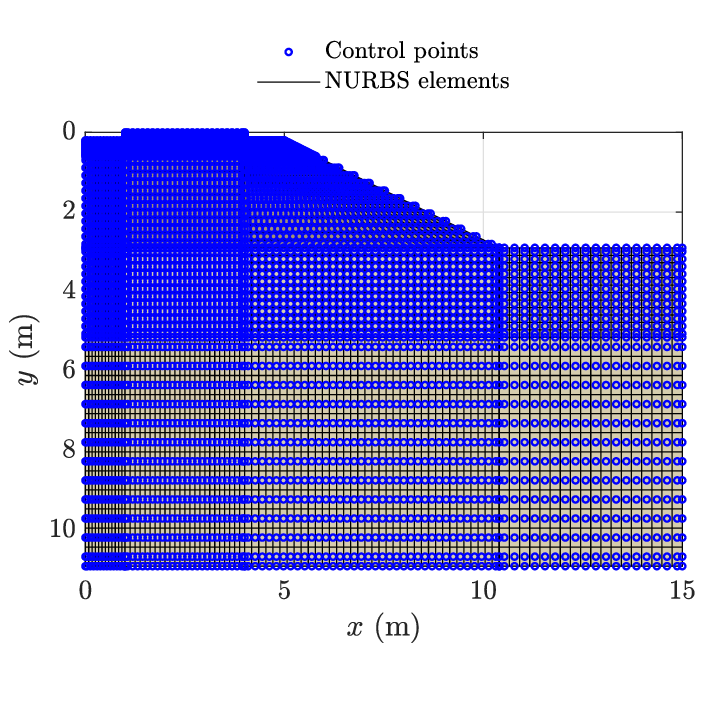}    
\caption{}  
\end{subfigure}  
  \caption{Multi-patch isogeometric meshes for the ballastless-track subgrade models.
  (a) and (b) show the baseline discretizations for the cutting- and embankment-type configurations, respectively.
  (c) and (d) show the corresponding locally \(h\)-refined meshes.
  Blue markers denote control points and black lines denote NURBS elements.}
  \label{fig:HSR_multipatch_mesh}
\end{figure}

Patch construction and refinement require particular care. The geometry includes sharp changes in the surface profile and a sloped transition zone, which require careful patch layout to avoid excessive element distortion. Compatible local \(h\)-refinement is also required across patch interfaces. This demands consistent knot insertion and interface matching so that field continuity and numerical stability are preserved when control points are densified in selected regions.
\fref{fig:HSR_multipatch_mesh} illustrates the IGA meshes for the two configurations, together with the corresponding locally refined discretizations.

\begin{figure}[!ht]
  \centering
  \begin{subfigure}[t]{0.40\textwidth}
    \centering
    \includegraphics[width=\linewidth]{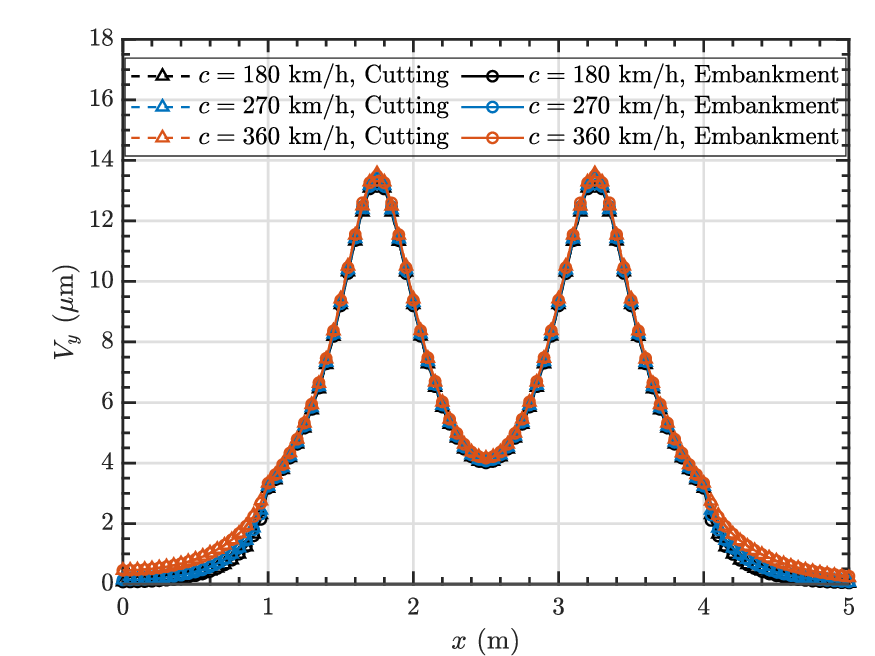}
    \caption{}
  \end{subfigure}\hspace{5mm}
  \begin{subfigure}[t]{0.40\textwidth}
    \centering
    \includegraphics[width=\linewidth]{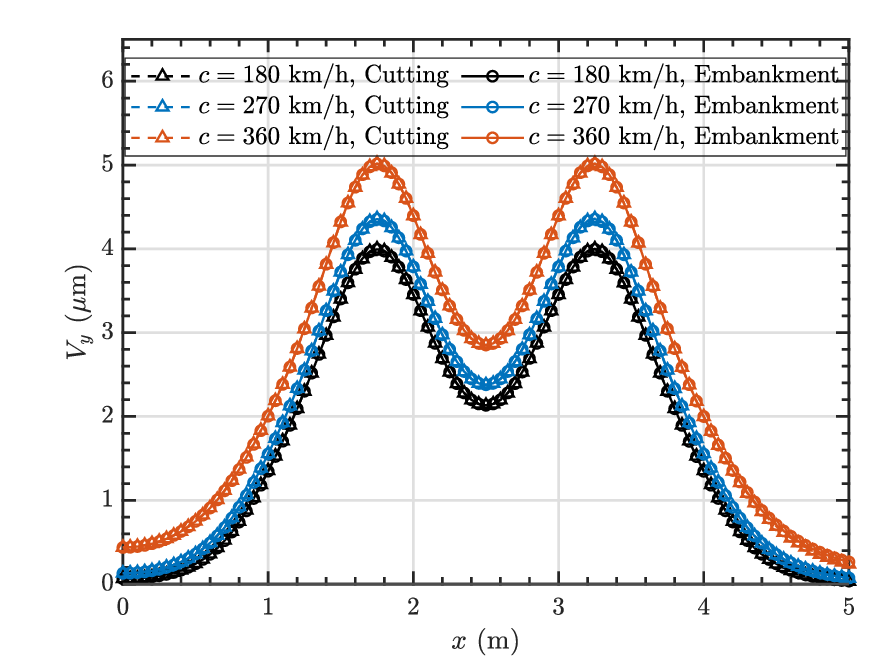}
    \caption{}
  \end{subfigure}
  \begin{subfigure}[t]{0.40\textwidth}
    \centering
    \includegraphics[width=\linewidth]{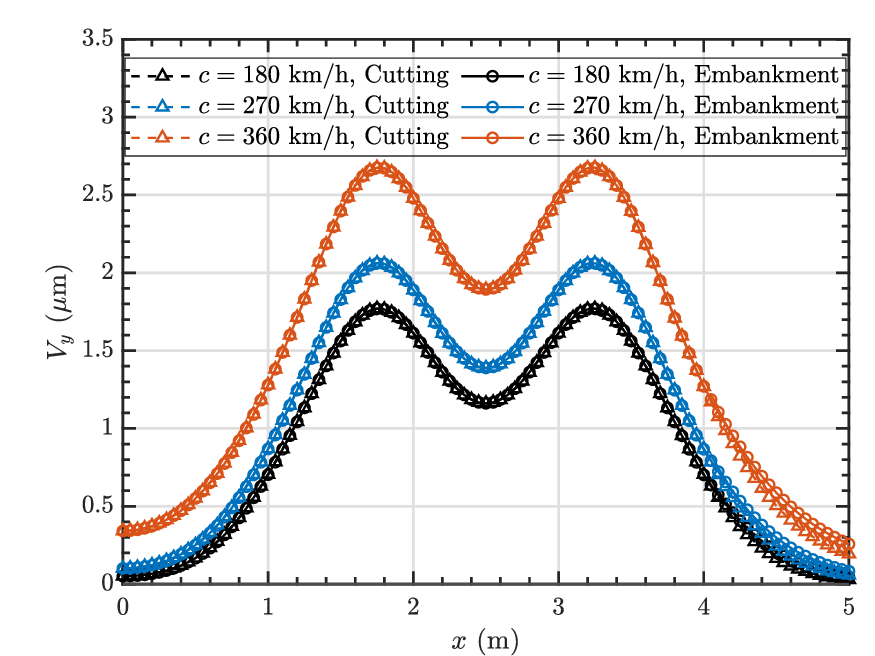}
    \caption{}
  \end{subfigure}\hspace{5mm}
  \begin{subfigure}[t]{0.40\textwidth}
    \centering
    \includegraphics[width=\linewidth]{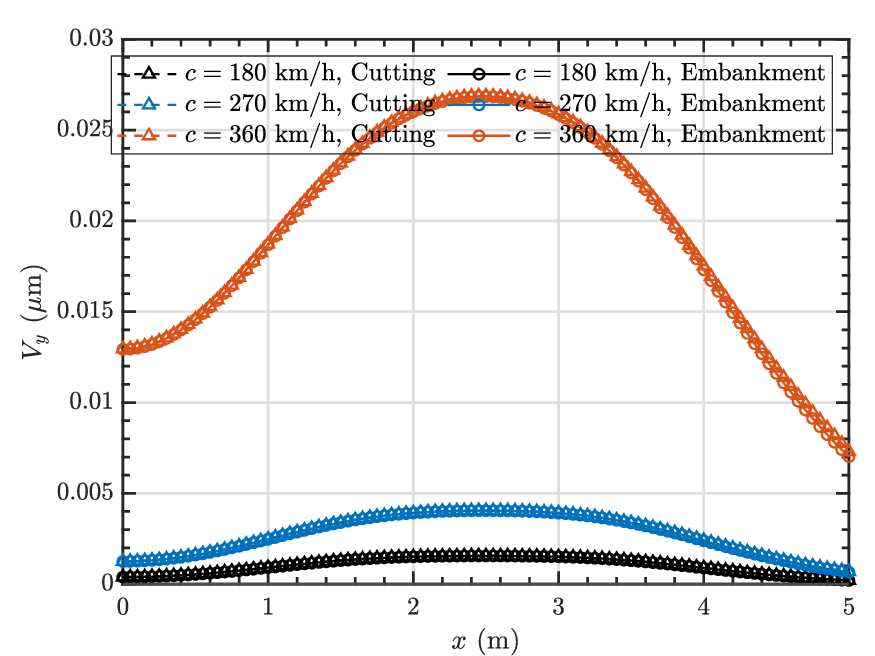}
    \caption{}
  \end{subfigure}
  \caption{Vertical displacement amplitude \(V_y\) along the \(x\)-direction at four representative depths for the cutting- and embankment-type ballastless-track subgrades under moving-load speeds \(c=180\), \(270\), and \(360~\mathrm{km/h}\).
  (a) Slab bottom, \(y=0.2~\mathrm{m}\);
  (b) Bottom of the subgrade surface layer, \(y=0.6~\mathrm{m}\);
  (c) Mid-depth of the subgrade bottom layer, \(y=1.0~\mathrm{m}\);
  (d) Bottom of the subgrade bottom layer (top of the underlying embankment/ground), \(y=2.9~\mathrm{m}\).}
  \label{fig:Vy_profiles_4depths}
\end{figure}

In this example, two rail lines are located at $x=1.75$ and $x=3.25~\mathrm{m}$, where concentrated loads moving along the rail directions, each with a total amplitude of $75~\mathrm{kN}$, are applied. For each train speed, a single analysis frequency is used according to \(f=c/\ell_s\), where $c$ is expressed in $\mathrm{m/s}$ and \(\ell_s=1.5~\mathrm{m}\) is the representative sleeper spacing; thus \(c=180\), \(270\), and \(360~\mathrm{km/h}\) correspond to \(f=33.3\), \(50.0\), and \(66.7~\mathrm{Hz}\), respectively, with \(f_0=0\). As a result, the transverse distributions of $V_y(x)$ in \fref{fig:Vy_profiles_4depths} exhibit a characteristic two-peak pattern at shallow depths, reflecting the superposition of the two rail excitations. With increasing depth, the profiles become broader and smoother, indicating progressive load spreading and wave-field redistribution in the layered subgrade.

The influence of the train speed depends strongly on depth. At the slab bottom ($y=0.2~\mathrm{m}$, \fref{fig:Vy_profiles_4depths}a), the curves for different speeds are close to each other, and the difference between the cutting and embankment configurations is negligible. This suggests that the response at this shallow level is mainly governed by the stiffness and load-spreading capacity of the upper track structure, which tends to attenuate the sensitivity to speed. As the depth increases into the subgrade ($y=0.6$ and $1.0~\mathrm{m}$, \fref{fig:Vy_profiles_4depths}b--c), the speed effect becomes more visible, while the cutting--embankment discrepancy remains minor throughout. The profiles still retain a clear two-peak shape, implying that the transverse response at these depths is largely controlled by the geometric positions of the rails and the layered transmission of the near-field wave components.

At the deepest interface ($y=2.9~\mathrm{m}$, i.e., the bottom of the subgrade bottom layer and the top of the underlying embankment/ground, \fref{fig:Vy_profiles_4depths}d), the response becomes distinctly \emph{speed-dominated}. For the low-speed cases ($c=180$ and $270~\mathrm{km/h}$), the magnitude has already decayed to an almost negligible level. In contrast, the high-speed case ($c=360~\mathrm{km/h}$) exhibits an order-of-magnitude increase in amplitude, accompanied by a clear change in spatial pattern: the shallow two-peak, rail-localized response transitions into a smoother, broad single-hump distribution. This behavior indicates that, under the selected high-speed excitation, a larger portion of the wave field reaches the deeper layers and modifies the response around the subgrade-bottom interface. The observed amplification is consistent with a reduced separation between the moving-load speed and the characteristic wave speeds of the layered ground, although a full critical-speed assessment would require a separate dispersion or velocity-ratio analysis. Consequently, the subgrade-bottom interface should be included when assessing deeper vibration and stress-related responses under high-speed operation, even when the surface-structure responses appear only weakly speed-dependent.

\begin{figure}[!ht]
    \centering
    \includegraphics[width=3in]{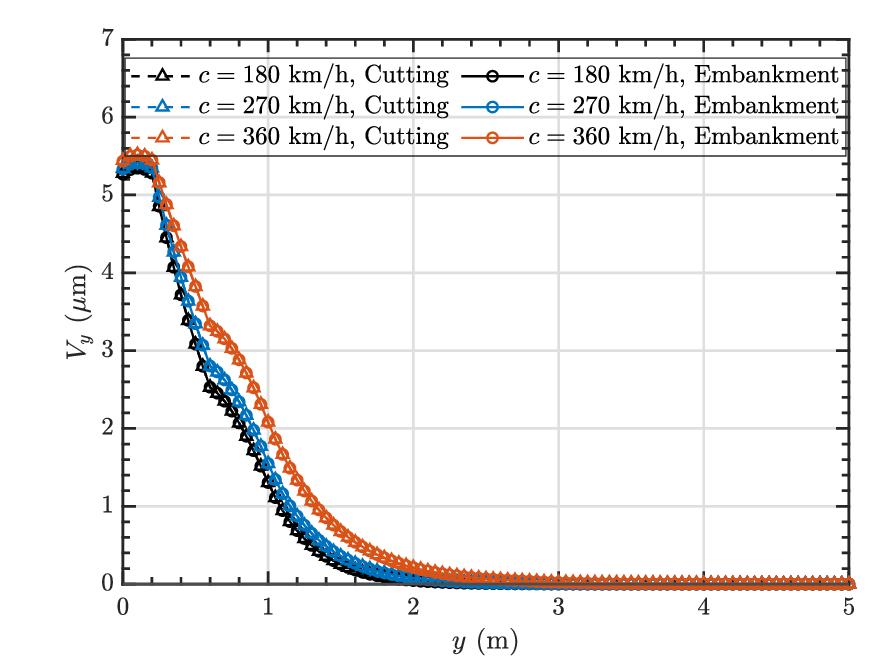}
\caption{Depth-wise variation of the \emph{vertical} displacement amplitude $V_y$ at the midpoint between the two rail load lines ($x=2.5~\mathrm{m}$) for different train speeds, comparing the cutting and embankment configurations.}
  \label{fig:Vy_depth_x2p5}
\end{figure}

\fref{fig:Vy_depth_x2p5} shows the depth-wise profile of the \emph{vertical} displacement amplitude $V_y$ at $x=2.5~\mathrm{m}$, i.e., the midpoint between the two rail load lines ($x=1.75$ and $3.25~\mathrm{m}$). A pronounced near-surface response occurs even though no load is applied directly at $x=2.5~\mathrm{m}$. This response indicates a slab bending/warping effect: the two rail excitations activate a flexural deformation mode of the concrete track slab, leading to a noticeable mid-span vertical motion.

The vertical-response amplitude depends strongly on speed. As the speed increases from $180$ to $360~\mathrm{km/h}$, the $V_y$ amplitude increases and its decay with depth becomes slower, indicating enhanced dynamic energy transmission into the subgrade under high-speed excitation. For all speeds, the response attenuates rapidly and becomes very small beyond approximately $y\approx 2~\mathrm{m}$, suggesting that the deformation demand is mainly concentrated within the upper $1$--$2~\mathrm{m}$ of the foundation in this case. The cutting and embankment curves remain very close over the entire depth range, implying that the depth-wise response at this midpoint is primarily speed-dominated for this metric.

\begin{figure}[!ht]
  \centering
  \begin{subfigure}[t]{0.40\textwidth}
    \centering
    \includegraphics[width=\linewidth]{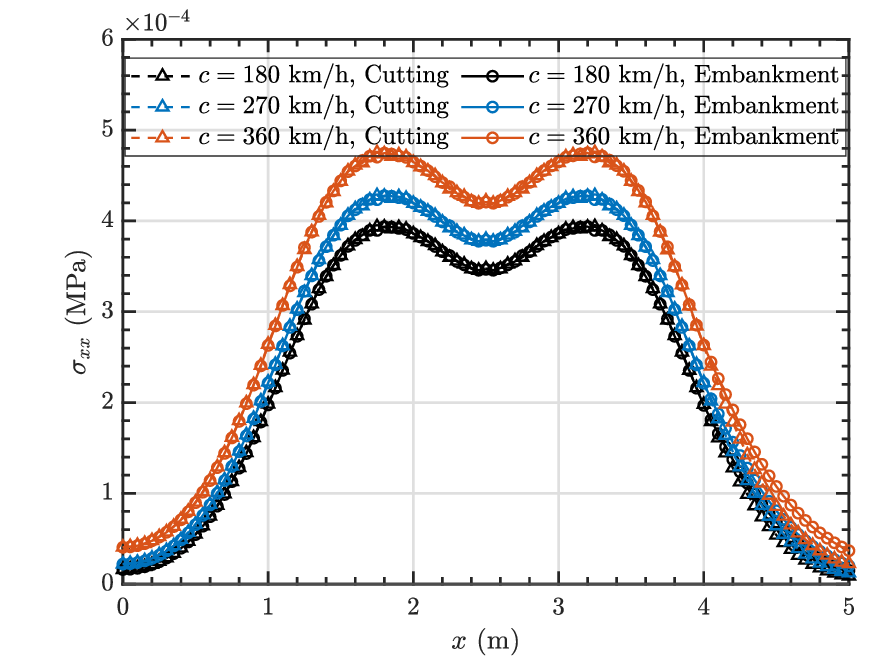}
    \caption{$\sigma_{xx}$}
    \label{fig:sxx_y1m}
  \end{subfigure}\hspace{3mm}
  \begin{subfigure}[t]{0.40\textwidth}
    \centering
    \includegraphics[width=\linewidth]{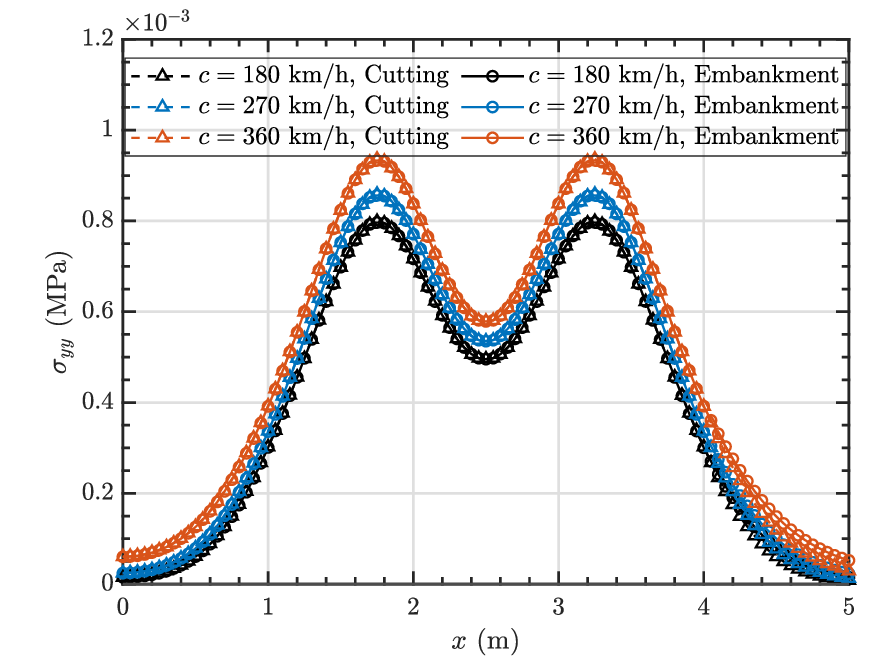}
    \caption{$\sigma_{yy}$}
    \label{fig:syy_y1m}
  \end{subfigure}\hspace{3mm}
  \begin{subfigure}[t]{0.40\textwidth}
    \centering
    \includegraphics[width=\linewidth]{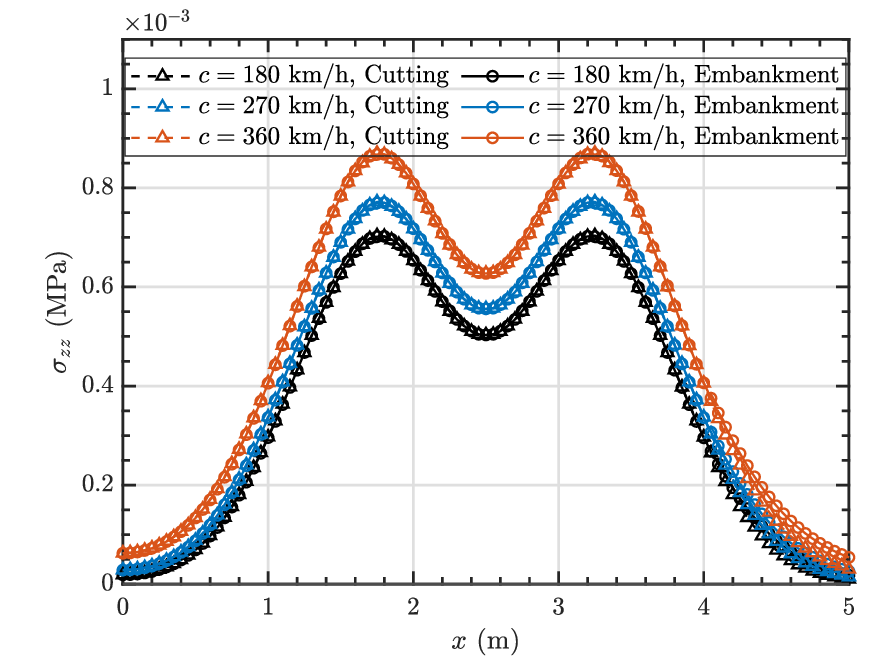}
    \caption{$\sigma_{zz}$}
    \label{fig:szz_y1m}
  \end{subfigure}
  \caption{Distributions of the normal stress components at depth $y=1~\mathrm{m}$ along the transverse coordinate $x$ for the cutting and embankment track sections under different train speeds ($c=180$, $270$, and $360~\mathrm{km/h}$).}
  \label{fig:normal_stress_y1m}
\end{figure}

In \fref{fig:normal_stress_y1m}, the dual-rail excitation produces two stress peaks aligned with the rail loads. Increasing $c$ amplifies the responses without materially shifting the peaks, while the cutting and embankment sections remain close at $y=1~\mathrm{m}$. The dominant component is $\sigma_{yy}$, followed by $\sigma_{zz}$ and $\sigma_{xx}$.

\begin{figure}[!ht]
\centering
\makebox[\linewidth][c]{%
\subcaptionbox{$U_y$ (Cutting)\label{fig:Uy_cut}}{%
\includegraphics[height=1.4in,keepaspectratio]{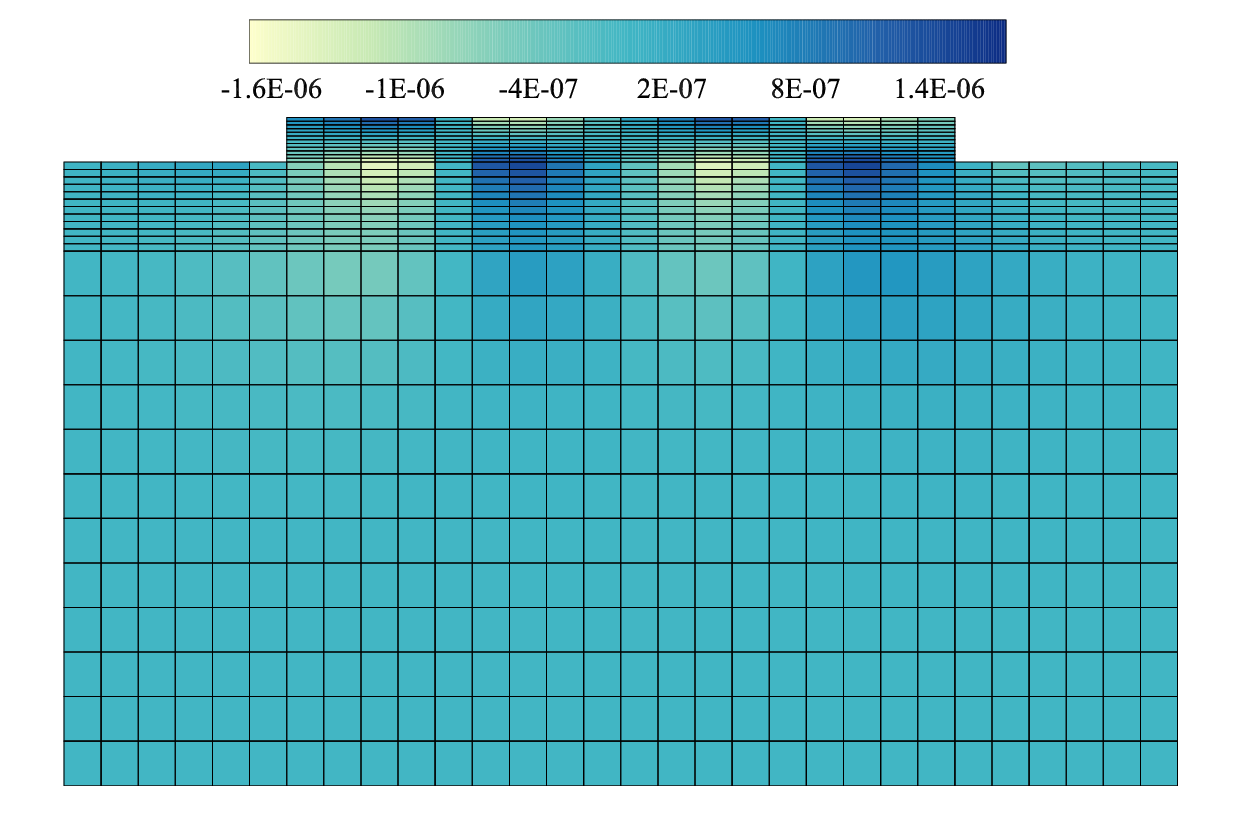}}%
\qquad
\subcaptionbox{$U_y$ (Embankment)\label{fig:Uy_emb}}{%
\includegraphics[height=1.4in,keepaspectratio]{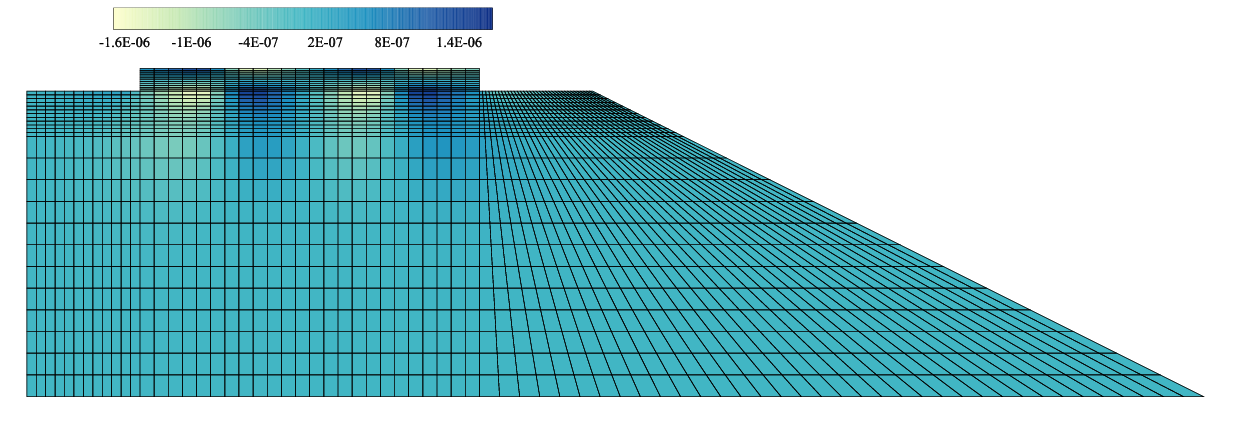}}%
}
\par\medskip
\makebox[\linewidth][c]{%
\subcaptionbox{$V_y$ (Cutting)\label{fig:Vy_cut}}{%
\includegraphics[height=1.4in,keepaspectratio]{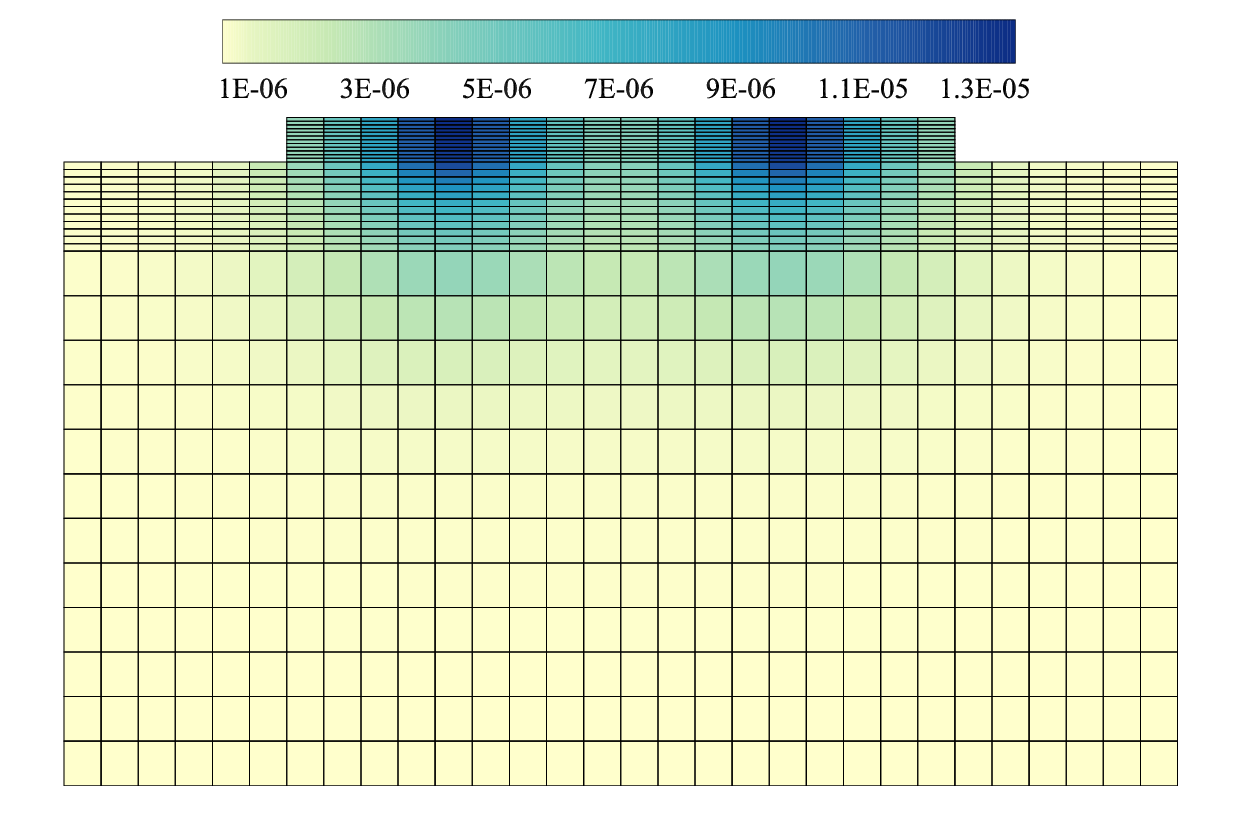}}%
\qquad
\subcaptionbox{$V_y$ (Embankment)\label{fig:Vy_emb}}{%
\includegraphics[height=1.4in,keepaspectratio]{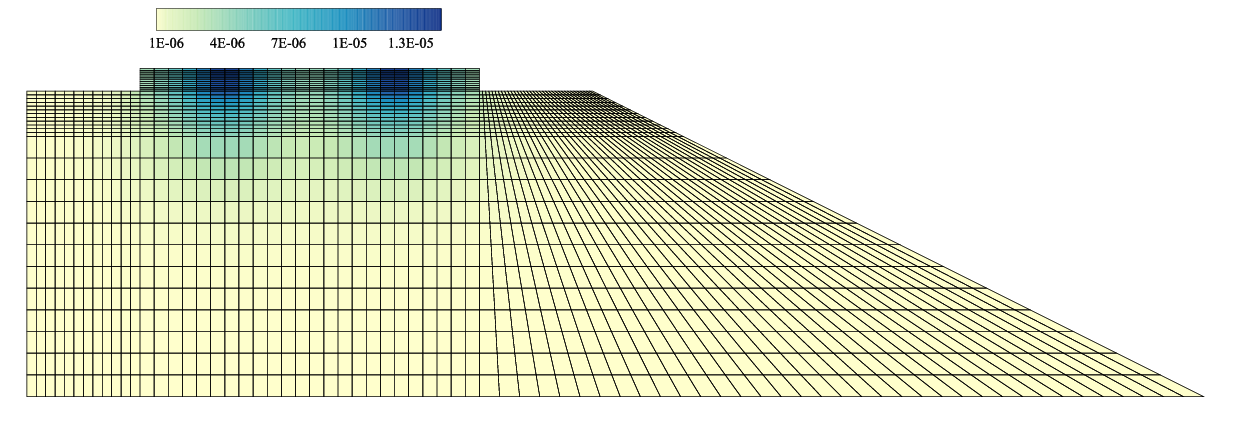}}%
}
\par\medskip
\makebox[\linewidth][c]{%
\subcaptionbox{$W_y$ (Cutting)\label{fig:Wy_cut}}{%
\includegraphics[height=1.4in,keepaspectratio]{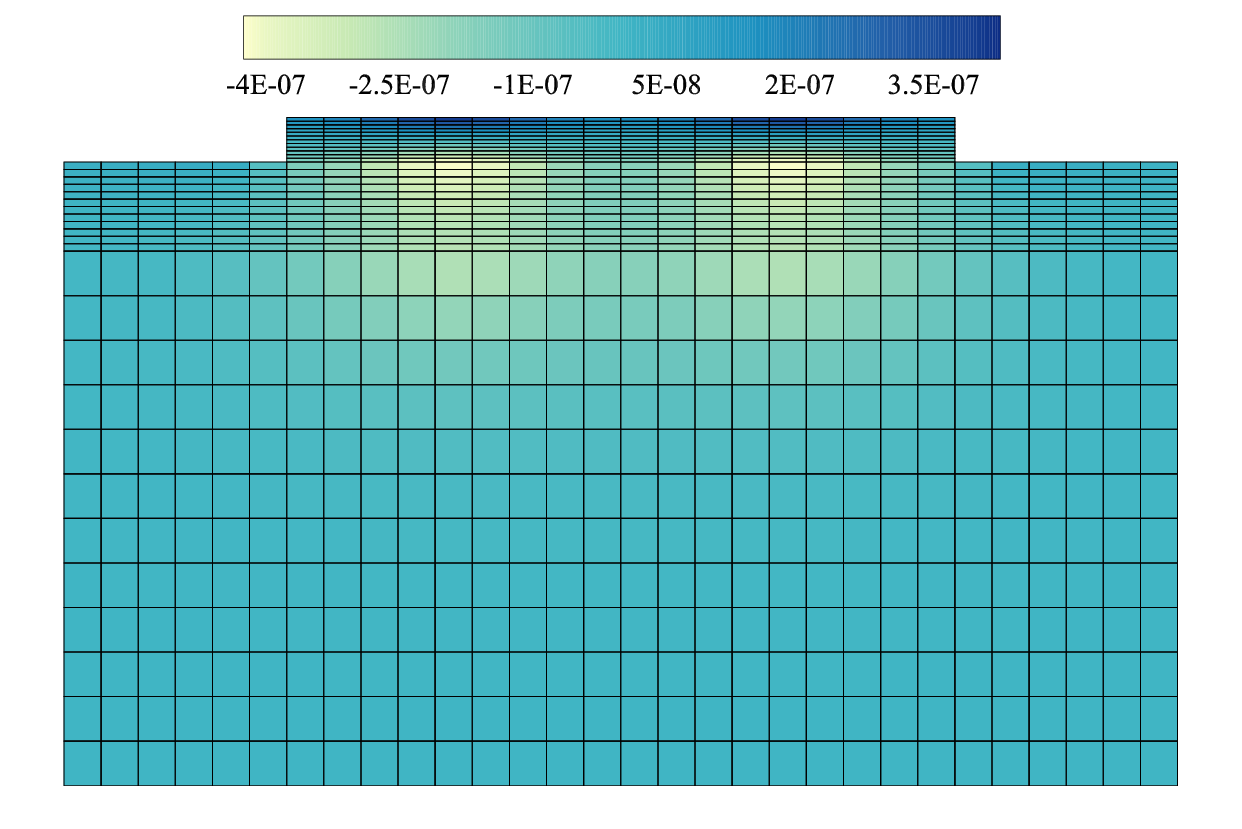}}%
\qquad
\subcaptionbox{$W_y$ (Embankment)\label{fig:Wy_emb}}{%
\includegraphics[height=1.4in,keepaspectratio]{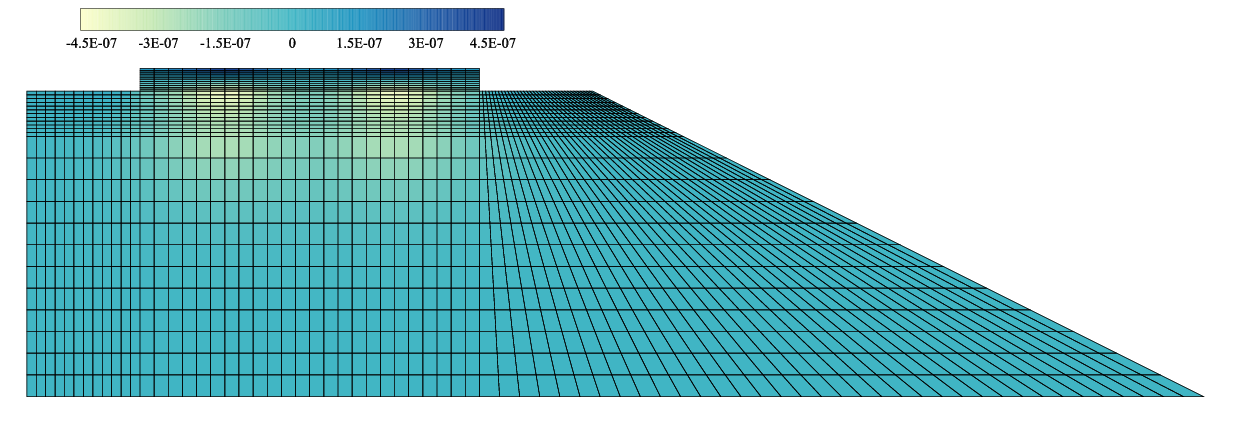}}%
}
\caption{Real-part contour plots of the displacement responses $(U_y,V_y,W_y)$ induced by a $y$-directed moving line load for the cutting (left column) and embankment (right column) configurations at $c=270~\mathrm{km/h}$, $f=50~\mathrm{Hz}$, and $f_0=0$.}
\label{fig:disp_contours}
\end{figure}

\begin{figure}[!htb]
\centering
\makebox[\linewidth][c]{%
\subcaptionbox{$\sigma_{xx}$ (Cutting)\label{fig:sxx_cut}}{%
  \includegraphics[height=1.15in,keepaspectratio]{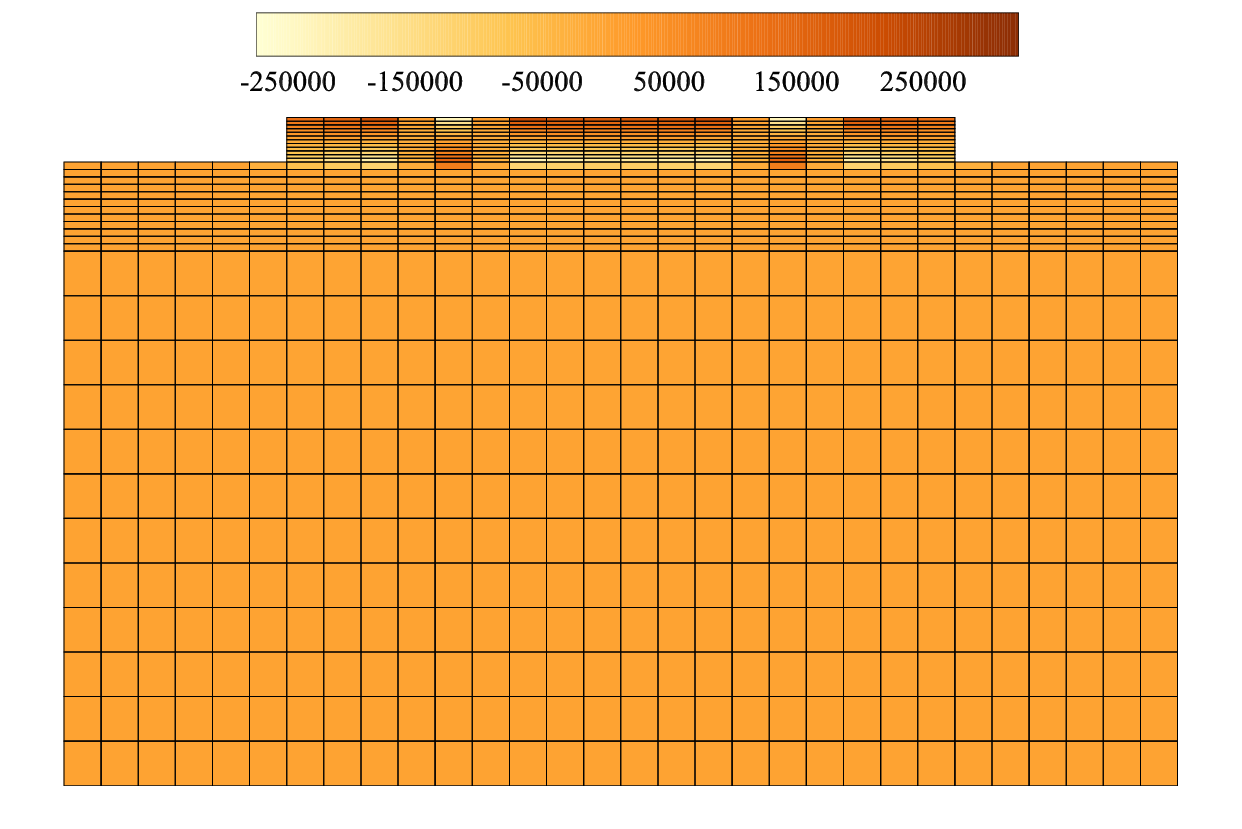}}%
\qquad
\subcaptionbox{$\sigma_{xx}$ (Embankment)\label{fig:sxx_emb}}{%
  \includegraphics[height=1.15in,keepaspectratio]{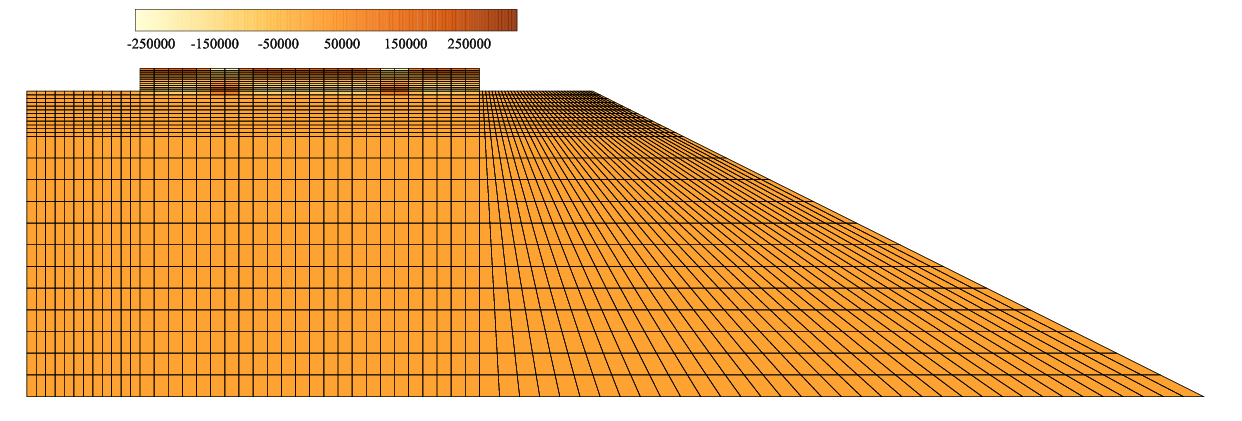}}%
}
\par\smallskip
\makebox[\linewidth][c]{%
\subcaptionbox{$\sigma_{yy}$ (Cutting)\label{fig:syy_cut}}{%
  \includegraphics[height=1.15in,keepaspectratio]{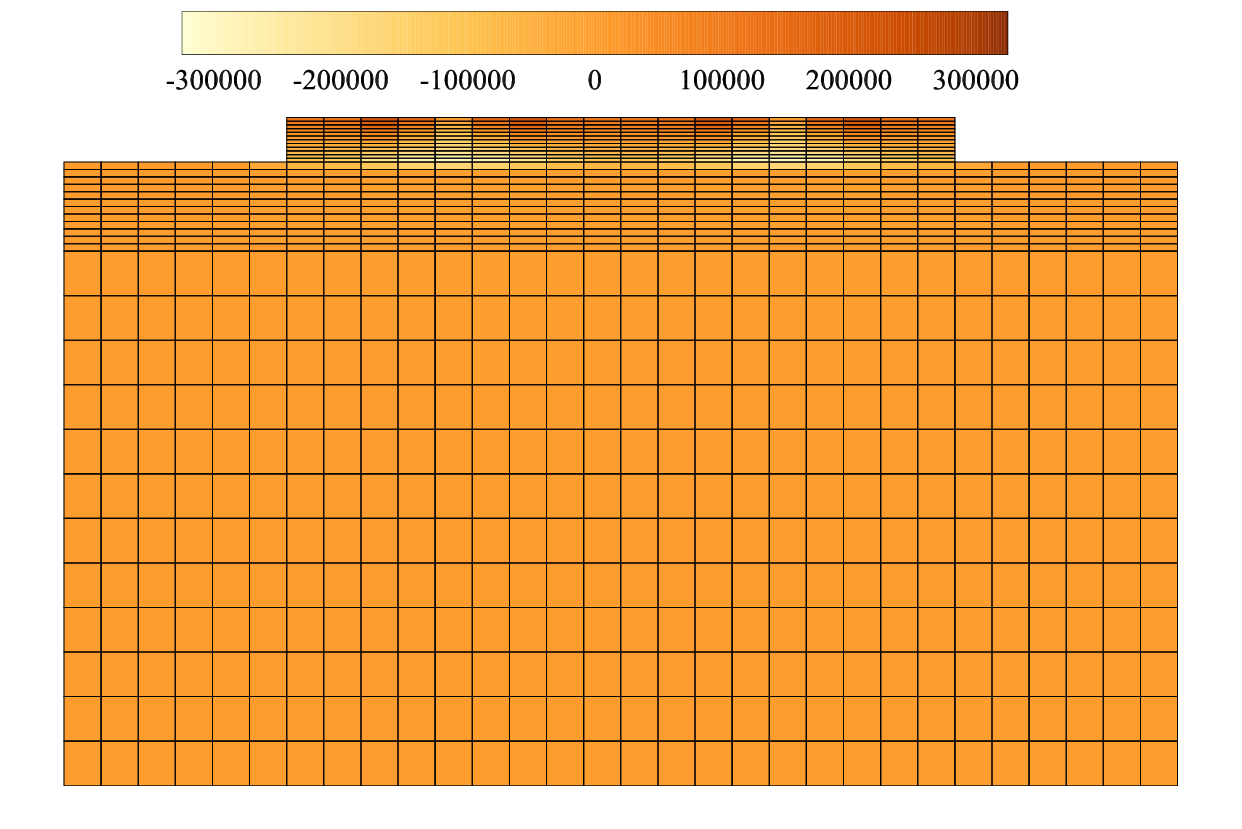}}%
\qquad
\subcaptionbox{$\sigma_{yy}$ (Embankment)\label{fig:syy_emb}}{%
  \includegraphics[height=1.15in,keepaspectratio]{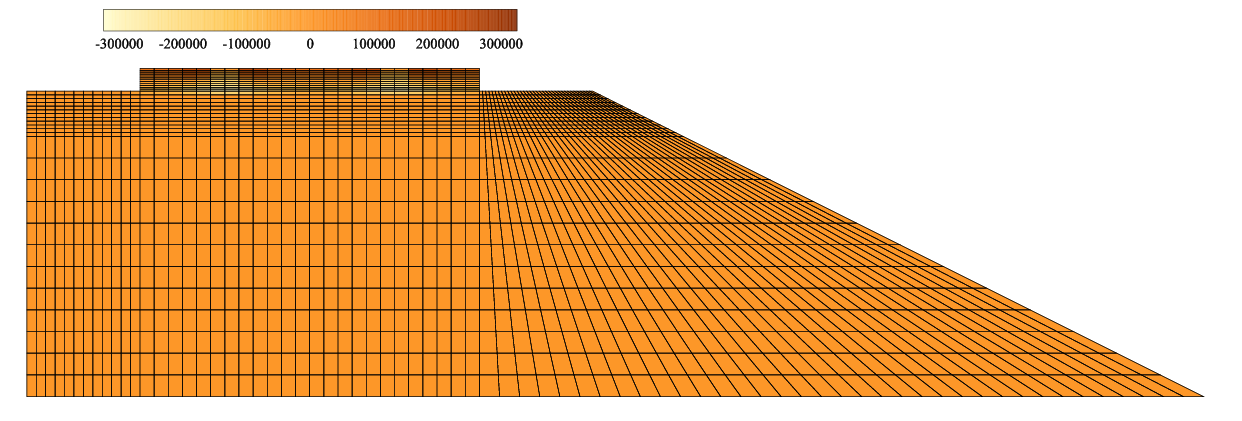}}%
}
\par\smallskip
\makebox[\linewidth][c]{%
\subcaptionbox{$\sigma_{zz}$ (Cutting)\label{fig:szz_cut}}{%
  \includegraphics[height=1.15in,keepaspectratio]{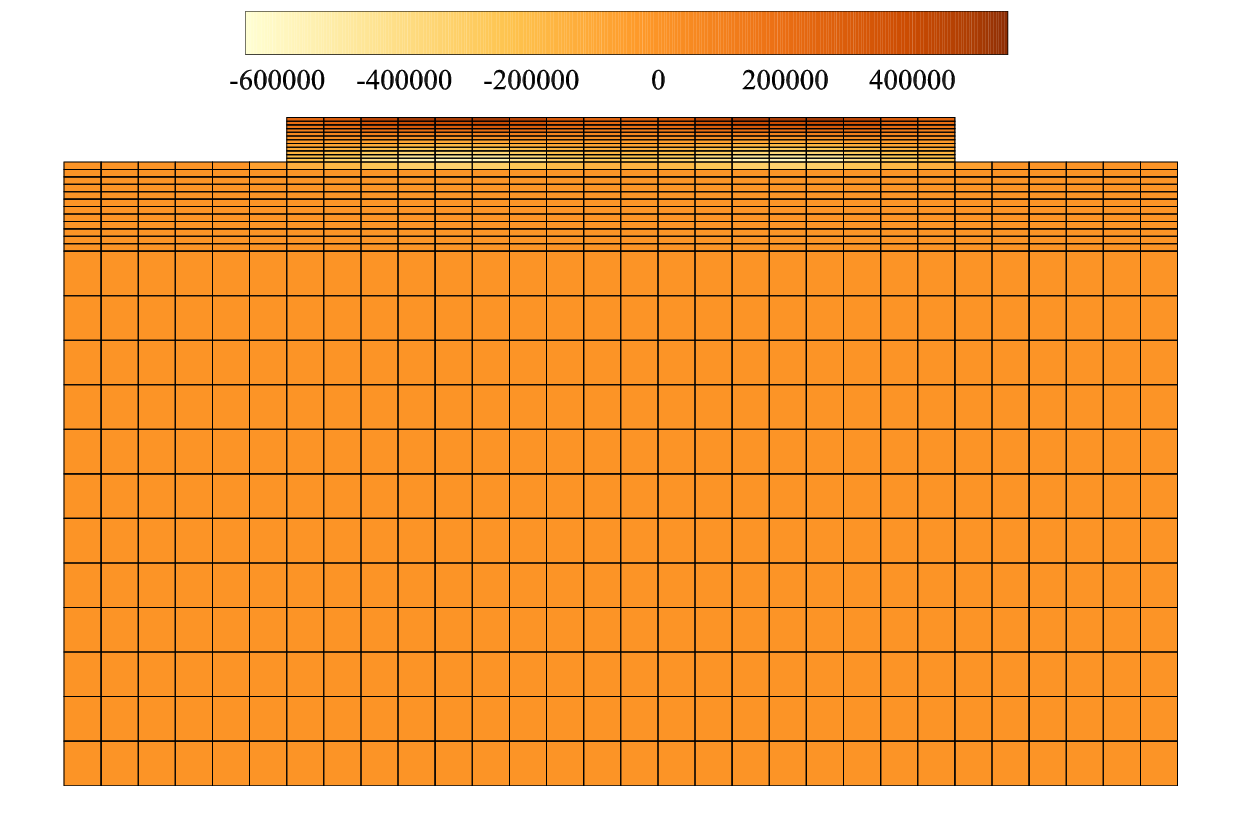}}%
\qquad
\subcaptionbox{$\sigma_{zz}$ (Embankment)\label{fig:szz_emb}}{%
  \includegraphics[height=1.15in,keepaspectratio]{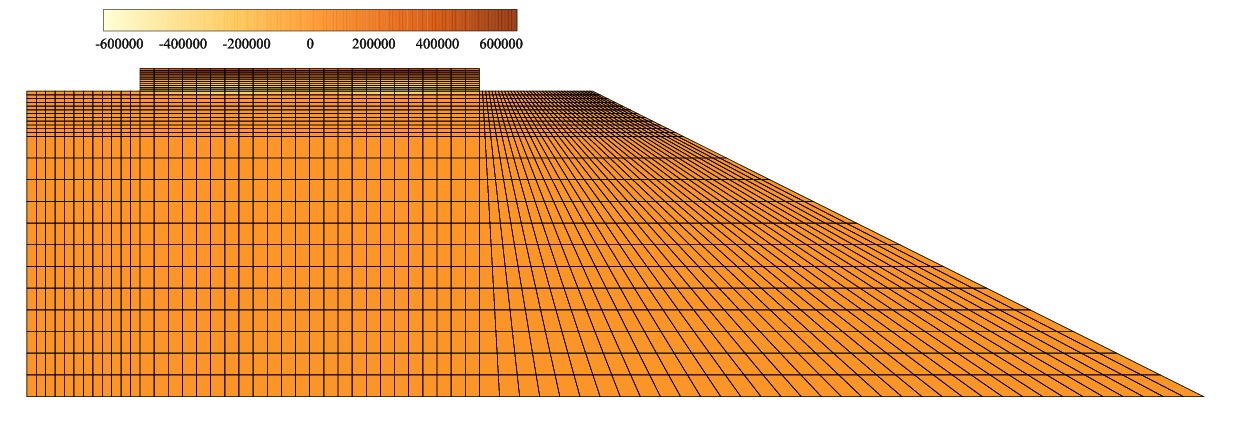}}%
}
\caption{Real-part contour plots of the normal stresses $(\sigma_{xx},\sigma_{yy},\sigma_{zz})$ induced by the $y$-directed moving line loads for the cutting (left column) and embankment (right column) trackbed configurations at $c=270~\mathrm{km/h}$, $f=50~\mathrm{Hz}$, and $f_0=0$. Stress values are reported in Pa.}
\label{fig:normal_stress_contours}
\end{figure}

Figs.~\ref{fig:disp_contours} and~\ref{fig:normal_stress_contours} show the in-phase displacement and normal-stress components for the representative $c=270~\mathrm{km/h}$ case ($f=50~\mathrm{Hz}$, $f_0=0$). These contours complement the amplitude profiles by showing the spatial redistribution with depth and near geometric or material transitions.

The contour plots show that the displacement field is primarily localized in the near-field region beneath the two rail loads, with a clear attenuation trend into the subgrade. Although the cutting and embankment cases exhibit similar overall patterns, subtle geometry-related differences occur near the slope transition and layer interfaces, where the wave paths are locally perturbed and the gradients become more concentrated. These full-field visualizations complement the one-dimensional profiles by illustrating how the speed-amplified response is redistributed spatially and where potential interface-sensitive regions may occur.

The stress contours provide a two-dimensional view of how the load-induced response redistributes with depth and across geometric transitions.
In all cases, $\sigma_{yy}$ exhibits the most pronounced concentration beneath the two rail lines and is the dominant compressive component transmitted into the subgrade. The component $\sigma_{zz}$ remains appreciable and follows a similar two-lobe footprint, whereas $\sigma_{xx}$ is comparatively smoother because of transverse constraint and redistribution.
Consistent with the one-dimensional comparisons, the embankment and cutting configurations show only minor differences in the near-track region, while geometry-related effects, when present, become more visible around the slope-transition zone where the stress field spreads and attenuates along the ground surface.

\subsection{\textit{Cover-depth-dependent wave attenuation in layered ground containing a buried tunnel}}
\label{tunnel_burial_depth}

\begin{figure}[!htbp]
  \centering
  \begin{minipage}[c]{0.46\textwidth}
    \centering
    \includegraphics[height=0.24\textheight,keepaspectratio]{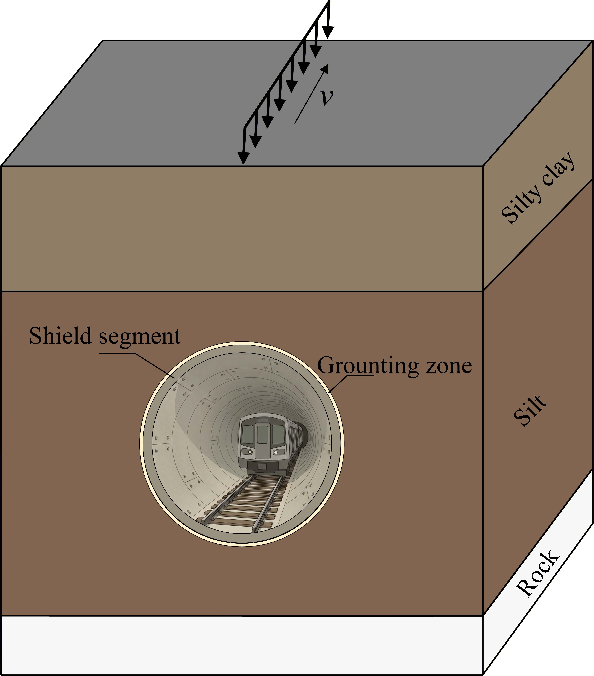}
    \subcaption{Schematic of the layered ground--tunnel configuration and the surface moving load.}
  \end{minipage}\hspace{4mm}
  \begin{minipage}[c]{0.46\textwidth}
    \centering
    \includegraphics[height=0.24\textheight,keepaspectratio]{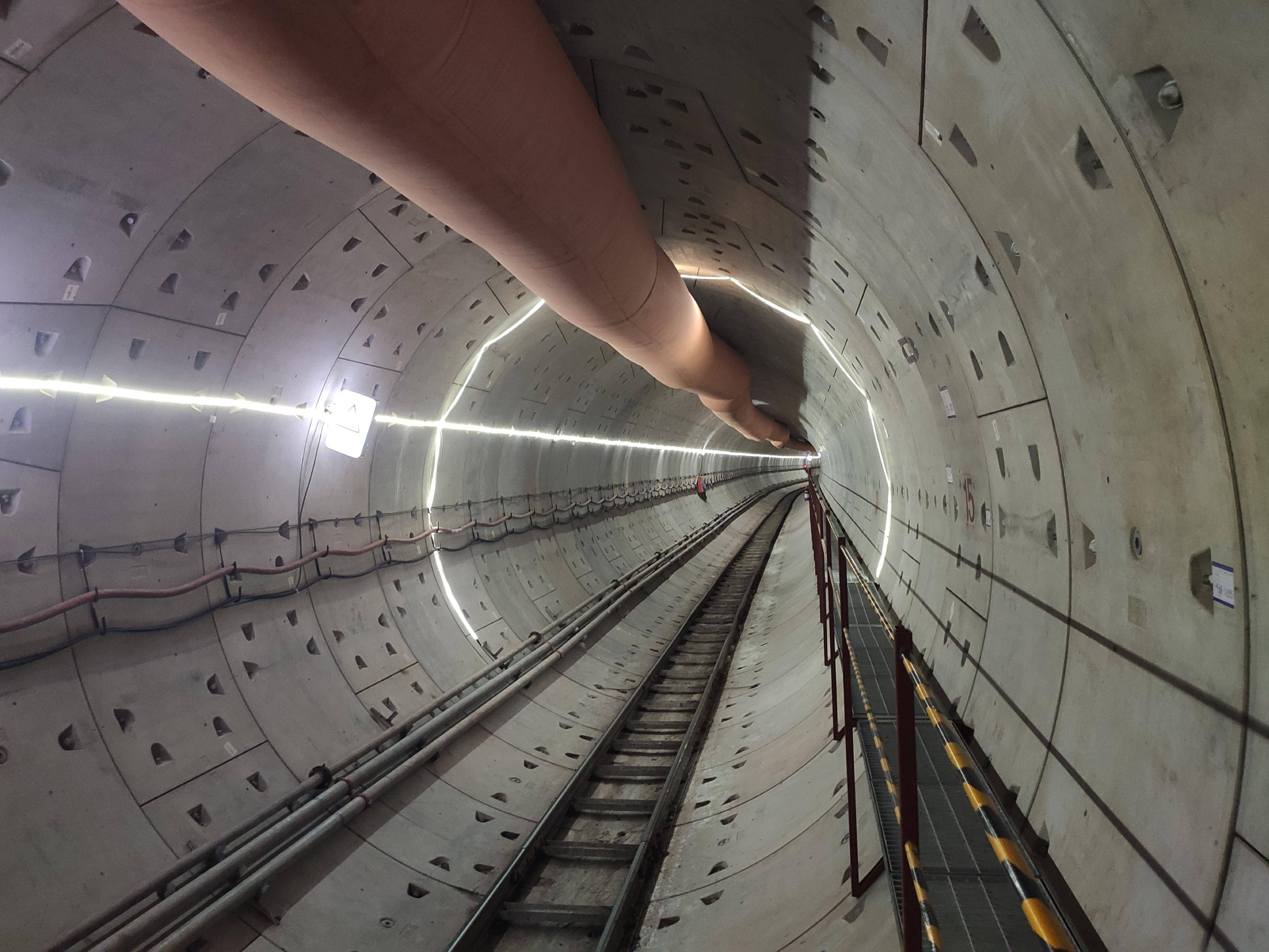}
    \subcaption{Metro tunnel photograph.}
  \end{minipage}
  \caption{Computational configuration of the metro tunnel example. The model consists of a segmental lining and an outer grouting zone embedded in a layered soil deposit, and is excited by a moving surface load applied above the tunnel axis.}
  \label{fig:tunnel_config}
\end{figure}

To quantify the influence of tunnel burial depth on load-induced wave attenuation, this example considers a circular metro tunnel embedded in layered ground and subjected to a vertical surface moving load applied directly above the tunnel axis (at $x=8~\mathrm{m}$ in the cross-sectional coordinate system). The tunnel configuration is based on Zhengzhou Metro Line~1. As illustrated in \fref{fig:tunnel_config}, the underground structure comprises a reinforced-concrete segmental lining surrounded by a grouting layer, while the geological profile is represented as a two-layer soil deposit consisting of silty clay over silt. The single-frequency response is evaluated for a nominal total surface-load amplitude \(P=150~\mathrm{kN}\), \(f=10~\mathrm{Hz}\), \(f_0=0\), \(c=35~\mathrm{m/s}\), and \(k=(\omega-\omega_0)/c\). In this setting, the burial depth controls the propagation and attenuation of vibration energy from the ground surface to the tunnel level. The loading and exterior-boundary construction are kept fixed to provide a consistent comparison of the cover-depth trends within the selected computational configuration. To characterize the computational requirements of the formulation in an engineering-scale setting, the wall-clock cost of each cover-depth configuration is also evaluated for the same frequency--wavenumber pair.

\fref{fig:tunnel_config} summarizes the ground--tunnel configuration adopted for the burial-depth study. The bounded near-field domain is 15 m wide in the transverse direction and 25 m deep, and the tunnel axis is located 8 m from the left boundary so that the surface moving load can be applied directly above the tunnel centerline at the same transverse location. The near-field depth keeps the grouting-zone invert inside the bounded near field before the lower infinite boundary is attached; in the deepest case, the grouting-zone invert is located at \(h_e+D_{\mathrm{grout}}=19.6~\mathrm{m}\), leaving a 5.4 m clearance to the lower near-field boundary. The ground profile, based on Zhengzhou Metro Line 1, consists of an upper silty-clay stratum with a fixed thickness of 6.8 m overlying a silt stratum, and the tunnel is embedded in the silt layer. The burial depth is characterized by the crown cover depth, defined as the vertical distance from the ground surface to the tunnel crown. Four burial-depth cases are considered, with crown cover depths of 8 m, 10 m, 12 m, and 14 m, which form the basis of the subsequent comparisons. The underground structure is modeled using a reinforced concrete segmental lining surrounded by a grouting zone. The lining inner diameter is taken as 5.4 m, the lining outer diameter as 5.5 m, and the grouting-zone diameter is \(D_{\mathrm{grout}}=5.6~\mathrm{m}\). 

\begin{figure}[!ht]
  \centering
  \begin{subfigure}[t]{0.40\textwidth}
    \centering
    \includegraphics[width=\linewidth]{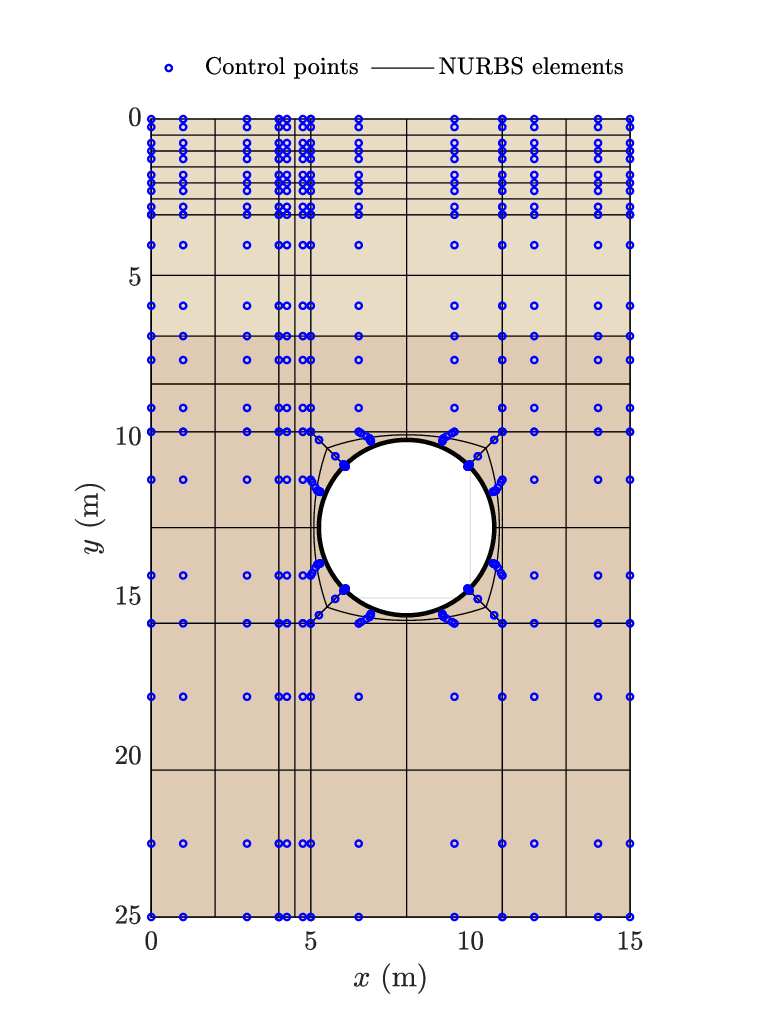}
    \caption{}
  \end{subfigure}\hspace{3mm}
  \begin{subfigure}[t]{0.40\textwidth}
    \centering
    \includegraphics[width=\linewidth]{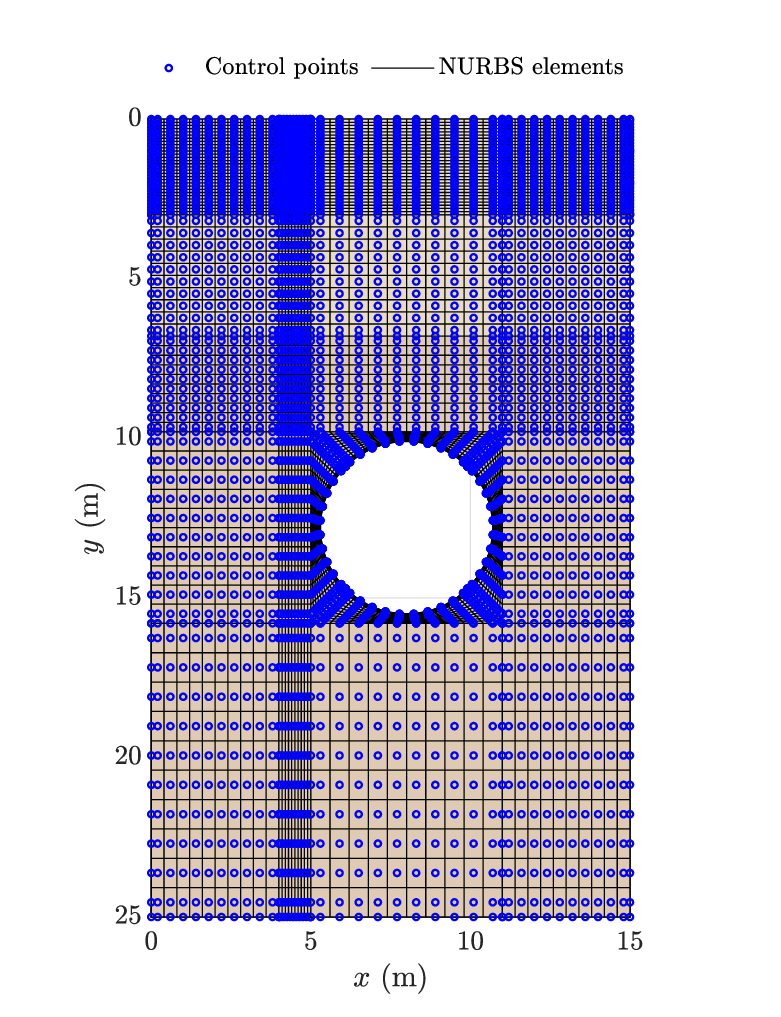}
    \caption{}
  \end{subfigure}
  \caption{IGA discretizations for the tunnel example. Blue dots denote control points and black lines denote NURBS elements. (a) Baseline multi-patch mesh used to represent the circular tunnel boundary and the surrounding ground. (b) Locally $h$-refined mesh with increased resolution around the tunnel lining and in the near-surface region along the load path to capture strong displacement/stress gradients.}
  \label{fig:tunnel_mesh}
\end{figure}

The corresponding IGA discretization is shown in \fref{fig:tunnel_mesh}, where a multi-patch construction is used to represent the circular interfaces accurately and to maintain a conforming connection between the lining, the grouting zone, and the surrounding ground. Local $h$-refinement is introduced in the near-field region, particularly along the surface load path and around the tunnel boundary, to resolve the strong displacement gradients and stress concentration induced by the moving excitation. Away from the tunnel, a coarser discretization is adopted to control the total degrees of freedom. The material parameters are summarized in Table~\ref{tab:tunnel_materials}, covering the silty-clay and silt strata, the shield segments, and the grouting region. Unless otherwise stated, the same viscoelastic damping setting as in the previous examples is used so that the parametric comparisons remain consistent across all cases.

\begin{table}[H]
\centering
\caption{Material properties adopted in the tunnel example.}
\label{tab:tunnel_materials}
\begin{tabular}{lcccc}
\hline
Layer/Region & $E$ (MPa) & $\rho$ (kg/m$^{3}$) & $\nu$ (-) & $\zeta$ (-) \\
\hline
Silty clay       & 23    & 2143 & 0.30 & 0.05 \\
Silt             & 17    & 1959 & 0.30 & 0.05 \\
Shield segment   & 30000 & 2551 & 0.20 & 0.05 \\
Grouting zone    & 500   & 2245 & 0.30 & 0.05 \\
\hline
\end{tabular}
\end{table}

Here $x$ denotes the transverse coordinate in the cross-section, and the tunnel axis is located at $x=8~\mathrm{m}$ so that the surface line load is applied directly above the tunnel centerline.
The crown cover depth is denoted by $h_e$ and is defined as the vertical distance from the ground surface to the tunnel crown. To examine the attenuation of the surface-load-induced response in the soil above the tunnel, response profiles are extracted at several elevations above the crown, quantified by $\Delta y$.
Specifically, a profile with a given $\Delta y$ corresponds to the horizontal line located at the depth $y=h_e-\Delta y$ in the adopted coordinate system (positive $y$ downward).

\begin{figure}[!ht]
    \centering
    \includegraphics[width=0.98\linewidth]{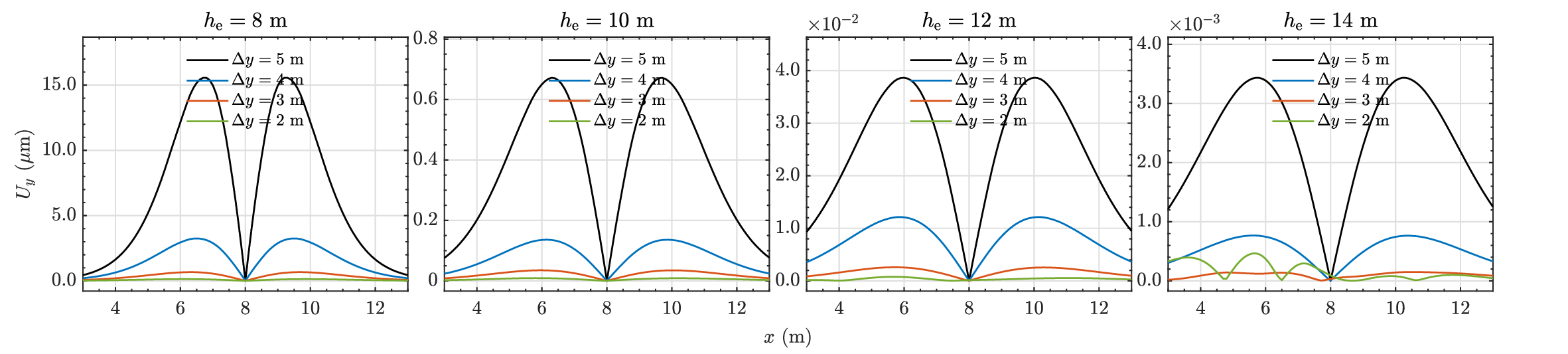}
\caption{Transverse distributions of the displacement component $U_y$ induced by a vertical surface line load, evaluated along $x$ at four elevations above the tunnel crown ($\Delta y=2$--$5~\mathrm{m}$) for different crown cover depths $h_e=8$, $10$, $12$, and $14~\mathrm{m}$. The load is applied at $x=8~\mathrm{m}$, vertically above the tunnel axis, with \(P=150~\mathrm{kN}\), $c=35~\mathrm{m/s}$, $f=10~\mathrm{Hz}$, and $f_0=0$.}
\label{fig:tunnel_Uy_Deltay_profiles}
\end{figure}

\begin{figure}[!ht]
    \centering
    \includegraphics[width=0.98\linewidth]{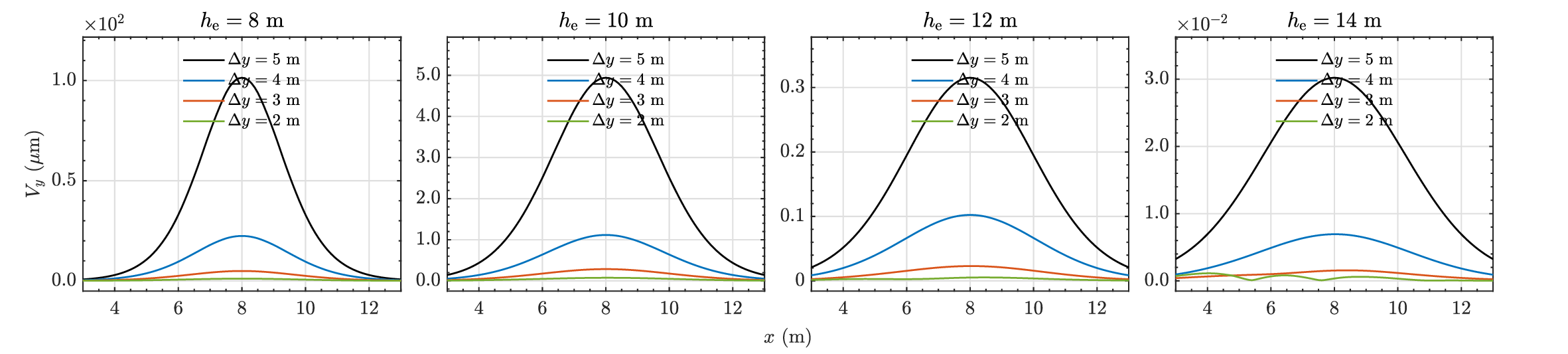}
\caption{Transverse distributions of the vertical displacement component $V_y$ induced by a vertical surface line load, evaluated along $x$ at four elevations above the tunnel crown ($\Delta y=2$--$5~\mathrm{m}$) for different crown cover depths $h_e=8$, $10$, $12$, and $14~\mathrm{m}$. The peak is centered at the loading position $x=8~\mathrm{m}$ and attenuates as $h_e$ increases. The loading setting is \(P=150~\mathrm{kN}\), $c=35~\mathrm{m/s}$, $f=10~\mathrm{Hz}$, and $f_0=0$.}
\label{fig:tunnel_Vy_Deltay_profiles}
\end{figure}

Figs.~\ref{fig:tunnel_Uy_Deltay_profiles} and~\ref{fig:tunnel_Vy_Deltay_profiles} show the transverse distributions of $U_y$ and $V_y$ along $x$ at four elevations above the tunnel crown, $\Delta y=2$, $3$, $4$, and $5~\mathrm{m}$, for different crown cover depths $h_e=8$, $10$, $12$, and $14~\mathrm{m}$.
For all cases, the responses are concentrated around the loading position ($x=8~\mathrm{m}$), and the amplitude decays rapidly away from the load, indicating a localized near-field influence in the transverse direction. As $h_e$ increases, both $U_y$ and $V_y$ decrease systematically at all considered $\Delta y$ levels, and the curves become progressively closer to the zero baseline. This trend indicates that a larger cover depth promotes stronger attenuation before the vibration field reaches the region above the tunnel crown.
The response increases with $\Delta y$: larger $\Delta y$ values correspond to shallower monitoring lines closer to the surface load and therefore produce larger displacement amplitudes and broader influence zones. Monitoring lines closer to the tunnel crown, corresponding to smaller $\Delta y$, exhibit weaker responses because of the longer propagation distance from the surface excitation.

From an engineering perspective, these trends indicate that increasing the crown cover depth reduces both the intensity and the spatial extent of the displacement disturbance reaching the crown-adjacent soil region. For the deeper configurations (e.g., $h_e=12$--$14~\mathrm{m}$), the responses extracted above the crown become markedly smaller than those for the shallow case ($h_e=8~\mathrm{m}$), and the incremental reduction from $h_e=12$ to $14~\mathrm{m}$ appears more limited, suggesting a tendency toward convergence in the attenuation effectiveness.
Under the selected material, loading, and geometric conditions, these trends indicate that increasing cover depth can reduce the vibration disturbance transmitted to the ground above the tunnel, with a smaller incremental change for the deeper cases.

\begin{figure}[!ht]
    \centering
    \includegraphics[width=0.98\linewidth]{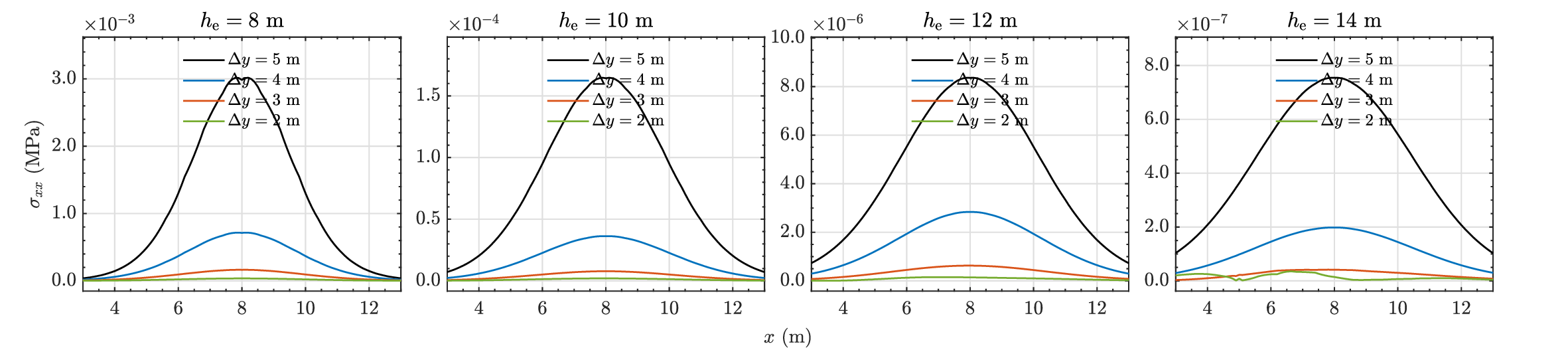}
\caption{Distributions of the normal stress $\sigma_{xx}$ along the transverse coordinate $x$ at monitoring lines located $\Delta y=2$--$5~\mathrm{m}$ above the tunnel crown, for four cover depths $h_e=8$, $10$, $12$, and $14~\mathrm{m}$ under \(P=150~\mathrm{kN}\), $c=35~\mathrm{m/s}$, $f=10~\mathrm{Hz}$, and $f_0=0$.}
\label{fig:sigmaxx_above_crown}
\end{figure}

\begin{figure}[!ht]
    \centering
    \includegraphics[width=0.98\linewidth]{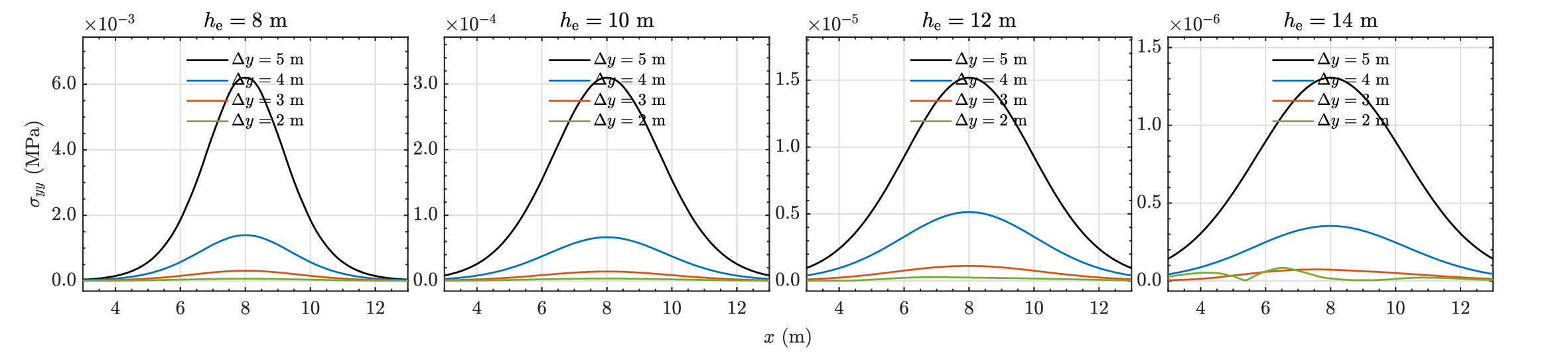}
\caption{Distributions of the normal stress $\sigma_{yy}$ along the transverse coordinate $x$ at monitoring lines located $\Delta y=2$--$5~\mathrm{m}$ above the tunnel crown, for four cover depths $h_e=8$, $10$, $12$, and $14~\mathrm{m}$ under \(P=150~\mathrm{kN}\), $c=35~\mathrm{m/s}$, $f=10~\mathrm{Hz}$, and $f_0=0$.}
\label{fig:sigmayy_above_crown}
\end{figure}

Figs.~\ref{fig:sigmaxx_above_crown} and~\ref{fig:sigmayy_above_crown} report the distributions of the normal stresses $\sigma_{xx}$ and $\sigma_{yy}$ along the transverse coordinate $x$ at a set of monitoring lines located $\Delta y=2$--$5~\mathrm{m}$ above the tunnel crown, for four cover depths $h_e=8$, $10$, $12$, and $14~\mathrm{m}$. Here $\Delta y$ denotes the vertical distance from the tunnel crown to the monitoring line, so a larger $\Delta y$ corresponds to a shallower observation line closer to the ground surface. For each $h_e$, both stress components exhibit a single dominant peak centered at $x\approx 8~\mathrm{m}$, which coincides with the surface load position directly above the tunnel axis. The peak location remains essentially unchanged across all $h_e$ and $\Delta y$, indicating that, within the present parameter range, the geometry primarily modulates the response magnitude and the lateral spreading rather than shifting the stress concentration zone laterally.

The curve ordering shows a systematic influence of $\Delta y$. For all $h_e$, the stress amplitude increases monotonically with $\Delta y$, with the $\Delta y=5~\mathrm{m}$ curve consistently being the largest and the $\Delta y=2~\mathrm{m}$ curve the smallest. In addition to the peak height, the peak width also increases as $\Delta y$ increases, and the post-peak tails decay more slowly, implying a broader lateral influence range at shallower depths. This behavior is consistent with depth-dependent attenuation and load-spreading in layered media. At smaller $\Delta y$ the observation line is closer to the tunnel and deeper in the soil deposit, whereas at larger $\Delta y$ the monitoring line approaches the near-surface region where the stress path is shorter and the wave field has undergone less geometric attenuation, leading to higher stress levels and a wider footprint.

The cover depth $h_e$ has a stronger effect on amplitude. From $h_e=8~\mathrm{m}$ to $10~\mathrm{m}$ the peak level decreases markedly, and further increases to $h_e=12$ and $14~\mathrm{m}$ drive the stress curves into progressively smaller orders of magnitude, as reflected by the changing axis scales in the subpanels. Meanwhile, the overall curve shapes remain similar, suggesting that increasing $h_e$ mainly enhances the attenuation between the surface excitation and the crown-adjacent region, rather than altering the fundamental load-transfer pattern. Practically, this indicates a rapid reduction of the computed stress response in the soil mass above the tunnel once the cover depth becomes sufficiently large, after which the marginal benefit of further burial becomes limited.

Comparing the two stress components, $\sigma_{yy}$ is consistently larger than $\sigma_{xx}$ for all $h_e$ and $\Delta y$, and its peak is sharper with steeper flanks. In contrast, $\sigma_{xx}$ shows a flatter crest and smoother transitions. $\sigma_{yy}$ is more directly aligned with the dominant vertical transmission path of the surface load and therefore retains stronger localization. $\sigma_{xx}$ is governed by transverse confinement and lateral redistribution, which act as a spatial smoothing mechanism and weaken the peak effect. Together, these patterns indicate that the crown-above soil region is primarily driven by vertical compression demand, while transverse normal stress plays a secondary role and is less sensitive to local concentration.

Under the present ground profile and structural configuration, the dynamic stress disturbance above the tunnel becomes substantially weaker once $h_e$ reaches about $12~\mathrm{m}$, because both $\sigma_{xx}$ and $\sigma_{yy}$ at $\Delta y=2$--$5~\mathrm{m}$ above the crown have already dropped to very small levels and the curves become less distinguishable. This trend should be interpreted as a model-based indication of diminishing cover-depth sensitivity for the selected stratified ground rather than as a general design threshold.

{
All four cover-depth configurations employ the same topology, comprising 624 NURBS elements, 1040 global control points, and 3120 total displacement degrees of freedom. Prescribing $u_x=0$ at 36 control points along $x=0$ leaves a $3084\times3084$ reduced complex system with 3084 free unknowns and 185,616 nonzero entries; the artificial boundary contains 56 trace control points over 44 infinite-element segments. For each configuration, one warm-up run was followed by nine repeated runs, with all matrices reassembled and the system re-solved without reusing matrices or factorizations. Table~\ref{tab:tunnel_computational_cost} reports the median near-field assembly ($t_{\mathrm{near}}$), infinite-element assembly including the closed-form radial moments ($t_{\mathrm{far}}$), sparse solution ($t_{\mathrm{solve}}$), and response-extraction ($t_{\mathrm{post}}$) times. Here $t_{\mathrm{core}}=t_{\mathrm{near}}+t_{\mathrm{far}}+t_{\mathrm{solve}}$ and $t_{\mathrm{total}}=t_{\mathrm{core}}+t_{\mathrm{post}}$. Geometry and input-file loading, plotting, console output, result-file writing, and software startup are excluded.
}

\begin{table}[H]
\centering

\small
\caption{Computational cost of the tunnel models for one frequency--wavenumber pair at $f=10$~Hz. The reported wall-clock times are medians of nine repeated runs following one warm-up run.}
\label{tab:tunnel_computational_cost}
\setlength{\tabcolsep}{4pt}
\begin{tabular}{ccccccc}
\hline
$h_e$ &
$t_{\mathrm{near}}$ &
$t_{\mathrm{far}}$ &
$t_{\mathrm{solve}}$ &
$t_{\mathrm{core}}$ &
$t_{\mathrm{post}}$ &
$t_{\mathrm{total}}$ \\
(m) & (s) & (s) & (s) & (s) & (s) & (s) \\
\hline
8  & 1.571 & 1.554 & 0.388 & 3.497 & 19.033 & 22.483 \\
10 & 1.576 & 1.548 & 0.330 & 3.425 & 14.708 & 18.133 \\
12 & 1.590 & 1.559 & 0.339 & 3.472 & 15.200 & 18.672 \\
14 & 1.855 & 1.636 & 0.433 & 3.947 & 16.489 & 20.438 \\
\hline
\end{tabular}
\par\smallskip
\footnotesize
\noindent Each reported entry is the median of the corresponding per-run quantity; therefore, the displayed component medians do not necessarily sum exactly to the reported median totals.
\end{table}

{
The median core analysis time is 3.425--3.947~s per frequency--wavenumber pair. Near-field and far-field assembly have comparable costs, while the reduced-system solution requires 0.330--0.433~s. Including response extraction, the median total time is 18.133--22.483~s. The response-extraction stage is dominated by receiver localization over the multi-patch geometry and depends on the requested receiver set, so it is reported separately from the core cost. Because the four models share the same topology and system dimension, these timings are interpreted as a representative computational range rather than as a cover-depth effect.
}

\section{\texorpdfstring{Discussion}{Discussion}}
\label{Discussion}

The numerical results indicate that NBIEM is an efficient boundary-attached exterior formulation for repeated frequency-domain analyses of wave propagation in semi-infinite ground. Under matched physical cross-sectional partitions, the different approximation spaces lead to different global numbers of degrees of freedom, and these differences are retained as part of the performance assessment. From the intermediate-frequency cases onward, the proposed formulation provides a favorable balance among displacement accuracy, phase accuracy, and computational cost. Relative to IGA/SBIGA, the semi-infinite exterior is represented directly through far-field stiffness and mass contributions assembled on the shared NURBS trace, without constructing a separate scaled-boundary dynamic-stiffness system. The engineering-scale tunnel examples further show that the proposed formulation remains computationally practical for layered, multi-patch configurations with curved interfaces, complementing the controlled relative efficiency comparisons with an absolute-cost assessment under more complex geometric and material conditions.

The formulation also differs from other exterior treatments used in geotechnical wave analysis. PML formulations typically represent the exterior through an absorbing layer and coordinate stretching, whereas FE--BE formulations describe radiation through problem-specific fundamental solutions and boundary-integral evaluation. In NBIEM, the exterior response is represented through outgoing or evanescent radial functions attached to the artificial boundary, and the associated radial moments are evaluated in closed form. The resulting construction preserves the geometric representation and approximation continuity of the bounded IGA domain while maintaining a compact exterior discretization.

For engineering applications, the all-S realization provides a consistent and explicitly defined default setting. Its decay parameter is determined from the exterior material properties and the declared geometric scale of each boundary segment, without frequency-by-frequency optimization or access to an analytical reference response. The indicators $I_{\beta}$ and $I_D$ complement this setting by quantifying the response sensitivity to the radial decay parameter and to the artificial-boundary location. The calibrated mixed-wave realization further demonstrates, in the benchmark setting, how additional wave information can improve low-frequency agreement.

The engineering examples use controlled single-frequency configurations to examine the influence of material layering, structural geometry, moving-load speed, treatment depth, and tunnel cover depth under clearly defined conditions. The multi-patch construction accommodates material interfaces and curved internal boundaries while retaining a common NURBS-based representation. These examples demonstrate how the proposed exterior treatment can be integrated into layered-ground, track--subgrade, and buried-structure models without changing the basic near--far-field coupling procedure.

{
Within this scope, the formulation is particularly suited to geotechnical systems whose cross-sectional geometry and material properties are invariant, or can be reasonably idealized as invariant, along the traveling direction, including layered or heterogeneous cross-sections and curved interfaces. The present implementation adopts small-strain linear viscoelastic behavior and resolves moving harmonic excitation through frequency--wavenumber-domain components; multi-harmonic linear loading can be represented by superposition. Extensions incorporating localized three-dimensional variations, nonlinear constitutive response, and nonlinear interface interaction would broaden the framework toward a still wider range of engineering conditions.
}

\section{Conclusions}
\label{Conclusions}

This study developed a three-component 2.5D formulation that couples a bounded isogeometric near field to NURBS-trace infinite elements for moving-load wave propagation and dynamic soil--structure interaction in semi-infinite geotechnical media. The bounded isogeometric domain and the semi-infinite exterior share the same discrete NURBS trace and boundary control-point unknowns, providing a direct and geometrically consistent near--far-field coupling. For the selected outgoing or evanescent exponential radial functions, the radial moments entering the far-field stiffness and mass matrices are evaluated in closed form, without introducing a finite radial truncation or radial numerical quadrature.

Analytical half-space benchmarks confirm that the formulation reproduces displacement and stress responses of a viscoelastic half-space over different moving-load speed regimes. The numerical studies characterize the working behavior of the exterior approximation over different frequencies, radial settings, and artificial-boundary configurations. The sensitivity indicators $I_{\beta}$ and $I_D$ provide practical diagnostics for assessing the selected exterior setting, while the method comparisons demonstrate the computational potential of the formulation for repeated frequency-domain analyses.

The engineering examples further demonstrate the applicability of the method to layered ground, track--subgrade systems, and buried structures involving material interfaces, compatible multi-patch discretizations, and curved boundaries. The demonstrated scope covers cross-sections that are longitudinally invariant or reasonably represented as such, with small-strain linear viscoelastic behavior and harmonic frequency--wavenumber components; multi-harmonic linear excitation follows by superposition. Overall, the results identify NBIEM as an efficient and geometry-consistent computational framework for 2.5D frequency-domain analyses of wave propagation and dynamic soil--structure interaction in semi-infinite ground.

\section*{Acknowledgements}
This research was supported by the National Natural Science Foundation of China (Grant Nos. 52478477 and 52578540), the China Postdoctoral Science Foundation (Grant No. 2024M752937), the Henan Provincial Science and Technology Research Project (Grant No. 262102221015), Zhengzhou University Young Student Basic Research Projects (PhD students) (Grant No. ZDBJ2026065), the Henan Provincial Natural Science Foundation (Grant No. 252600421827), and the Key Research Projects of Higher Education Institutions in Henan Province (Grant No. 25A560006).

\bibliographystyle{elsarticle-harv} 
\bibliography{myRef}

\end{document}